\documentclass[aps,physrev,groupedaddress,twocolumn,amsmath,amssymb]{revtex4-2}

\usepackage{amsmath}
\usepackage{centernot}
\usepackage{amsmath, amssymb}
\usepackage{amsthm}
\usepackage{mathrsfs}
\usepackage{braket,mleftright}
\usepackage{dsfont}
\usepackage{bbm}
\usepackage{float}
\usepackage{graphicx}% Include figure files
\usepackage{dcolumn}% Align table columns on a decimal point
\usepackage{bm}% bold math
\usepackage[colorlinks=true, urlcolor=blue, linkcolor=blue, citecolor=blue]{hyperref}% add hypertext capabilities
\usepackage{xcolor}
\usepackage{url}
\usepackage{lipsum}
\usepackage{mathtools}
\usepackage{easybmat}

\newcommand{\RNum}[1]{\expandafter{\romannumeral #1\relax}}

\begin{document}

\title{The Weighted Hartree-Fock-Bogoliubov method for interacting fermions: \\
An application to ultracold Fermi superfluids}

\author{Nikolai Kaschewski}
\affiliation{Physics Department and Research Center OPTIMAS, Rhineland-Palatinate Technical University Kaiserslautern-Landau, 67663 Kaiserslautern, Germany}

\author{Axel Pelster}
\affiliation{Physics Department and Research Center OPTIMAS, Rhineland-Palatinate Technical University Kaiserslautern-Landau, 67663 Kaiserslautern, Germany}

\author{Carlos A. R. {S\'a} de Melo}
\affiliation{School of Physics, Georgia Institute of Technology, Atlanta, 30332, USA}
\date{\today}

\begin{abstract}
For several decades it has been known that divergences arise in the ground-state energy and chemical potential of unitary superfluids, where the scattering length diverges, due to particle-hole scattering. Leading textbooks and research articles recognize that there are serious issues but ignore them due to the lack of an approach that can regularize these divergences.
We find a solution to this difficulty by proposing a general method, called the Weighted Hartree-Fock-Bogolibubov theory, to handle multiple decomposition channels 
originating from the same interaction.
We distribute the interaction in weighted channels determined by minimization of the action, 
and apply this idea to unpolarized Fermi superfluids.
Using our method, we solve a long-standing difficulty in the partitioning of the interaction into Hartree, Fock, and Bogoliubov channels for Fermi superfluids
and we obtain a phase diagram at the saddle-point level, which contains multichannel non-perturbative corrections. In particular, we find a previously overlooked superfluid phase for weak interactions, which is dominated by particle-hole processes, in addition to the usual superfluid phase only containing particle-particle physics.
\end{abstract}

\maketitle

\section{Introduction}
\label{sec:one}

The subject of ultracold Fermi gases is of great interest to the atomic, condensed matter, and nuclear physics communities, because it can explore the evolution from weakly to strongly correlated regimes by tuning interaction strengths via Fano-Feshbach resonances \cite{Feshbach-1958,Fano-1961,Radzihovsky-2007,Tiesinga-2010, Zwerger-2011,Pitaevskii-2016} or modifying the density \cite{Melo-2000, Melo-2022,Liang-2022,Kei-2022}. In condensed matter physics the effective interactions between fermions (electrons and/or holes) in a solid are largely unknown, and for each material one has to rely on guesses of the type and range of the Fermi-Fermi interactions to establish the phase diagrams of solids \cite{Matsuda-2019,Herrero-2021,Iwasa-2021,Iwasa-2022,Shibauchi-2023,Loon-2024,Levin-2024}.
In nuclear physics, the interactions between fermions (neutrons and/or protons) are typically short-ranged and their effective range is known to play a role in determining the phase diagram of nuclear matter \cite{Soloviev-1958,Schaefer-2008,Schuck-2010,Urban-2018,Clark-2019,Urban-2020,Schuck-2021,Watanabe-2022}. 
In atomic physics, the interactions between neutral ultracold fermions (atoms) is also short-ranged and for experimental systems such as ${^6}{\rm Li}$ and $^{40}{\rm K}$ it is commonly believed that the effective range plays no role in determining the 
phase diagram of Fermi gases~\cite{Legget-2009}.

Despite a substantial amount of experimental work in 	
$^{6}{\rm Li}$ and $^{40} {\rm K}$ \cite{Ketterle-2003,Jin-2003,Grimm-2004,Jin-2004,Ketterle-2004,Ketterle-2005,Ketterle-2006,Hulet-2006,Jochim-2016,Widera-2018,Pan-2022,Grimm-2023}, there is no reliable thermometry that can be used to determine, with good precision, the critical temperature of the superfluid phase of these ultracold fermions, as interactions are tuned from the Bardeen-Cooper-Schrieffer (BCS) to the Bose-Einstein Condensation (BEC) regimes. This experimental difficulty arises for both harmonic traps \cite{Ketterle-2003,Jin-2003,Grimm-2004,Jin-2004,Ketterle-2004,Ketterle-2005,Ketterle-2006,Hulet-2006,Jochim-2016,Widera-2018,Pan-2022,Grimm-2023,Widera-2023-2} and for the more recent box traps \cite{Moritz-2020,Moritz-2022,Pan-2024}.
However, some regions of the phase diagram are accessible to direct thermometry; most prominently is the unitary region, believed to have universal thermal behavior~\cite{Mukaiyama-2010,Salomon-2010,Zwierlein-2012}. In the BCS regime, the method of adiabatic sweeping to the BEC side is commonly used as an indirect method of thermometry \cite{Jin-2004,Grimm-2004,Castin-2004,Levin-2005}. However, in the BCS regime, a direct method was suggested recently, allowing to unravel interaction and thermal contributions in the measured densities of unpolarized trapped fermionic atoms \cite{Widera-2023}. Furthermore, some creative techniques, using machine learning, were used to attempt the determination of the critical temperature for superfluidity of $^{6}{\rm Li}$~\cite{Koehl-2023-1,Koehl-2023-2} in three dimensions. Nevertheless, precise and reliable direct thermometric experimental methods over the entire BEC-BCS crossover are still lacking.
	
Early functional integral theoretical efforts provided a basic understanding of the phase diagram 
of ultracold fermions with short-ranged s-wave interactions, but only included the effects of the Bogoliubov (pairing) 
channel~\cite{Engelbrecht-1993, SadeMelo-1997}, that is, only particle-particle fluctuations were investigated and particle-hole effects were neglected. However, it is known theoretically that particle-hole fluctuations renormalize the critical temperature
in the weak-coupling BCS regime, as demonstrated by Gorkov-Melik-Bakhudarov (GMB)~\cite{Gorkov-1961}.
Other early theoretical investigations, using diagramatic methods~\cite{SchmittRink-1985, Haussmann-1993, Strinati-1998}, 
only include the pairing channel and its fluctuations. 
In a homogeneous system, the typical argument for ignoring Hartree contributions is that they can be absorbed into the chemical potential, however, for a fixed number of particles, the actual value of the chemical potential with respect to the band minimum can decide whether the Fermi system is degenerate or not, and thus has physical consequences. Furthermore, as GMB showed, particle-hole fluctuations can reduce the BCS critical temperature by a factor of $2.2$, which is a substantial effect. We note that GMB did not include Hartree corrections to the BCS mean-field state, rather they computed particle-hole fluctuations on top of the BCS theory, meaning that their approach is incomplete.
	
Some authors neglect Hartree contributions for simplicity~\cite{Kinnunen-2010,Torma-2014,Paredes-2020}; others include them by partitioning half of the interaction energy into the Bogoliubov (particle-particle) channel and the other half into the Hartree (particle-hole) channel~\cite{Timmermans-2011,Widera-2023}. 
In both cases, there is a degree of arbitrariness that needs to the addressed, because different choices can lead to different qualitative and quantitative answers.
Since the physics associated with s-wave pairing cannot depend on the choice of the decomposition, a higher principle needs to be invoked to decide how much of the interaction energy goes into each channel. Furthermore, the equally or arbitrarily weighted inclusion of Hartree (particle-hole) and Bogoliubov (particle-particle) channels for short-ranged interactions leads to divergent Hartree energy and chemical potential at unitarity~\cite{Timmermans-2011,Torma-2014,Pitaevskii-2016,Paredes-2020},
a difficulty that cannot be ignored in the context of short-ranged or zero-ranged interactions, as applied to ultracold gases. 
    
In this paper, we present a general solution to the arbitrariness of separation between Hartree (direct), Fock (exchange), and Bogoliubov (pairing) channels for interacting fermions. 
We introduce a method that weights the Hartree, Fock, and Bogoliubov partitions, in which the interaction can be decomposed, with the constraint that the sum of the respective weights is one. The contribution of each partition is obtained by minimizing the action with respect to the weights, thus eliminating the arbitrary separation of the channels, and distributing the interaction energy into each sector without bias. We apply this idea to the case of short-ranged interactions, where our method also provides a solution to the unphysical divergence of the Hartree energy and chemical potential at unitarity~\cite{Pitaevskii-2016,Pethick-2008}, when the
interaction energy is equally distributed between the Hartree and Bogoliubov channels. 
Our method eliminates the miscounting or double-counting of states that contribute to each channel in standard approaches, and thus removes the aforementioned unphysical results.
	
We emphasize that our approach can, in principle, be used for any type of interactions, where two or more channels compete for the partitioning of the same interaction term, potentially leading to two or more order parameters describing spontaneously broken symmetries. The method can be used irrespective of the underlying type of interaction, which could be s-wave, p-wave, d-wave, Coulomb, dipolar, or spin-spin. However, we illustrate our technique in the simplest possible case: Fermi systems with s-wave short-ranged attractive interactions, identical masses and equal populations. There are two important consequences of introducing weighting factors in the partitioning of interactions.
The first consequence is that the simultaneous regularization of order parameter equations, for the Bogoliubov and Hartree channels, requires the introduction of a many-body effective range. The second consequence is that the Hartree shift acquires the status of an order parameter, which vanishes before the unitarity regime is reached, thus removing the unphysical (singular) behavior in the Hartree energy or the chemical potential at unitarity, as described in textbooks~\cite{Pitaevskii-2016}. The vanishing of the Hartree shift leads to the emergence of a new phase that we call the Hartree superfluid, where the Hartree and Bogoliubov order parameters are nonzero, in contrast to the standard superfluid where the Hartree order parameter is zero and the Bogoliubov order parameter does not vanish. 
	
The remainder of the paper is organized as follows.
In Sec.~\ref{sec:two}, we present our Weighted Hartree-Fock-Bogoliubov Theory for the case of contact s-wave interactions. We decompose the interaction into Hartree, Fock, and Bogoliubov channels, then we particularize to equal masses and balanced populations, where only the Hartree and Bogoliubov channels are important. In addition, we introduce Hartree and Bogoliubov Hubbard-Stratonovich fields and derive the system's effective action. 
In Sec.~\ref{sec:three}, we perform a saddle-point analysis of the Weighted Hartree-Bogoliubov Theory and obtain the corresponding self-consistency relations for the Hartree and Bogoliubov order parameters. Furthermore, when the Hartree and Bogoliubov channels are considered simultaneously, we show that a many-body effective range is required to regularize the theory. Moreover, we demonstrate that the 
Hartree and  Bogoliubov channels are nonperturbatively coupled already at the saddle-point level. In Sec.~\ref{sec:four}, we discuss the resulting ground-state properties in detail. First, we describe the phase diagram in the interaction range versus interaction parameter plane revealing the Hartree and the standard superfluid phases. Second, we analyze the behavior of the order parameters, the chemical potentials, and the ground-state energies in each of the phases. In particular, we determine asymptoptic behaviors in both weak and strong coupling, as well as at unitarity.	In Sec.~\ref{sec:PairSize}, we obtain an analytic expression for the pair size, which serves as an indicator of the evolution from weak to strong coupling. We also describe the various asymptotic limits of the pair size with respect to the interaction parameter and range.
In Sec.~\ref{sec:FiniteTemperature}, we reveal the finite-temperature phase diagram
at the saddle point approximation, as well as the behaviors of the Hartree and Bogoliubov order parameters, and of their weighting factors. Finally, 
we make a quick comparison between the pairing temperature calculated at the saddle point with Hartree corrections and the pairing temperature without the Hartree term, but including particle-hole fluctuations, as performed by GMB. 
In Sec.~\ref{sec:conclusions}, we concisely summarize our findings. Finally, in Sec.~\ref{sec:outlook}, we outline important next steps for simultaneously including both particle-hole and particle-particle fluctuations, which are essential for determining the finite-temperature phase diagram throughout the evolution from weak to strong coupling. 
    \section{Weighted Hartree-Fock-Bogoliubov Theory}
	\label{sec:two}
In the following discussion, we present the Weighted Hartree-Fock-Bogoliubov (WHFB) theory 
to describe Fermi superfluids with contact s-wave interactions. We use the functional integral method to introduce the weighting constraint and to determine the contribution of each channel via a minimization procedure of the system's action. The development of such a theory is important because it solves long-standing theoretical issues regarding channel decompositions \cite{Kleinert-2011,Kleinert-2016} and divergences \cite{Pitaevskii-2016}. Moreover, the utilization of our method is motivated by the increasing number of experimental platforms that allow for more precise comparisons between theory and experiment in three spatial dimensions \cite{Zwierlein-2017,Widera-2023}. Furthermore, our methodology has also important ramifications in addressing related issues in two-dimensional systems, that were experimentally investigated recently~\cite{Yefsah-2024,Zwierlein-2024,Yefsah-2024-2}.

\subsection{Interaction decomposition}
We discuss the pairing theory of fermionic superfluidity, from BCS to BEC for two fermionic species or states labelled by ${\rm s} = (\uparrow,\downarrow)$ representing, for instance, two hyperfine states of $^6\text{Li}$ or $^{40}\text{K}$. We explore the three-dimensional (3D) dilute regime of this system using the Hamiltonian density
\begin{equation}
\label{eqn:hamiltonian-density}
\hspace*{-1mm}       
\mathcal{H}(x) = \sum_{\rm s} \overline{\psi}_{\rm s}(x)\mathcal{K}_{\rm s}\psi_{\rm s}(x) + \mathcal{V}(x) \text{,}
\end{equation}
where $\overline{\psi},\psi$ are anticommuting Grassmann fields. Here, the kinetic energy operator is
\begin{equation}
\mathcal{K}_{\rm s} = -\frac{\mbox{\boldmath $\nabla$}^2}{2m_{\rm s}} - \mu_{\rm s}
\end{equation}
with respect to the chemical potential $\mu_{\rm s}$, and $m_{\rm s}$ stand for 
the mass characterizing the species or state 
${\rm s}$. The interaction term 
\begin{equation}
\label{interaction-term}
\mathcal{V}(x) = -g\overline{\psi}_\uparrow(x)\overline{\psi}_\downarrow(x)\psi_\downarrow(x)\psi_\uparrow(x)   
\end{equation}
corresponds to a s-wave contact attractive interaction, where the strength $g$ is considered to be positive, i.e., we assume $g > 0$. The interaction in Eq.~(\ref{interaction-term}) is SU(2) invariant with respect to the label ${\rm s} = (\uparrow,\downarrow)$ and is written in normal order using the four-vector notation $x = (\mathbf{x}, \tau)$, where 
$\mathbf{x}$ is real space position and $\tau$ denotes imaginary time.
The corresponding action associated with the Hamiltonian in Eq.~(\ref{eqn:hamiltonian-density}) is 
\begin{equation}
\label{action-original}
\mathcal{S}[\overline{\psi},\psi] =  
\int\text{d}x
\left[ \sum_{\rm s}\overline{\psi}_{\rm s}(x)\partial_\tau\psi_{\rm s}(x) + \mathcal{H}(x) \right].
\end{equation}
Here, we used the notation 
$
\int\text{d} x = \int_0^\beta\text{d}\tau\int\text{d}\mathbf{x},
$
where $\beta = T^{-1}$ is 
the inverse temperature in natural units, 
that is, $\hbar = k_{\rm B} = 1$. 

Up to date, theories of interacting fermions explore one of the following options: a) particle-hole channel only, b) particle-particle channel only, or c) an equal mixture of the two, where the particle-hole and particle-particle channels have equal weights. In a), one explores instabilities driven by the particle-hole channel when direct and exchange interactions are present~\cite{Tempere-2018}. In b), one investigates instabilities driven by particle-particle (hole-hole) interactions resulting in pairing~\cite{Bardeen-1957,Leggett-1975,SchmittRink-1985,Engelbrecht-1993,SadeMelo-1997,Strinati-1998,SaDeMelo-2005-a,SaDeMelo-2005-b}.
In c), one analyzes instabilities with equal weights in the particle-hole and particle-particle channels~\cite{Stoof-2008,Czycholl-2015,Widera-2023}. Although this last option includes Hartree, Fock, and Bogoliubov terms, it treats the channels using equal weights. As a result of this arbitrary choice, the equal weight method miscounts contributions by overestimating one channel and underestimating another.

To remove the arbitrariness of the choices a), b), or c), and the miscounting that they introduce, 
we use a Weighted Hartree-Fock-Bogoliubov theory by partitioning the interaction into the Hartree, Fock, and Bogoliubov channels with weights $\{h,f,b\}$, respectively, satisfying the constraint $h + f + b = 1$. We implement this procedure by partitioning the interaction 
$\mathcal{V}(x)$, shown at Eq.~(\ref{interaction-term}), into
\begin{equation}
\mathcal{V}(x) = \mathcal{V}_{\rm H}(x) + \mathcal{V}_{\rm F}(x) + \mathcal{V}_{\rm B}(x)\, ,
\end{equation}
where the Hartree (H), Fock (F), and Bogoliubov (B) 
terms are
\begin{subequations}
\begin{align}
\mathcal{V}_{\rm H}(x) & = -g_{\rm H} \overline{\psi}_\uparrow(x)\psi_\uparrow(x)\overline{\psi}_\downarrow(x)\psi_\downarrow(x) \, , \label{hartree}\\
\mathcal{V}_{\rm F}(x) & = +g_{\rm F} \overline{\psi}_\uparrow(x)\psi_\downarrow(x)\overline{\psi}_\downarrow(x)\psi_\uparrow(x) \, ,
\label{eqn:Fock-I}\\
\mathcal{V}_{\rm B}(x) & = -g_{\rm B} \overline{\psi}_\uparrow(x)\overline{\psi}_\downarrow(x)\psi_\downarrow(x)\psi_\uparrow(x)\, .
\label{bogoliubov}
\end{align}
\end{subequations}
Notice that the interactions written in terms of the Grassmann fields satisfy the constraint 
$g = g_{\rm H} + g_{\rm F} + g_{\rm B}$. 
This is equivalent to attributing weights to each interaction channel through the relations $g_{\rm H} = hg$, $g_{\rm F} = fg$, and $g_{\rm B} = b g$, with $h+f+b=1$ and $\{h,f,b\} \in [0,1]$. With these constraints, the weights $\{h,f,b\}$ represent the probability of participation of 
each channel in the interaction decomposition.
Using this partitioning, we rewrite the action in Eq.~(\ref{action-original}) as
\begin{equation}
\label{eqn:action-HSB}
\mathcal{S}[\overline{\psi},\psi] = \mathcal{S}_{\rm kin}[\overline{\psi},\psi] + \mathcal{S}_{\rm H}[\overline{\psi},\psi] + \mathcal{S}_{\rm F}[\overline{\psi},\psi] + \mathcal{S}_{\rm B}[\overline{\psi},\psi]\text{,}
\end{equation}
where the kinetic contribution reads
\begin{equation}
\label{eqn:action-kinetic}
\mathcal{S}_{\rm kin}[\overline{\psi},\psi] 
= \int \text{d} x \sum_{\rm s} \overline{\psi}_{\rm s}(x) \left( \partial_\tau +\mathcal{K}_{\rm s} \right)\psi_{\rm s}(x)
\end{equation}
and the interaction corresponding to each separate channel is
\begin{equation}
\label{eqn:action-all-channels}
\mathcal{S}_{\rm J}[\overline{\psi},\psi] 
= \int \text{d} x 
\hspace{.1cm} \mathcal{V}_{\rm J}(x),
\end{equation}
with ${\rm J = \{H, F, B\}}$ labeling the the Hartree, Fock, and Bogoliubov channels, respectively.
The specific value of the weights $\{h,f,b\}$ is obtained via the minimization of the action $\mathcal{S}[\overline{\psi},\psi]$ given in Eq.~(\ref{eqn:action-HSB}).

To understand the impact on thermodynamic properties, when including all three channel simultaneously, we need to analyze the grand-canonical partition function 
\begin{equation}
\mathcal{Z} = 
\oint \mathcal{D}\overline{\psi}\mathcal{D}\psi \exp(-\mathcal{S}[\overline{\psi},\psi])\, ,
\label{partition-function}
\end{equation}
where the symbol $\oint$ represents functional integration over the Grassmann fields, which are anti-periodic with respect to imaginary time.
This yields the grand-canonical potential
\begin{equation}
\Omega(V,T,\mu) = - T\, \ln \mathcal{Z} \, .
\label{eqn:potential}
\end{equation}
Therefore, the weights $\{h,f,b\}$ are determined either by minimizing the action in Eq.~(\ref{eqn:action-HSB}) or the grand-canonical potential in Eq.~(\ref{eqn:potential}), according to the principle of minimal sensitivity \cite{Stevenson-1981,Kleinert-2009}.

The discussion above shows that our approach treats the Hartree, Fock, and Bogoliubov channels without biases and at a non-perturbative level, unlike earlier attempts of including particle-hole effects via the Hartree and Fock channels as perturbations about the Bogoliubov channel~\cite{Gorkov-1958, Gorkov-1961}.
Furthermore, our procedure removes the arbitrariness of assigning equal weights to different channels, which leads to a miscounting (overcounting or undercounting) of contributions of different states to each interaction sector. Next on we introduce the Hubbard-Stratonovich fields separately in the different channels to decouple the interaction terms.

\subsection{Hubbard-Stratonovich transformations}
In investigating the functional integral (\ref{partition-function}), the next step relies on factorizing the integrand into different exponentials corresponding to the respective interaction channels. For this purpose, we apply to each exponential factor a specific Hubbard-Stratonovich transformation (HST), which decomposes the interaction with four fermionic fields into an auxiliary bosonic and two fermionic fields. In this section, we concentrate on the Hartree and Bogoliubov decompositions, because our main interest in this manuscript is the application of our method for equal mass and balanced populations, as discussed in Sec.~\ref{sec:Balanced-population-equal-masses}
and beyond. 

To tackle the Hartree channel we decompose the interaction as
\begin{equation}
\mathcal{V}_{{\rm H}}(x) = - g_{\rm H,\uparrow\downarrow} \rho_\uparrow(x)\rho_\downarrow(x) - g_{\rm H,\downarrow\uparrow} \rho_\downarrow(x)\rho_\uparrow(x),
\end{equation}
where both contributions are weighted independently by $ g_{\rm H,\uparrow\downarrow}$ and $g_{\rm H,\downarrow\uparrow}$ with the constraint $g_{\rm H} = g_{\rm H,\uparrow\downarrow} + g_{\rm H\downarrow\uparrow}$.
For Hermitian and Reciprocal systems the two interactions 
$g_{{\rm H},\uparrow\downarrow}$ and 
$g_{{\rm H},\downarrow\uparrow}$ are indeed the same quantity, that is, equal to $g_{\rm H}/2$. Even though non-Hermitian and non-Reciprocal interactions are not considered in this paper, we kept a more general notation to allow for future research in this direction. In addition, we use this notation as a book-keeping device for the matrix elements shown below in Eq.~(\ref{eqn:Hartree-Matrix}).

Further, we set the spin-resolved density to $\rho_{\rm s}(x) = \overline{\psi}_{\rm s}(x)\psi_{\rm s}(x)$ and introduce a real valued bosonic field
\begin{equation}
\boldsymbol{\Delta}_{\rm H}^{\rm T}(x) 
= \begin{pmatrix}
\Delta_{\rm H,\downarrow}(x) 
& \Delta_{\rm H,\uparrow}(x)
\end{pmatrix}
\end{equation}
coupling to the Hartree channel source term
\begin{equation}
\boldsymbol{j}_{\rm H}(x) = \begin{pmatrix}
    \rho_\uparrow(x)  \\ \rho_\downarrow(x) 
\end{pmatrix},
\end{equation}
which is associated to particle-hole processes for the different densities. The Hubbard-Stratonovich transformation for the Hartree channel action in Eq.~(\ref{eqn:action-all-channels}) then reads
\begin{equation}
	\label{eqn:exponential-action-hartree}
	e^{-\mathcal{S}_{\rm H}
		[{\overline \psi}, \psi]} 
	= {\cal N}_{\rm H} \oint 
	\mathcal{D} \boldsymbol{\Delta}_{\rm H}
	e^{- \mathcal{S}^{\rm H}_{\rm aux}[\boldsymbol{\Delta}_{\rm H}; {\overline\psi}, \psi]},
\end{equation}
where the auxiliary action is
\begin{eqnarray}
\mathcal{S}^{\rm H}_{\rm aux}[\boldsymbol{\Delta}_{\rm H}; {\overline\psi}, \psi] = \int\text{d} x && \hspace{.1cm} \bigg[ \frac{1}{2}\,\boldsymbol{\Delta}_{\rm H}^{\rm T}(x)
\mathbf{\mathcal{M}}_{\rm H}\boldsymbol{\Delta}_{\rm H}(x) \nonumber \\ && 
+ \boldsymbol{\Delta}_{\rm H}^{\rm T}(x) \boldsymbol{j}_{\rm H}(x) \bigg]\,\text{.} \label{eqn:action-auxiliary-hartree}
\end{eqnarray}
Here the matrix that couples 
$\boldsymbol{\Delta}_{\rm H}^{\rm T}(x)$
and $\boldsymbol{\Delta}_{\rm H}(x)$ 
is
\begin{equation}\label{eqn:Hartree-Matrix}
    \mathbf{\mathcal{M}}_{\rm H} = \frac{1}{2} \begin{pmatrix}
			0 & 1/g_{\rm H, \uparrow\downarrow} \\ 1/g_{\rm H,\downarrow\uparrow} & 0
		\end{pmatrix}.
\end{equation}
Thus,
the auxiliary action in Eq.~(\ref{eqn:action-auxiliary-hartree}) is explicitly given by evaluating the scalar products
\begin{widetext}
\begin{equation}
\label{eqn:action-auxiliary-hartree2}
\begin{split}
    \mathcal{S}^{\rm H}_{\rm aux}[\Delta_{{\rm H}}; {\overline\psi}, \psi]
= \int\text{d} x \hspace{.1cm} \bigg\{ \frac{\Delta_{\rm H,\downarrow}(x)\Delta_{\rm H,\uparrow}(x)}{4g_{\rm H,\uparrow\downarrow}} + \frac{\Delta_{\rm H,\uparrow}(x)\Delta_{\rm H,\downarrow}(x)}{4g_{\rm H,\downarrow\uparrow}} + \Delta_{\rm H,\uparrow}(x)\overline{\psi}_{\downarrow}(x)\psi_{\downarrow}(x) + \Delta_{\rm H,\downarrow}(x)\overline{\psi}_{\uparrow}(x)\psi_{\uparrow}(x)  \bigg\}.
\end{split}
\end{equation}
\end{widetext}
This transforms the direct contribution of the interaction, after which we will focus on the pairing terms.
As the Bogoliubov channel is represented by complex scalar fields, we rewrite its action in  Eq.~(\ref{eqn:action-all-channels}) by means of the Hubbard-Stratonovich transformation
\begin{equation}
\label{eqn:exponential-action-fock-bogoliubov}
e^{-{\mathcal S}_{\rm B}
[{\overline \psi}, \psi]}
= 
\mathcal{N}_{\rm B} \oint
{\mathcal D}{\overline \Delta}_{\rm B}
\mathcal D \Delta_{\rm B}
e^{-{\mathcal{S}_{\rm aux}^{\rm B}[\overline{\Delta}_{\rm B},\Delta_{\rm B}, \overline{\psi},\psi}]}.
\end{equation}
The explicit form of the auxiliary action is  
\begin{align}
\mathcal{S}_{\rm aux}^{\rm B}& [\overline{\Delta}_{\rm B},\Delta_{\rm B}, \overline{\psi},\psi
] = \int\text{d} x \bigg[ \overline{\Delta}_{\rm B}(x)\mathcal{M}_{\rm B}\Delta_{\rm B}(x) \nonumber \\ &  + \overline{j}_{\rm B} (x)\Delta_{\rm B}(x) + j_{\rm B} (x)\overline{\Delta}_{\rm B}(x)  \bigg], \label{eqn:action-auxiliary-fock-bogoliubov}
\end{align}
when expressed in terms of the auxiliary fields ${\overline \Delta}_{\rm B} (x)$, $\Delta_{\rm B} (x)$ and the source term $j_{\rm B} (x)$.
The Bogoliubov channel source term
\begin{equation}
\label{eqn:current-bogoliubov}
j_{\rm B} (x) = \psi_\downarrow(x)\psi_\uparrow(x)
\end{equation}
is associated to singlet pairing and we identify $\mathcal{M}_{\rm B} = g_{\rm B}^{-1}$.
Defining
$\mathcal{Z}_{\rm H} [{\overline \psi}, \psi] = e^{-{\mathcal S}_{\rm H} [{\overline \psi, \psi}]}$ and using Eq.~(\ref{eqn:exponential-action-hartree}), as well as
defining
$\mathcal{Z}_{\rm B} [{\overline \psi}, \psi] = e^{-{\mathcal S}_{\rm B} [{\overline \psi, \psi}]}$ and using Eq.~(\ref{eqn:exponential-action-fock-bogoliubov}), 
we rewrite the grand-canonical partition function shown in Eq.~(\ref{partition-function}) in terms of the auxiliary fields 
$\{\Delta_{\rm H,s},{\overline\Delta}_{\rm B},\Delta_{\rm B}\}$ as
\begin{equation}
\label{partition-function2}
\mathcal{Z} = 
\oint \mathcal{D}\overline{\psi}\mathcal{D}\psi \exp(-\mathcal{S}_{\rm kin}[\overline{\psi},\psi])
\mathcal{Z}_{\rm HB} [\overline\psi, \psi],
\end{equation}
where the decomposition into the two interaction channels is described by the product
\begin{equation}
\label{partition-function3}
\mathcal{Z}_{\rm HB} [\overline\psi, \psi]
=
\mathcal{Z}_{\rm H} [\overline\psi, \psi]
\mathcal{Z}_{\rm B} [\overline\psi, \psi].
\end{equation}
As a consequence, using 
Eq.~(\ref{partition-function3}), the 
Hubbard-Stratonovich transformations in 
Eq.~(\ref{eqn:exponential-action-hartree}) for the Hartree 
sector and in Eq.~(\ref{eqn:exponential-action-fock-bogoliubov}) for the Bogoliubov channel, transforms 
the partition function in Eq.~(\ref{partition-function2})
to
\begin{equation}
\label{partition-function4}
\mathcal{Z}        
= \oint \mathcal{D}\overline{\psi}\mathcal{D}\psi \oint \mathcal{D} \{\Delta \} \exp(-\mathcal{S}_{\rm HB}[\overline{\psi},\psi;\{\Delta\}])\text{.}
\end{equation}
Here, the notation $\{\Delta\}$ abbreviates the set $\{ \Delta_{\rm H,s }, {\overline \Delta}_{\rm B}, \Delta_{\rm B} \}$ of auxiliary fields and $\mathcal{D}\{\Delta\}$ represents their combined functional integral measure 
$\mathcal{D} \Delta_{\rm H,s}
\mathcal{D} {\overline\Delta}_{\rm B}
\mathcal{D} \Delta_{\rm B}$. The resulting Hartree-Bogoliubov action 
\begin{align}
\mathcal{S}_{\rm HB}[\overline{\psi},\psi;\{\Delta\}]& = \mathcal{S}_{\rm kin}[\overline{\psi},\psi]
+ \mathcal{S}_{\rm aux}^{\rm H}[\Delta_{{\rm H,s}}; \overline{\psi},\psi] \nonumber \\ & + \mathcal{S}_{\rm aux}^{\rm B}[\overline{\Delta}_{\rm B},\Delta_{\rm B}; \overline{\psi},\psi] \label{eqn:action-hartree-bogoliubov}
\end{align}
contains the kinetic contribution 
$\mathcal{S}_{\rm kin}[\overline{\psi},\psi]$ 
given in Eq.~(\ref{eqn:action-kinetic}) for equal masses and balanced populations, the auxiliary Hartree action
$\mathcal{S}_{\rm aux}^{\rm H}[\Delta_{\rm H,s}, \overline{\psi},\psi]$ described in Eq.~(\ref{eqn:action-auxiliary-hartree}) and the auxiliary Bogoliubov
action $\mathcal{S}_{\rm aux}^{\rm B}[\overline{\Delta}_{\rm B},\Delta_{\rm B}, \overline{\psi},\psi]$ from 
Eq.~(\ref{eqn:action-auxiliary-fock-bogoliubov}).
Writing the Hartree-Bogoliubov (HB) action explicitly
\begin{widetext}
\begin{eqnarray}
&& \hspace*{-1cm}	\mathcal{S}_{\rm HB} [\overline{\psi},\psi;\{\Delta\}] = \int \text{d} x  \bigg\{ \sum_{\rm s} \overline{\psi}_{\rm s}(x)\left( \partial_\tau + \mathcal{K}_{s} \right)\psi_{\rm s}(x) + \Delta_{\rm H,\uparrow}(x)\overline{\psi}_{\downarrow}(x)\psi_{\downarrow}(x) + \Delta_{\rm H,\downarrow}(x)\overline{\psi}_{\uparrow}(x)\psi_{\uparrow}(x) \nonumber\\
&& \hspace*{-1cm} + \overline{\Delta}_{\rm B}(x)\psi_\downarrow(x)\psi_\uparrow(x) + \Delta_{\rm B}(x)\overline{\psi}_\uparrow(x)\overline{\psi}_\downarrow(x) +   \frac{\overline{\Delta}_{\rm B}(x)\Delta_{\rm B}(x)}{g_{\rm B}} +   \frac{\Delta_{\rm H,\downarrow}(x)\Delta_{\rm H,\uparrow}(x)}{4g_{\rm H,\uparrow\downarrow}}  + \frac{\Delta_{\rm H,\uparrow}(x)\Delta_{\rm H,\downarrow}(x)}{4g_{\rm H,\downarrow\uparrow}} \bigg\} \label{eqn:action-hartree-bogoliubov-explicit-imbalance}
\end{eqnarray}
\end{widetext}
reveals that it is quadratic in the fermionic as well as in the bosonic auxiliary fields.
We emphasize that the Hartree-Bogoliubov action 
in Eq.~(\ref{eqn:action-hartree-bogoliubov-explicit}),
expressed in terms of the Hubbard-Stratonovich fields, represents a generalization of 
the cases, where either only the Hartree sector~\cite{Tempere-2018} or
only the Bogoliubov channel~\cite{SaDeMelo-2005-a,SaDeMelo-2005-b,Engelbrecht-1993,SadeMelo-1997,Urban-2018,Bardeen-1957,Strinati-1998,Leggett-1975,SchmittRink-1985} occurs
and for now does not have any restrictions to the two spin-populations $n_{\rm s} = \langle \overline{\psi}_{\rm s}(x) \psi_{\rm s}(x) \rangle$.
A similar approach can be applied also for the Fock term in Eq.~(\ref{eqn:Fock-I}), but since we focus next on the case of equal masses and balanced populations, the Fock terms are not relevant, 
as explained in Sec.~\ref{sec:Balanced-population-equal-masses}.
Now, we are ready to use the method outlined above to investigate non-perturbative effects of the Hartree (particle-hole) sector on the Bogoliubov (particle-particle) channel. These non-perturbative effects reflect the coupling between the Hartree and Bogoliubov fields that result from integrating out the fermionic degrees of freedom and keeping track of the interaction partitioning between the two channels via $g_{\rm H,\uparrow\downarrow}$, $g_{\rm H,\downarrow\uparrow}$ and $g_{\rm B}$. These steps lead to an effective action, which only includes the bosonic auxiliary fields and the weighting parameters.

Thus, we discuss next the example of balanced populations and equal masses, as the simplest example of the application of general weighted Hartree-Fock-Bogoliubov theory discussed above.

\subsection{Balanced populations and equal masses}\label{sec:Balanced-population-equal-masses}
To understand the effects of the competing 
channels $\{ {\rm H, F, B}\}$ in systems of balanced populations and equal masses, we notice that the interaction term of the Hamiltonian density given in 
Eq.~(\ref{eqn:hamiltonian-density}) is local, includes only the singlet s-wave component, and preserves SU(2) symmetry. Also the kinetic energy term is spin-diagonal and proportional to the identity matrix in spin space, and thus is also SU(2) invariant. Due to these symmetries, no spin-flip processes are allowed and the Fock source field is absent. 
The irrelevance of the Fock term is not directly connected to the range of the singlet s-wave interaction used, but rather to the preservation of SU(2) symmetry.  
For balanced populations and equal masses, we consider only the case of spontaneously broken U(1) symmetry, but SU(2) symmetry is preserved. Fock channel source terms involve spin flips, breaking SU(2) symmetry explicitly, and thus they do not appear for spontaneously broken U(1) but preserved SU(2) symmetry. On the other hand, for imbalanced populations and equal masses, when U(1) is spontaneously broken and SU(2) is explicitly broken, a non-trivial solution for the Fock order parameter may exist, because the system can develop a magnetization. However, in the limit of balanced populations and equal masses, the Fock order parameter is identically zero, because it is proportional to the population imbalance. The generalization of our theory to include population and/or mass imbalances will be the topic of our next publication. In the present paper, our intention is to discuss the simplest example possible, where the weighted Hartree-Fock-Bogoliubov method needs to be applied.

Thus, for balanced populations and equal masses, the Fock contribution does not emerge at a saddle-point level, that is, $g_{\rm F} = 0$.
Since we focus on the 
example of a single atomic species with balanced populations, that is, $\mu_{\uparrow} = \mu_{\downarrow} = \mu$ and $m_{\uparrow} = m_{\downarrow} = m$,
the kinetic terms simplify to 
${\cal K}_{\uparrow} = {\cal K}_{\downarrow} =
{\cal K} =  -\boldsymbol{\nabla}^2/(2m) - \mu$.
Furthermore, equal populations implies that 
$\Delta_{\rm H,\uparrow} = \Delta_{\rm H,\downarrow} 
= \Delta_{\rm H}$ 
and using that $g_{\rm H,\uparrow\downarrow}$ = $g_{\rm H,\downarrow\uparrow} = g_{\rm H}/2$ leads to two contributions: one originating from the Hartree channel with weight
$h$, yielding $g_{\rm H} = h g$ and the other from the Bogoliubov channel with weight $b$ included in $g_{\rm B} = b g$, such that we have $h + b = 1$ or $g_{\rm H} + g_{\rm B} = g$.
Under those simplifications the overall HB action from  Eq.~(\ref{eqn:action-hartree-bogoliubov-explicit-imbalance})
becomes
\begin{widetext}
\begin{eqnarray}
&& \hspace*{-1cm} \mathcal{S}_{\rm HB} [\overline{\psi},\psi;\{\Delta\}] = \int \text{d} x  \bigg\{ \sum_{\rm s} \overline{\psi}_{\rm s}(x)\left( \partial_\tau + \mathcal{K}\right)\psi_{\rm s}(x) + \overline{\Delta}_{\rm B}(x)\psi_\downarrow(x)\psi_\uparrow(x) + \Delta_{\rm B}(x)\overline{\psi}_\uparrow(x)\overline{\psi}_\downarrow(x) \nonumber\\
&& \hspace*{-1cm} + \Delta_{\rm H}(x)\left[\overline{\psi}_{\downarrow}(x)\psi_{\downarrow}(x) + \overline{\psi}_{\uparrow}(x)\psi_{\uparrow}(x) \right] + \frac{\overline{\Delta}_{\rm B}(x)\Delta_{\rm B}(x)}{g_{\rm B}}  + \frac{\Delta_{\rm H}(x)\Delta_{\rm H}(x)}{g_{\rm H}} \bigg\}, \label{eqn:action-hartree-bogoliubov-explicit}
\end{eqnarray}
\end{widetext}
covering contributions from the Hartree and 
Bogoliubov channels. The action in Eq.~(\ref{eqn:action-hartree-bogoliubov-explicit}) is the starting point for investigating the effects of particle-hole fluctuations not only in the BCS regime~\cite{Gorkov-1961} but
throughout the BCS-BEC crossover. Thus, 
we construct the effective action of the Hartree-Bogoliubov sector by integrating out the fermions,
as discussed next. 

\subsection{Effective action}
To obtain the effective action of our system, 
we write the fermion fields as a 
Nambu spinor
$\overline{\Psi}(x) = 
\begin{pmatrix} \overline{\psi}_\uparrow(x) & \psi_\downarrow(x)
	\end{pmatrix}$
and express the HB action from  
Eq.~(\ref{eqn:action-hartree-bogoliubov-explicit}) 
as
\begin{eqnarray}
\mathcal{S}_{\rm HB}
[\overline{\Psi}, \Psi,
\{\Delta\}]
=
\int \text{d} x
\bigg[ && 
\overline{\Psi}(x) 
{\bf A}\Psi (x) + \frac{\overline{\Delta}_{\rm B}(x)\Delta_{\rm B}(x)}{g_{\rm B}} \nonumber\\
&& \textbf{} + \frac{\Delta_{{\rm H}} (x)\Delta_{{\rm H}} (x)}{g_{\rm H}} 
\bigg]\,,
\end{eqnarray}
where matrix ${\bf A}$ has the structure
\begin{equation}
{\bf A} =
\begin{pmatrix}
\partial_{\tau} + {\cal K} + \Delta_{\rm H} (x)
& \Delta_{\rm B} (x)\\
{\overline \Delta}_{\rm B} (x) &  \partial_{\tau} - {\cal K} - \Delta_{\rm H} (x)
\end{pmatrix},
\end{equation}
with $
{\cal K} = - \nabla^2/2m - \mu,
$
as discussed earlier for equal masses and 
balanced populations. The integration over fermionic Grassmann fields is performed by converting the measure 
${\cal D}{\overline\psi}
{\cal D}{\psi}$ into 
${\cal D}{\overline\Psi}{\cal D}\Psi$ leading to the effective action
\begin{eqnarray}
&& \hspace*{-1cm} \mathcal{S}_{\rm eff} (\{\Delta\})
= 
-\ln \left[
{\rm Det}{(\beta {\bf A})}
\right] \nonumber\\
&& \hspace*{-1cm} + \int \frac{\text{d} x}{\beta V}
\bigg[{\beta V}\frac{\vert \Delta_{\rm B}(x) \vert^2}{g_{\rm B}}
+ {\beta V}\frac{\Delta_{{\rm H}} (x)^2}{g_{\rm H}} 
\bigg]\,, \label{Hartree-Bogoliubov-Effective-Action}
\end{eqnarray}
where ${\rm Det} (\beta {\bf A})$ means the product of the eigenvalues of the operator matrix 
$\beta {\bf A}$, including spins. This represents the exact effective action for the
Hartree-Bogoliubov decomposition. 
Therefore, integration over fermion fields in Eq.~(\ref{partition-function4}) leads to the grand-canonical partition function
	\begin{align}
		\mathcal{Z}     = \oint \mathcal{D}\{\Delta\} \exp(-\mathcal{S}_{\rm eff}[\{\Delta\}])\, .
	\end{align}

	To make progress towards a saddle point description, we write $\Delta_{{\rm B}}(x) = \Delta_{{\rm B}, 0} + \eta_{\rm B} (x)$ for the superfluid order parameter and 
	$\Delta_{{\rm H}}(x) = \Delta_{{\rm H}, 0} + \eta_{\rm H} (x)$ for the Hartree order parameter. Here, $\Delta_{{\rm B},0}$ and 
	$\Delta_{{\rm H}, 0}$ represent saddle points, 
	while $\eta_{\rm B} (x)$ and $\eta_{\rm H} (x)$ correspond to fluctuations. Such a representation leads to two contributions to the effective action. 
	The first one is the saddle point action $\mathcal{S}_0[\{\Delta_0\}]$ discussed 
	below in detail and the fluctuation action 
	$\mathcal{S}_{\rm fluct}[\{\Delta_0\};\{\eta\}]$
	discussed briefly in Section~\ref{sec:outlook}, which represents the outlook of the paper.
	
	For the rest of this section, we ignore fluctuations and consider only the saddle point contribution leading to the grand-canonical partition function
	\begin{align}
		\mathcal{Z}_0 = \exp(-\mathcal{S}_{0}[\{\Delta_0\}])\, ,
	\end{align}
	with the saddle point action
	\begin{eqnarray}
		\mathcal{S}_{0}[\{\Delta_0\}]
		= &&  -\int \frac{\text{d} x}{\beta V}
		\ln \left[
		\det{(\beta {\bf A}_0)} 
		\right] \nonumber \\
		&&  + 
		\beta V 
		\left( \frac{\vert \Delta_{{\rm B}, 0} \vert^2}{g_{\rm B}} + 
		\frac{\Delta_{{\rm H},0}^2}{g_{\rm H}}
		\right)\,.\label{eqn:saddle-point-action}
	\end{eqnarray}
	The notation $\det{(\beta {\bf A}_0)}$
	refers to the determinant in the spin subspace only, while the space-time part of the determinant was converted into the integral $\int {\rm d}x$, and 
\begin{equation}
\label{eqn:inverse-propagator-matrix}
{\bf A}_0 =
\begin{pmatrix}
\partial_{\tau} + {\cal K}  + \Delta_{{\rm H},0}
& \Delta_{{\rm B},0}\\
{\overline \Delta}_{{\rm B},0} &  \partial_{\tau} - {\cal K} - \Delta_{{\rm H},0}
\end{pmatrix}
\end{equation}
represents the inverse propagator 
matrix.
Performing a Fourier transformation into momentum $({\bf k})$ and Matsubara $(i k_n)$ space with the fermionic Matsubara frequencies $k_n = (2n+1)\pi /\beta$, where $n$ is an integer, we obtain
\begin{eqnarray}
		\mathcal{S}_{0}[\{\Delta_0\}]
		= && -\sum_{k}
		\ln[\det (\beta {\widetilde{\bf A}}_{0, k})] \nonumber \\
		&& +\beta V\left(\frac{|\Delta_{{\rm B},0}|^2}{g_B} + \frac{\Delta_{{\rm H},0}^2}{g_H} \right)\,. \label{eqn:saddle-point-action-momentum-space}
\end{eqnarray}
Here, we use the four-momentum notation
$k = (ik_n, {\bf k})$, as well as the transformation
$\partial_\tau \to -ik_n$,  
$-i\boldsymbol{\nabla} \to {\bf k}$,
and $\mathcal{K} \to {\bf k}^2/2m - \mu$ to write the Fourier transform of 
	Eq.~(\ref{eqn:inverse-propagator-matrix})
	as 
\begin{eqnarray}
{\widetilde {\bf A}}_{0,k} = && -ik_n \textbf{I} + \left(\xi_{\mathbf{k}} + \Delta_{{\rm H},0} \right)\boldsymbol{\sigma}_z \nonumber\\
&&  + \Delta_{{\rm B},0}\boldsymbol{\sigma}^+ + \overline{\Delta}_{{\rm B},0} \boldsymbol{\sigma}^-\,,
\end{eqnarray}
	where $\xi_\mathbf{k} = \mathbf{k}^2/(2m) - \mu$ is the kinetic energy with
	respect to the chemical potential $\mu$, 
	$\boldsymbol{\sigma}_j$ stands for the Pauli matrices with $j = \{x, y, z\}$, and $\boldsymbol{\sigma}^\pm = (\boldsymbol{\sigma}_x \pm i\boldsymbol{\sigma}_y)/2$ represent the spin raising and lowering operators.
	In Eq.~(\ref{eqn:saddle-point-action-momentum-space}), the determinant 
	\begin{equation}
		\label{eqn:determinant}
		\det (\beta {\widetilde{\bf A}}_{0, k})
		= 
		\beta^2 ( ik_n - E_{\bf k} ) 
		( ik_n + E_{\bf k} ),
	\end{equation}
	is the product of the eigenvalues of 
	$\beta {\widetilde {\bf A}}_{0, k}$,
	with the quasiparticle dispersion being
	\begin{equation}
		\label{eqn:Bogoliubov-dispersion}
		E_\mathbf{k} = \sqrt{(\xi_\mathbf{k} + \Delta_{{\rm H},0})^2 + |\Delta_{{\rm B},0}|^2}\,.
	\end{equation}
Notice that $\mathcal{S}_0$ in Eq.~(\ref{eqn:saddle-point-action-momentum-space}) contains a branch cut due to the logarithm, which needs to be carefully handled when recovering the correct zero-point energy and the saddle-point equations to be discussed next.

\section{Saddle-Point Analysis}
\label{sec:three}
	
In this section, we discuss the saddle point equations derived from our WHFB approach. We show that the Hartree and Bogoliubov channels exhibit a non-analytic and non-perturbative coupling, and that the inclusion of both contributions requires a many-body renormalization scheme to regularize 
the order parameter equations. Our approach also solves a long-standing issue with the Hartree contribution near unitarity~\cite{Pitaevskii-2016}.
So, let us start our analysis by discussing next the self-consistency relations.
	
	\subsection{Self-consistency equations}\label{sec:Self-consistency-equations}
	
	To establish self-consistency 
	for $\Delta_{{\rm B},0}$ and $\Delta_{{\rm H},0}$, we extremize the action 
	$S_0$ given in Eq.~(\ref{eqn:saddle-point-action-momentum-space}), that is, we set
	$\partial\mathcal{S}_{0}/\partial\Delta_{{\rm H},0} = 0$ and 
	$\partial\mathcal{S}_{0}/\partial\overline{\Delta}_{{\rm B},0} = 0$. 
	Evaluating these partial derivatives leads to the saddle-point conditions
	\begin{subequations}
		\begin{align}
			\Delta_{{\rm H},0} & = - \frac{g_{\rm H}}{2\beta V} \sum_k \left[c_{\rm H,+}(k) - c_{\rm H,-}(k)\right]\text{,}\label{eqn:saddle-point-delta-h} \\
			\Delta_{{\rm B},0} & = - \frac{g_{\rm B}}{\beta V} \sum_k c_{\rm B}(k)\text{.}
			\label{eqn:saddle-point-delta-b}
		\end{align}
	\end{subequations}
	The relation given in Eq.~(\ref{eqn:saddle-point-delta-h})
	represents the order parameter for the Hartree (particle-hole) channel, while Eq.~(\ref{eqn:saddle-point-delta-b})
	refers to the order parameter in the Bogoliubov (particle-particle) channel.
	The functions appearing on the right-hand side of Eqs.~(\ref{eqn:saddle-point-delta-h}) and~(\ref{eqn:saddle-point-delta-b}) are given by
	\begin{subequations}
		\begin{align}
			c_{\rm H,\pm}(k) & = - \frac{ik_n \pm (\xi_\mathbf{k} + \Delta_{{\rm H},0})}{(ik_n)^2 - E_\mathbf{k}^2} e^{ik_n 0^{\pm}}\text{,}\label{cH} \\
			c_{\rm B}(k) & = \frac{\Delta_{{\rm B},0}}{(ik_n)^2 - E_\mathbf{k}^2}\text{.}\label{eqn:cB}
		\end{align}
	\end{subequations}
	Note that the exponentials $e^{ik_n 0^{\pm}}$
	capture the existence of a branch cut in the logarithm of Eq.~(\ref{eqn:saddle-point-action-momentum-space})
	due to the analytical structure
	of $\det(\beta {\widetilde {\bf A}}_{0,k})$ shown in Eq.~(\ref{eqn:determinant}).
	This extra care is necessary for recovering the zero-point energy in the action $S_0$ and the grand-canonical thermodynamic potential $\Omega_0(V,T,\mu) = - \beta^{-1}\ln {\cal Z}_0$. This is a well-known point that can be found in textbooks~\cite{Abrikosov-1975,Kleinert-2009,Altland-2010}. 
	
	In Eq.~(\ref{eqn:saddle-point-delta-h}) the interaction $g_{\rm H} = h g$ appears, while in Eq.~(\ref{eqn:saddle-point-delta-b}) the interaction
	$g_{\rm B} = b g$ emerges. As highlighted below in Sec.~\ref{sec:four}, we remove the arbitrariness of assigning equal weights to the particle-hole and to the particle-particle channels by 
	preventing the miscounting of  states involved in the Bogoliubov (particle-particle) and Hartree (particle-hole) partitions. For this purpose, we extremize the action 
	${\cal S}_0$ with respect to $b$, that is, 
	set $\partial \mathcal{S}_0/\partial b = 0$, subject to the constraint
	$h + b = 1$ $(g_{\rm H} + g_{\rm B} = g)$ 
	and the physical requirement that $0 \le b \le 1$,
	which guarantees that ${\cal S}_0$ is minimal with respect to $b$, that
	is, $\partial^2 {\cal S}_0/\partial b^2 \ge 0$.
	This procedure leads to the saddle point solution
	\begin{subequations}
		\begin{align}
			h_0 & = \frac{|\Delta_{{\rm H},0}|}{|\Delta_{{\rm B},0}| + |\Delta_{{\rm H},0}|}\text{,}\label{eqn:h0} \\
			b_0 & = \frac{|\Delta_{{\rm B},0}|}{|\Delta_{{\rm B},0}| + |\Delta_{{\rm H},0}|}\text{,}\label{eqn:b0}
		\end{align}
	\end{subequations}
where $\Delta_{{\rm H}, 0}$ and $\Delta_{{\rm B}, 0}$ are defined by Eqs.~(\ref{eqn:saddle-point-delta-h}) and (\ref{eqn:saddle-point-delta-b}), respectively. Notice that $h_0 + b_0 = 1$.

At the saddle point, we have checked that the global minimum always occurs for $b = b_0$ and $h = h_0$ between $[0, 1]$, when variables like temperature, effective range and scattering parameter change. When the minimum occurs at endpoints of the domain, the derivative of the action with respect to $b$ is still zero.

In general, given that we have a constrained system where 
$b + h = 1$, the minimization of the action with respect to any value of $b$ gives always a global minimum between $[0, 1]$. 
This result provides the mathematical basis for the physical interpretation that $b$ and $h$ are distribution weights of the interaction, meaning that both $b$ and $h$ are always in the interval $[0, 1]$ and thus can be viewed as the probability of participation of each interaction channel. For instance, minima of the action with $b < 0$ $(h > 1)$ or $b > 1$ $(h < 0)$ do not arise mathematically in our problem, and if they did, these types of solutions would be considered unphysical, since they would change the nature of the interaction in one the channels from attractive to repulsive. 
	
The evaluation of the Matsubara sums in Eqs.~\eqref{eqn:saddle-point-delta-h} and \eqref{eqn:saddle-point-delta-b} is performed by using Cauchy's residue theorem~\cite{Stalker-1998}, leading to 
	\begin{subequations}
		\begin{align}
			\Delta_{{\rm H},0} & = - \frac{g_{{\rm H},0}}{2V}\sum_\mathbf{k} \left[ 1 - \frac{\widetilde{\xi}_\mathbf{k}}{E_\mathbf{k}}\tanh\left(\frac{\beta E_\mathbf{k}}{2}\right) \right]\text{,}\label{eqn:hartree-order-parameter} \\
			\Delta_{{\rm B},0} & = g_{{\rm B},0} \Delta_{{\rm B},0} \frac{1}{V}\sum_\mathbf{k} \frac{\tanh\left(\frac{\beta E_\mathbf{k}}{2}\right)}{2E_\mathbf{k}}\text{.}\label{eqn:bogoliubov-order-parameter}
		\end{align}
	\end{subequations}
Here, the shifted free particle dispersion $\widetilde{\xi}_\mathbf{k} = \xi_\mathbf{k} + \Delta_{\rm H,0}$ was used.
The expression in Eq.~(\ref{eqn:hartree-order-parameter}) is the Hartree order parameter equation, and the expression 
in Eq.~(\ref{eqn:bogoliubov-order-parameter}) describes the Bogoliubov order 
parameter equation. We additionally used the optimized interaction strengths $g_{{\rm H},0}=h_0g$ and $g_{{\rm B},0}=b_0g$ with $g_{{\rm H},0} + g_{{\rm B},0} = g$.
	
	To obtain the number equation that fixes the chemical potential $\mu$, 
	we perform the Matsubara sums in Eq.~(\ref{eqn:saddle-point-action-momentum-space}) 
	and calculate the saddle-point grand-canonical potential 
\begin{eqnarray}
\frac{\Omega_{0}}{V} = &&  \frac{|\Delta_{{\rm B},0}|^2}{g_{{\rm B},0}} + \frac{\Delta_{{\rm H},0}^2}{g_{{\rm H},0}} - \frac{1}{V}\sum_\mathbf{k}\left(E_\mathbf{k} -\widetilde{\xi}_\mathbf{k} \right)
\nonumber\\
&&  - \frac{1}{V} \sum_\mathbf{k} \frac{2}{\beta}\ln\left(1 + e^{-\beta E_\mathbf{k}}\right)
, \label{eqn:saddle-point-thermodynamic-potential} 
\end{eqnarray}
	where we used $\Omega_ 0 = \beta^{-1} S_0$. The particle density $n = N/V$, where $N$ is the number of particles, is obtained from the thermodynamic relation
	$N = -\partial \Omega/\partial \mu\vert_{T,V}$. Thus, at the saddle-point level, the number of particles is 
	$N_0 = -\partial \Omega_0/\partial \mu\vert_{T,V}$ giving 
	\begin{equation}
		\label{eqn:saddle-point-number}
		n_0 = \frac{1}{V}\sum_\mathbf{k}\left[ 1 - \frac{\widetilde{\xi}_\mathbf{k}}{E_\mathbf{k}}\tanh\left(\frac{\beta}{2}E_\mathbf{k}\right)\right]
	\end{equation}
	for the saddle-point number density equation. 
	
	The saddle-point Hartree order parameter equation results from combining Eq.~(\ref{eqn:saddle-point-delta-h}) with Eq.~(\ref{eqn:saddle-point-number}) yielding
	\begin{equation}
		\label{eqn:hartree-order-parameter-density}
		\Delta_{{\rm H},0} = -\frac{g_{{\rm H},0} n_0}{2},
	\end{equation}
	where the factor $1/2$ arises from two spin states, in contrast with the Hartree shift for spinless bosons, where the factor of two is absent~\cite{Pitaevskii-2016}. 
	
Notice that $\Delta_{{\rm H}, 0}$ is proportional to $n_0$, and is always non-positive since $g \ge 0$. Substituting the expression for $h_0$ from Eq.~(\ref{eqn:h0})
into Eq.~(\ref{eqn:hartree-order-parameter-density}) leads
to 
\begin{equation}\label{eqn:hartree-equation-explicit}
		\Delta_{\rm H,0} = -\frac{|\Delta_{\rm H,0}|}{|\Delta_{\rm H,0}| + |\Delta_{\rm B,0}|}\frac{gn_0}{2}.
\end{equation}
Since the interaction is attractive or zero $(g \ge 0)$, the only physically acceptable solutions for $\Delta_{{\rm H},0}$ are negative or zero, that
is, $\Delta_{\rm H,0} \le 0$. 
Using $\Delta_{{\rm H},0} 
= - \vert \Delta_{{\rm H}, 0} \vert$
we see that Eq.~(\ref{eqn:hartree-equation-explicit}) has two possible solutions. The first is the trivial solution $\Delta_{\rm H,0} = 0$, and the second is 
\begin{equation}\label{eqn:Final-Hartree-Eqn}
		\Delta_{\rm H,0} = -\frac{gn_0}{2} + |\Delta_{\rm B,0}|
		\le 0.
\end{equation}
Notice that as soon as $\vert \Delta_{{\rm B},0} \vert \ge g n_0/2 $, 
$\Delta_{{\rm H}, 0}$ must vanish, meaning that when the interaction strength $g$ is sufficiently strong, that is, 
$ g  \ge 
2 \vert \Delta_{{\rm B}, 0} \vert/n_0$, 
there are no Hartree corrections.
Therefore, we arrive to the closed analytical form for the Hartree order parameter
\begin{equation}
\label{eqn:self-consistency-equation-hartree-full}
\Delta_{{\rm H},0} = \min\left(0; -\frac{g n_0}{2} + |\Delta_{{\rm B},0}|\right).
\end{equation}

Next, we turn our attention to the superfluid order parameter given in Eq.~(\ref{eqn:bogoliubov-order-parameter}). Restricting ourselves to three spatial dimensions $(d = 3)$, where the saddle point solutions are a reasonable starting point, we take the thermodynamic limit $\{ N,V \}\to \infty$ with finite density $n = N/V$, and transform the summations over ${\bf k}$ into three-dimensional integrals using the prescription
$\sum_\mathbf{k} \to V\int {\rm d}^3\mathbf{k}/(2\pi)^3$. This procedure
leads to 
\begin{equation}
\label{eqn:unregularized-bogoliubov-order-parameter-equation}
\Delta_{\rm B,0}\left[ \frac{1}{g_{\rm B,0}} - \int \frac{\mathrm{d}^3k}{(2\pi)^3} \frac{\tanh\left( \frac{\beta E_{\mathbf{k}}}{2} \right)}{2E_{\mathbf{k}}} \right]
= 0.
\end{equation}

Naturally, there are two types of solutions for this equation. The first is the trivial one with
$\Delta_{{\rm B}, 0} = 0$ and the second is the solution of 
\begin{equation}
\label{eqn:unregularized-bogoliubov-order-parameter-equation-v2}
\frac{1}{g_{\rm B,0}} - \int \frac{\mathrm{d}^3k}{(2\pi)^3} \frac{\tanh\left( \frac{\beta E_{\mathbf{k}}}{2} \right)}{2E_{\mathbf{k}}}
= 0.
\end{equation}
This order parameter equation is similar to the standard one where the Hartree term is ignored~\cite{Engelbrecht-1993}, in that
case we have $g_{{\rm B},0} \to g$ since $b_0 \to 1$.
However, in the presence of the Hartree term $g_{{\rm B}, 0} = b_0 g$ additional care is necessary. The integral over momenta has an ultraviolet divergence that needs to be regularized, but the regularization procedure is slightly different from the standard one, because the interaction in the pairing channel $g_{{\rm B},0}$ is no longer the bare interaction $g$. Thus, we outline next the standard regularization procedure and show how it needs to be modified to regularize 
Eq.~(\ref{eqn:unregularized-bogoliubov-order-parameter-equation-v2}).   

\subsection{Two-body scattering renormalization}
	
For $g_{{\rm B}, 0} \to g $ ($b_0 \to 1$), the ultraviolet (UV) divergence in Eq.~(\ref{eqn:unregularized-bogoliubov-order-parameter-equation-v2}) is resolved 
	by taking advantage of two-body scattering theory. Since we are considering contact interactions,  we can use the Lippmann-Schwinger equation (LSE) ~\cite{Engelbrecht-1993, SadeMelo-1997,Radzihovsky-2007,Stoof-2008}
	to obtain the scattering phase shift
	\begin{equation}
		\label{eqn:phase-shift}
		q\cot\delta (\mathbf{q}) = \frac{4\pi}{mg} - \frac{2k_{\rm c}}{\pi} + \frac{2}{\pi k_{\rm c}}q^2 + \mathcal{O}(q^4)\text{,}
	\end{equation}
	where ${\bf q}$ is the 
	center of mass momentum of the scattering fermions, and $k_{\rm c}$ plays the role of the UV cutoff. 
	A direct comparison of
	Eq.~(\ref{eqn:phase-shift}) 
	with the phase shift 
	\begin{equation}
		\label{eqn:phase-shift-spherical-potential}
		q\cot\delta (\mathbf{q}) = -\frac{1}{a_{\rm s}} + \frac{r_{\rm e}}{2}q^2 + \mathcal{O}(q^4)
	\end{equation}
	for spherically symmetric potentials,
	where $a_{\rm s}$ denotes the s-wave scattering length and $r_{\rm e}$ stands for the effective range \cite{Bethe-1949,Madsen-2002},
	leads to the relation 
	\begin{equation}
		\label{eqn:LSE}
		\frac{1}{g(k_{\rm c})} = -\frac{m}{4\pi a_{\rm s}} + \frac{m}{2\pi^2}k_{\rm c}
	\end{equation}
for the leading $q^0$ term.
	This is the well-known renormalization condition for the bare coupling strength $g$ \cite{Radzihovsky-2007,Engelbrecht-1993,SadeMelo-1997}. Comparing the coefficients of the $q^2$ term in 
    Eqs.~(\ref{eqn:phase-shift}) and~(\ref{eqn:phase-shift-spherical-potential}) provides a direct connection between
	the UV cutoff $k_{\rm c}$ and the effective range $r_{\rm e}$ \cite{Ramanan-2021} given by
	\begin{equation}
		\label{eqn:two-body-cutoff}
		k_{\rm c}(r_{\rm e}) = \frac{4}{\pi r_{\rm e}}.
	\end{equation}
	Introducing the resonance value of the coupling strength $g_\ast(r_{\rm e})= \pi^3 r_{\rm e}/(2m)$, we use Eqs.~(\ref{eqn:LSE}) and ~(\ref{eqn:two-body-cutoff}) to 
    rewrite the s-wave scattering length as
	\begin{equation}
		\label{eqn:Scattering-length-explicit}
		a_{\rm s} = \frac{\pi^2}{8} r_{\rm e}\frac{g/g_\ast(r_{\rm e})}{g/g_\ast(r_{\rm e}) - 1}.
	\end{equation}
This shows a finite background scattering length for infinite attractive interactions $g \to \infty$, that is, 
$a_s (g \to \infty) = r_e \pi^2/8$.

\begin{figure}[t]
\centering\includegraphics[width=1\linewidth]{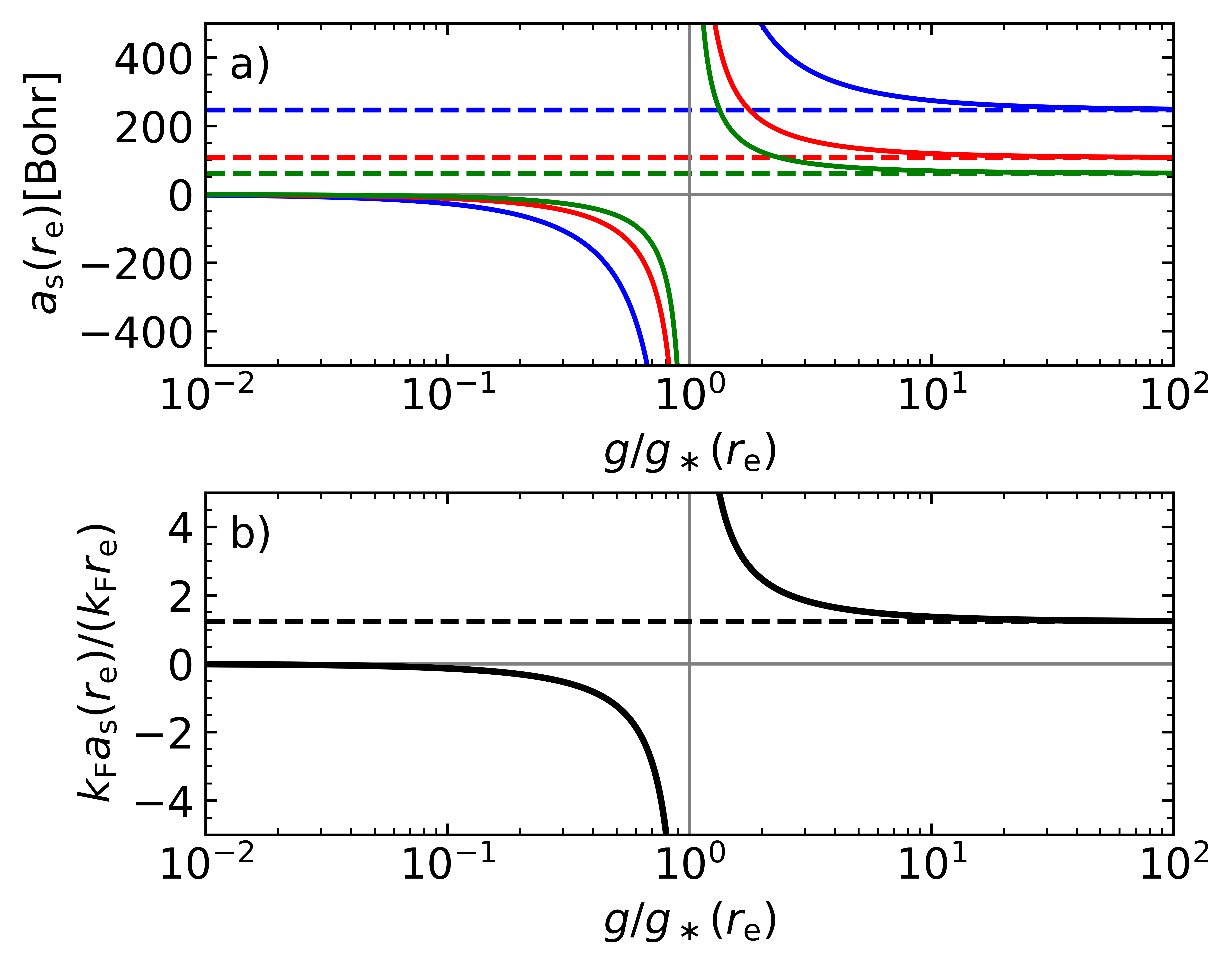}
\caption{Relation between scattering length 
$a_{\rm s}$ and bare interaction strength 
$g$ in units of the resonance interaction 
strength $g_\ast (r_{\rm e})$. Panel a) shows $a_s$, in units of the Bohr radius $a_{\rm Bohr}$, versus $g/g_* (r_{\rm e})$ for
different effective ranges 
$r_{\rm e} = 200\,a_{\rm Bohr}$ (solid blue line), 
$87\,a_{\rm Bohr}$ (solid red line) and 
$50\,a_{\rm Bohr}$ (solid green line). 
The dashed lines indicate the 
asymptotes for infinite interactions 
$(g \to \infty)$ corresponding to the background scattering length associated to that effective range. 
Panel b) displays the universal behavior of the inverse scattering parameter $k_{\rm F} a_{\rm s}$, in units of the effective range $k_{\rm F} r_{\rm e}$, versus 
$g/g_* (r_{\rm e})$. 
}
\label{fig:Scattering-lengths}
\end{figure}

In Fig.~\ref{fig:Scattering-lengths}, we visualize the behavior of the scattering length $a_{\rm s}$ versus the rescaled coupling strength $g/g_\ast(r_{\rm e})$ in two panels. In Fig.~\ref{fig:Scattering-lengths}a, we show that $a_{\rm s}$, in units of the Bohr radius $a_{\rm Bohr}$,  versus $g/g_\ast(r_{\rm e})$ strongly depends on the effective range $r_{\rm e}$;
see plots for effective ranges 
$r_{\rm e} = 200\,a_{\rm Bohr}$ (solid blue line), 
$87\,a_{\rm Bohr}$ (solid red line) and 
$50\,a_{\rm Bohr}$ (solid green line).
In Fig.~\ref{fig:Scattering-lengths}b, we display the universal behavior of the inverse scattering parameter $k_{\rm F} a_{\rm s}$, in units of the effective range $k_{\rm F} r_{\rm e}$, versus 
$g/g_* (r_{\rm e})$. Here, $ k_{\rm F} = (3 \pi^2n)^{1/3}$ is the Fermi momentum defined by the 
particle density $n$. The analytical expression for the universal behavior is
\begin{equation}\label{eqn:Rescaled-Scattering}
\frac{k_{\rm F}a_{\rm s}}{k_{\rm F}r_{\rm e}} = \frac{\pi^2}{8} \frac{g/g_\ast(r_{\rm e})}{g/g_\ast(r_{\rm e}) - 1}.
\end{equation}
For $g_{{\rm B}, 0} \to g$, that is, $b_0 \to 1$, the 
UV cutoff $k_{\rm c}$ and the zero-ranged interaction strength $g$ can be directly eliminated in favor of the s-wave scattering length $a_{\rm s}$ only. 
However, for $g_{{\rm B},0} \ne g$, that is, $b_0 \neq 1$, we cannot simultaneous eliminate in 
Eq.~(\ref{eqn:unregularized-bogoliubov-order-parameter-equation-v2}) both $g$ and $k_{\rm c}$ in favor of $a_{\rm s}$. 
	
Since we are interested in scattering processes simultaneously involving particle-particle (Bogoliubov) and particle-hole (Hartree) channels, it is necessary to modify standard scattering theory, described above,  to provide a suitable regularization when both sectors are present. Thus, next, we discuss how to implement such a procedure within our approach.
	
\subsection{Effective many-body scattering renormalization}\label{sec:many-body-scattering}
	
Since the interaction $g_{{\rm B},0} = b_0 g$
is a fractionalization of
the bare interaction $g$ into the Bogoliubov channel due to the existence of the Hartree order parameter, 
we need to renormalize Eq.~(\ref{eqn:unregularized-bogoliubov-order-parameter-equation-v2}) to reflect 
this many-body effect. This is achieved
by using the LSE equation for $g_{{\rm B},0}$ as
\begin{equation}
\label{eqn:Many-Body-Renormalization}
\frac{1}{g_{{\rm B},0}(k_{\rm c, B})} = -\frac{m}{4\pi a_{\rm s}} + \frac{m}{2\pi^2}k_{\rm c, B}\,,
\end{equation} 
and writing the many-body UV cutoff 
	\begin{equation}
		k_{\rm c,B} = \frac{4}{\pi r_{\rm e, {\rm B}}}\label{eqn:Many-Body-Cutoff}
	\end{equation}
	in terms of the many-body effective range $r_{\rm e, B}$,
    in analogy to Eq.~(\ref{eqn:two-body-cutoff}).
	Using $g_{\rm B, 0} = b_0 g$ and the expression for $1/g$ from Eq.~(\ref{eqn:LSE}), we obtain the many-body cutoff
	\begin{equation}
 \label{express}
		k_{\rm c, B} = \frac{k_{\rm c}}{b_0} - \frac{\pi}{2a_{\rm s}} \frac{h_0}{b_0}\,,    
	\end{equation}
	which must be used to remove the UV divergence in Eq.~(\ref{eqn:unregularized-bogoliubov-order-parameter-equation-v2}).
	We can rewrite the expression in Eq.~(\ref{express})
	as 
	\begin{equation}\label{eqn:cutoff-comparison}
		\frac{k_{\rm c, B}}{k_{\rm c}}
		= 
		\frac{1}{b_0} 
		\left(
		1 - \frac{\pi h_0}{2k_{\rm c} a_{\rm s}}
		\right)
		= 
		\frac{1}{b_0}
		\left(
		1 - h_0\frac{\pi^2}{8} \frac{k_{\rm F} r_{\rm e}}
		{k_{\rm F} a_{\rm s}}
		\right),
	\end{equation}
	where $k_{\rm c}$, from Eq.~(\ref{eqn:two-body-cutoff}), was used on the right-hand side.
	Since we have $0 \le h_0 \le 1$ and $0 \le b_0 \le 1$ with the constraint $h_0 + b_0 =1$, it is immediately apparent that $k_{\rm c,B} \ge k_{\rm c}$, when the interaction parameter lies in the interval
	$-\infty < 1/(k_{\rm F} a_{\rm s}) \le 0$.  Physically this means that the interaction $g_{\rm B}$ has shorter effective range 
	than the bare interaction $g$, that is, the
	many-body effective range $r_{\rm e, B}$ is shorter than the two-body effective range $r_{\rm e}$. The ratio between the effective ranges is 
	\begin{equation}\label{eqn:range-comparison}
		\frac{r_{\rm e, B}}{r_{\rm e}} =
		\frac{b_0}{			1 - h_0\frac{\pi^2}{8} \frac{k_{\rm F} r_{\rm e}}
			{k_{\rm F} a_{\rm s}}}.
	\end{equation}
\begin{figure}[t]
\centering\includegraphics[width=.95\linewidth]{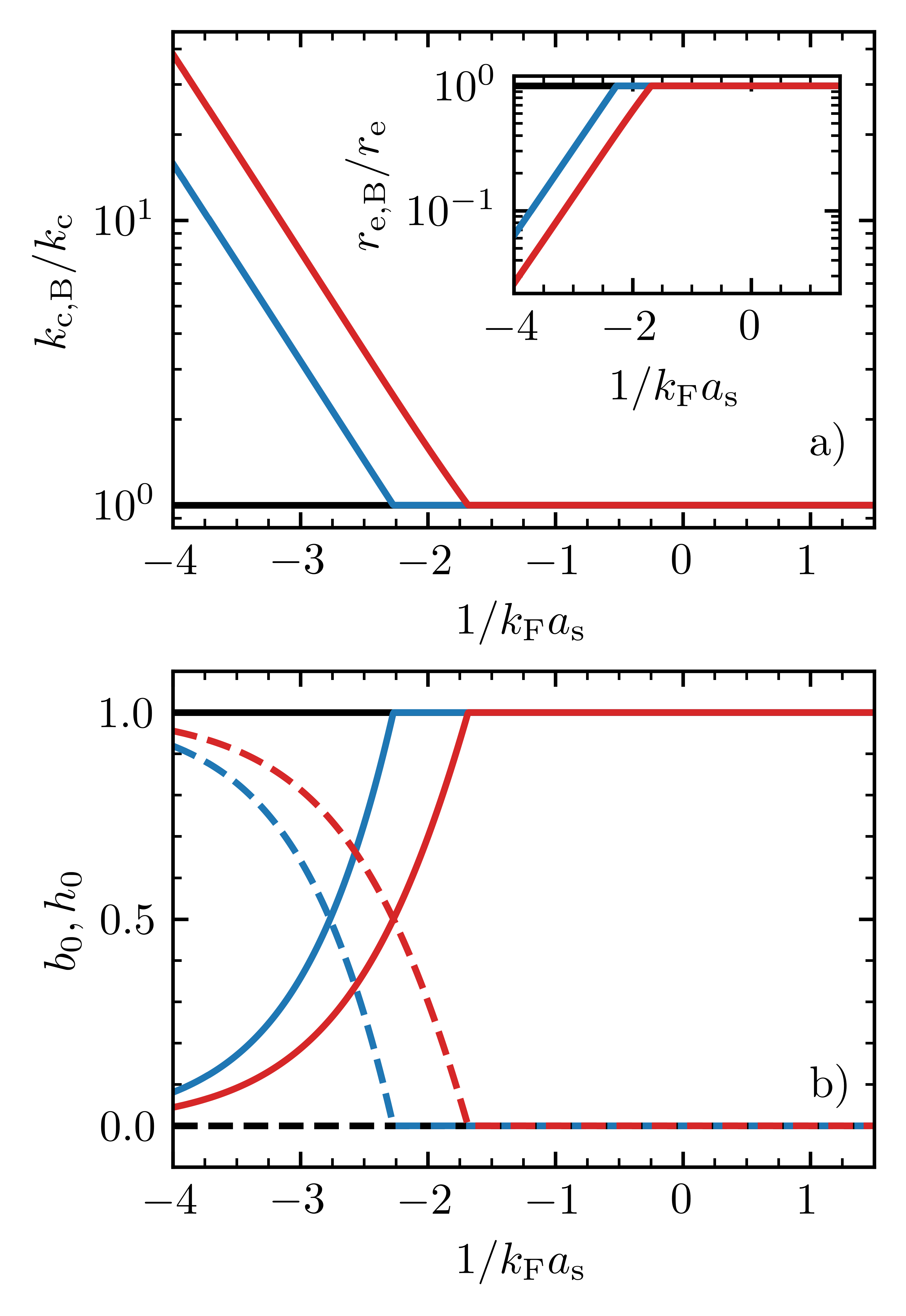}
\caption{The many-body UV cut-off $k_{\rm c, B}$ in units of the two-body cut-off $k_{\rm c}$, the many-body effective range $r_{\rm e, B}$ in units of the two-body effective range $r_{\rm e}$ and the Hartree $(h_0)$ and Bogoliubov $(b_0)$ weights are plotted versus the scattering parameter $1/k_{\rm F} a_{\rm s}$.
The effective range parameters used are $k_{\rm F}r_{\rm e} = 0$ (black lines), $0.0625$ (blue lines) and $0.1535$ (red lines).
In panel a), $k_{\rm c, B}/k_{\rm c}$ 
(main figure) and $r_{\rm e, B}/r_{\rm e}$ (inset) 
versus $1/k_{\rm F} a_{\rm s}$ are shown, where the $y$-axis is scaled logarithmically.
In panel b), the weights $h_0$ (dashed lines) and $b_0$ (solid lines) versus $1/k_{\rm F} a_{\rm s}$ are displayed.
}
\label{fig:cutoff-range-compare}
\end{figure}

Note that $r_{\rm e, B}/r_{\rm e} \le 1$ in the interval $-\infty <  1/k_{\rm F} a_{\rm s} \le 0$, becoming one only when $h_0 \to 0$, i.e.~$b_0 \to 1$.
Further insight into
this behavior is gained by analyzing the 
first two terms on the right-hand side of Eq.~(\ref{eqn:saddle-point-thermodynamic-potential}), as we shall see next.
	
\subsection{Non-analytic coupling between Bogoliubov and Hartree channels}
	
The first two terms on the right-hand side of Eq.~(\ref{eqn:saddle-point-thermodynamic-potential}) involve the order parameters $\Delta_{{\rm B},0}$ and 
$\Delta_{{\rm H},0}$ and the weighted interaction strengths $g_{{\rm B},0} = b_0g$ and $g_{{\rm H},0} = h_0g$ representing the Hartree and Bogoliubov channels, respectively. 
Using the expressions for $h_0$ and $b_0$ in Eqs.~(\ref{eqn:h0}) and (\ref{eqn:b0}), we get 
\begin{equation}
\label{eqn:non-analytical-coupling}
		\frac{|\Delta_{{\rm B},0}|^2}{g_{{\rm B},0}} + \frac{\Delta_{{\rm H},0}^2}{g_{{\rm H},0}} = \frac{|\Delta_{{\rm B},0}|^2}{g} + \frac{\Delta_{{\rm H},0}^2}{g} + \frac{2}{g}|\Delta_{{\rm B},0}||\Delta_{{\rm H},0}|\,,
\end{equation}
where we used the identity $\vert \Delta_{{\rm H},0} \vert^2 = \Delta_{{\rm H}, 0}^2$, whenever needed, since $\Delta_{{\rm H}, 0}$ is real. The result
in Eq.~(\ref{eqn:non-analytical-coupling}) shows explicitly a non-analytic
and a non-perturbative
coupling between the Hartree and Bogoliubov channels already at the saddle-point level via
the term $2\vert \Delta_{{\rm B},0} \vert \vert \Delta_{{\rm H}, 0} \vert/g$. 
	
Considering the non-analytic coupling and the effective many-body scattering renormalization, the Hartree's order parameter in Eq.~(\ref{eqn:self-consistency-equation-hartree-full}) becomes
\begin{equation}
		\label{eqn:self-consistency-equation-hartree-full-v2}
		\Delta_{{\rm H},0} = \min\left(0; \frac{n_0}{2}\left[ \frac{m}{4\pi a_{\rm s}} - \frac{2m}{\pi^3 r_{\rm e}} \right]^{-1} + |\Delta_{{\rm B},0}|\right)\,, 
\end{equation}
while the Bogoliubov's order parameter 
in Eq.~(\ref{eqn:unregularized-bogoliubov-order-parameter-equation}) reduces to 
	\begin{equation}
		\label{eqn:self-consistency-equation-bogoliubov-full}
		\Delta_{{\rm B},0}\bigg\{\frac{m}{4\pi a_{\rm s}} - \int\limits_{I} \frac{\text{d}^3k}{(2\pi)^3} \bigg[ \frac{m}{\mathbf{k}^2} - \frac{\tanh(\frac{\beta}{2}E_\mathbf{k})}{2E_\mathbf{k}} \bigg]\bigg\} = 0,
	\end{equation}
    where $I = \{|\mathbf{k}| \le k_{\rm c, B}\}$ defines the integration volume.
	A direct consequence of the analysis above is that the expressions for the Hartree and Bogoliubov order parameters shown in Eqs.~(\ref{eqn:self-consistency-equation-hartree-full-v2}) and~(\ref{eqn:self-consistency-equation-bogoliubov-full}), together
	with the particle density 
	in Eq.~(\ref{eqn:saddle-point-number}), elliminate a well-known divergence that emerges when these channels are not properly considered. Ignoring particle-hole effects and a divergence at unitarity, as is done in textbooks~\cite{Pitaevskii-2016}, is not a solution for the difficulty but rather an avoidance of the issue.
	
	Without considering the proper counting (partitioning) of states 
	and regularization introduced here, prior attempts of including the simultaneous effects of particle-particle and particle-hole channels led to ultraviolet divergences in the Hartree order parameter Eq.~(\ref{eqn:hartree-order-parameter})~\cite{Kleinert-2011,Pitaevskii-2016,Widera-2023}. Thus, next, we discuss results that explicitly show the fixing of this well-known issue while including simultaneously properly partitioned and regularized particle-particle (Bogoliubov) and particle-hole (Hartree) sectors.

    \subsection{Self-consistency and implications of many-body renormalization}

All the relations derived above, Eqs.~(\ref{eqn:self-consistency-equation-hartree-full-v2}) and \ref{eqn:self-consistency-equation-bogoliubov-full}) for the order parameters, Eq.~(\ref{eqn:saddle-point-number}) for the number density, Eqs.~(\ref{eqn:h0}) and~(\ref{eqn:b0}) for the weight parameters, and Eq.~(\ref{eqn:Many-Body-Cutoff}) for the effective many-body range, form a set of transcendental equations that has to be solved self-consistently. 
For instance, the many-body effective range $r_{\rm B,e}$ depends explicitly not only on the two-body effective range $r_{\rm e}$ and the s-wave scattering length $a_{\rm s}$, but also on the weights $h_0$ and $b_0$ as 
seen in Eq.~(\ref{eqn:range-comparison}). However, $h_0$ and $b_0$ also 
dependent on the order parameters, which are explicit functions of
 $r_{\rm B,e}$ and $r_{\rm e}$, thus closing the self-consistency conditions. 
In order to obtain the full solution, we solve all those equations simultaneously by a numerical algorithm, where a standard iterative procedure is applied. Among the solutions obtained there are two distinct families of quantities determined. The first are auxiliary quantities, that is, the many-body effective range $r_{\rm B,e}$ and the weight factors $h_0$ and $b_0$, which are discussed below in this section, and the second are thermodynamic quantities, that is, $\Delta_{\rm B, 0}$, $\Delta_{\rm H, 0}$ and $\mu$ discussed in Sec.~\ref{sec:four}.
    
In Fig.~\ref{fig:cutoff-range-compare}, we plot auxiliary quantities, that is, the many-body UV cut-off $k_{\rm c, B}$ in units of the two-body cut-off $k_{\rm c}$, the many-body effective range $r_{\rm e, B}$ in units of the two-body effective range $r_{\rm e}$, and the Hartree $(h_0)$ and Bogoliubov $(b_0)$ weights calculated at $T=0$ versus the scattering parameter $1/k_{\rm F} a_{\rm s}$. The effective range parameters used are $k_{\rm F}r_{\rm e} = 0$ (black line), $0.0625$ (blue line) and $0.1535$ (red line). Further discussions about the effective range parameter $k_{\rm F}r_{\rm e}$ are found at the beginning of Sec.~\ref{sec:four}. 
    
In Fig.~\ref{fig:cutoff-range-compare}a, we display 
$k_{\rm c, B}/k_{\rm c}$ 
(main figure) and $r_{\rm e, B}/r_{\rm e}$ (inset) 
versus $1/k_{\rm F} a_{\rm s}$ in a semilog plot. Note that 
$k_{\rm c, B}/k_{\rm c}$ $(r_{\rm e, B}/r_{\rm e})$ decreases (increases) exponentially towards 1, as 
$1/k_{\rm F} a_{\rm s}$ grows. 
In the weakly interacting regime, an exponential behavior occurs because the Hartree and Bogoliubov channels are competing for the interaction energy. However, beyond a critical value of $1/k_{\rm F} a_{\rm s}$, where the Hartree weight factor $h_0$ vanishes, 
the many-body $k_{\rm c, B}$ and the two-body $k_{\rm c}$
UV cut-offs coincide, that is, the system is fully determined by two-body properties.

In Fig.~\ref{fig:cutoff-range-compare}b, we show the Hartree $h_0$ (dashed lines) and Bogoliubov $b_0$ (solid lines) weights versus $1/k_{\rm F} a_{\rm s}$ for effective range parameters $k_{\rm F}r_{\rm e} = 0$ (black lines), $0.0625$ (blue lines) and $0.1535$ (red lines). Note that $h_0$ $(b_0)$ converges to $0$ $(1)$ beyond a critical value of $1/k_{\rm F} a_{\rm s}$, which moves closer to unitarity $(1/k_{\rm F}a_{\rm s} = 0)$ with increasing effective range parameter $k_{\rm F}r_{\rm e}$. Beyond this critical value, the physical properties are determined by the particle-particle channel $(b_0 = 1, h_0 = 0)$, while for weaker interactions the particle-hole and particle-particle channels compete for the interaction energy
$(b_0 \ne 0, h_0 \ne 0)$.
    
Having discussed this general behavior at $T=0$, we discuss next ground state properties including phase diagrams, order parameters, chemical potential and ground state energy.  
 
	\section{Ground-State Properties}
	\label{sec:four}
	
	In this section, we specialize our theory to zero temperature and
    discuss a few ground-state properties based on the self-consistency relations mentioned in the previous section. Even though quantum fluctuations are known to play a large role, ground-state analysis, especially in the weakly interacting limit, is a suitable first approximation already at the saddle-point level.
	Our results are expressed in Fermi units, that is,
    our unit of energy is $\varepsilon_{\rm F} = k_{\rm F}^2/2m$ and our unit of momentum is $k_{\rm F} = (3\pi^2n)^{1/3}$, since we are using $\hbar = k_{\rm B} = 1$. For instance, the dimensionless scattering parameter is given by $1/k_{\rm F}a_{\rm s}$ and the dimensionless effective range is $k_{\rm F}r_{\rm e}$.
	
	The two-particle effective range $r_{\rm e}$ is a property of the interaction potential. For a specific particle species, $r_{\rm e}$ is a constant over a broad Feshbach resonance~\cite{Hutson-2014}. 
	All dimensionless thermodynamic quantities depend on the dimensionless effective range $k_{\rm F}r_{\rm e}$,
	which describes the ratio between the two-particle effective range $r_{\rm e}$ and the typical interparticle spacing $k_{\rm F}^{-1}$ fixed by the density $n$.
	Quite naturally, the larger $k_{\rm F} r_{\rm e}$ the stronger is the deviation from universality of ultracold Fermi gases at unitarity.

In what follows, we discuss first the emergence of a new phase in the ground state (Sec.~\ref{SubSec:4A}) before we analyze thermodynamic properties of this new phase as well as the effects of the interaction partitioning on the standard superfluid phase (Secs.~\ref{SubSec:4B}-\ref{SubSec:4E}). 

In Sec.~\ref{SubSec:4A}, we show a plot of our $T = 0$ phase diagram, and in Secs.~\ref{SubSec:4B}-\ref{SubSec:4E}, the main plots show results using our partitioning method, while the insets show the standard approach~\cite{Pitaevskii-2016,Kleinert-2011} that weights both channels by a factor of $1$: $g_{\rm H,0} = g = g_{\rm B,0}$.
In these figures we use the effective range $k_{\rm F} r_{\rm e} = 0$ (black line), $k_{\rm F} r_{\rm e} = 0.0625$ (blue line), and $k_{\rm F} r_{\rm e} = 0.1535$ (red line).
As the effective range is a quantity fixed by the interaction potential, one can change the dimensionless effective range by adjusting the density of the system.
For example, $^{6}{\rm Li}$ has a two-body effective range $r_{\rm e} = 87\,a_0$ throughout the broad s-wave Feshbach resonance centered at $832~{\rm Gauss}$~\cite{Hutson-2014}, with $a_0$ being the Bohr radius.
For $^{6}{\rm Li}$, the value of $k_{\rm F} r_{\rm e} = 0.0625$ (blue line) 
corresponds to the density $n = 8\times10^{13}/{\rm cm}^3$, while for $k_{\rm F} r_{\rm e} = 0.1535$ (red line) the density is $n = 1\times10^{15}/{\rm cm}^3$.
The latter value represents an exaggerated density, which has not yet been realized experimentally.
However, for an effective range of $r_{\rm e} = 214\,a_0$ the value
$k_{\rm F}r_{\rm e} = 0.1538$ (red line) corresponds to the realistic density
of $n \approx 8\times10^{13}/\mathrm{cm}^3$.
The same color code holds also for the insets, where we additionally visualize the approach used in Ref.~\cite{Pitaevskii-2016, Kleinert-2011, Widera-2023}. There the replacement $g \propto a_{\rm s}$ is shown by the magenta line.

\subsection{Phase Diagram}\label{SubSec:4A}
    
The main consequence of the WHFB theory is the emergence of the 
non-analytic coupling between Hartree $\Delta_{\rm H,0}$ and Bogoliubov $\Delta_{\rm B,0}$ order parameters, which allows for the possibility of a vanishing $\Delta_{\rm H,0}$, which otherwise would be impossible, if we had chosen an arbitrary partitioning as done in standard theories. In other words, if the WHFB theory is not used, $\Delta_{\rm H,0}$ becomes simply a non-vanishing Hartree shift due to an arbitrary partitioning of the interactions. Thus, the WHFB theory exposes the existence of two superfluid phases at $T = 0$,
where $\Delta_{\rm B,0} \ne 0$. In one phase, which we call the Hartree superfluid (HSF), the Hartree order parameter is non-zero, that is, $\Delta_{\rm H,0} \neq 0$; while in the other phase, 
which we call the standard superfluid (SSF), the Hartree order parameter vanishes, that is, $\Delta_{\rm H,0} = 0$. 
\begin{figure}[t]
\centering
\includegraphics[width=1\linewidth]{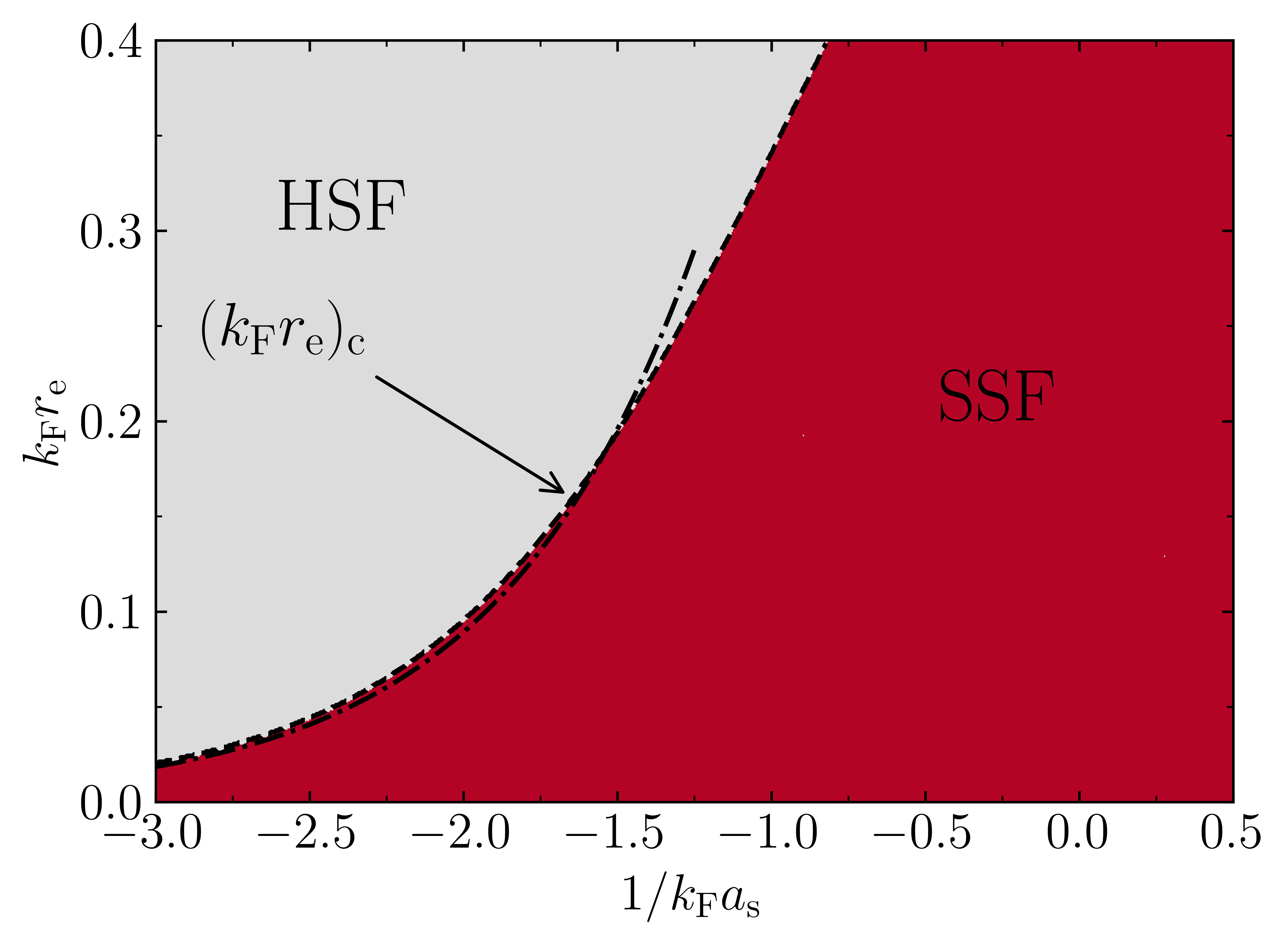}
\caption{The ground-state ($T=0$) phase diagram in the plane of $k_{\rm F} r_{\rm e}$ versus $k_{\rm F} r_{\rm e}$, showing Hartree superfluid (HSF) and standard superfluid (SSF) phases. The dashed line indicates the numerically determined phase boundary, and the dash-dotted line represents an analytical asymptotic 
result. 
}
\label{fig:ZeroTPhase}
\end{figure}

In Fig.~\ref{fig:ZeroTPhase}, we show the resulting ground-state phase diagram $(T = 0)$ in the plane of $k_{\rm F} r_{\rm e}$ versus $1/k_{\rm F} a_{\rm s}$. 
The dashed black line describes the numerical phase boundary between the HSF phase (grey region) and the SSF phase (red region). 
This phase boundary is established via the condition given in 
Eq.~(\ref{eqn:self-consistency-equation-hartree-full-v2}),
which can be expressed as
\begin{equation}\label{eqn:HSF-SSF-condition}
\left(k_{\rm F}r_{\rm e}\right)_{\rm c} = \frac{8}{\pi^2}\left[ \frac{1}{k_{\rm F}a_{\rm s}} + \frac{4\varepsilon_{\rm F}}{3\pi\vert \Delta_{\rm B,0}\vert} \right]^{-1}, 
\end{equation}
when written in Fermi units.
The dash-dotted line represents an analytic result for the phase boundary, derived later in Eq.~(\ref{eqn:critical_scattering_parameter}).
Note that, when the scattering parameter $1/k_{\rm F} a_{\rm s}$ is very negative 
(weak coupling), it is easier to reach the HSF phase at fixed interactions since 
the critical $k_{\rm F} r_{\rm e}$ is smaller. However, for larger 
$1/k_{\rm F} a_{\rm s}$, towards unitary and beyond,
reaching the HSF phase requires larger 
$k_{\rm F} r_{\rm e}$. In summary, for fixed $1/k_{\rm F} a_{\rm s}$, the SSF phase is favored at lower effective range parameter $k_{\rm F} r_{\rm e}$, while 
the HSF phase is energetically more favorable
at larger $k_{\rm F} r_{\rm e}$.
	
Before discussing the quantitative differences between the HSF and the SSF phases, we briefly outline the general characteristics of the two order parameters to create an all-encompassing picture about the general trends.

\begin{figure}[tb]
\centering\includegraphics[width=.95\linewidth]{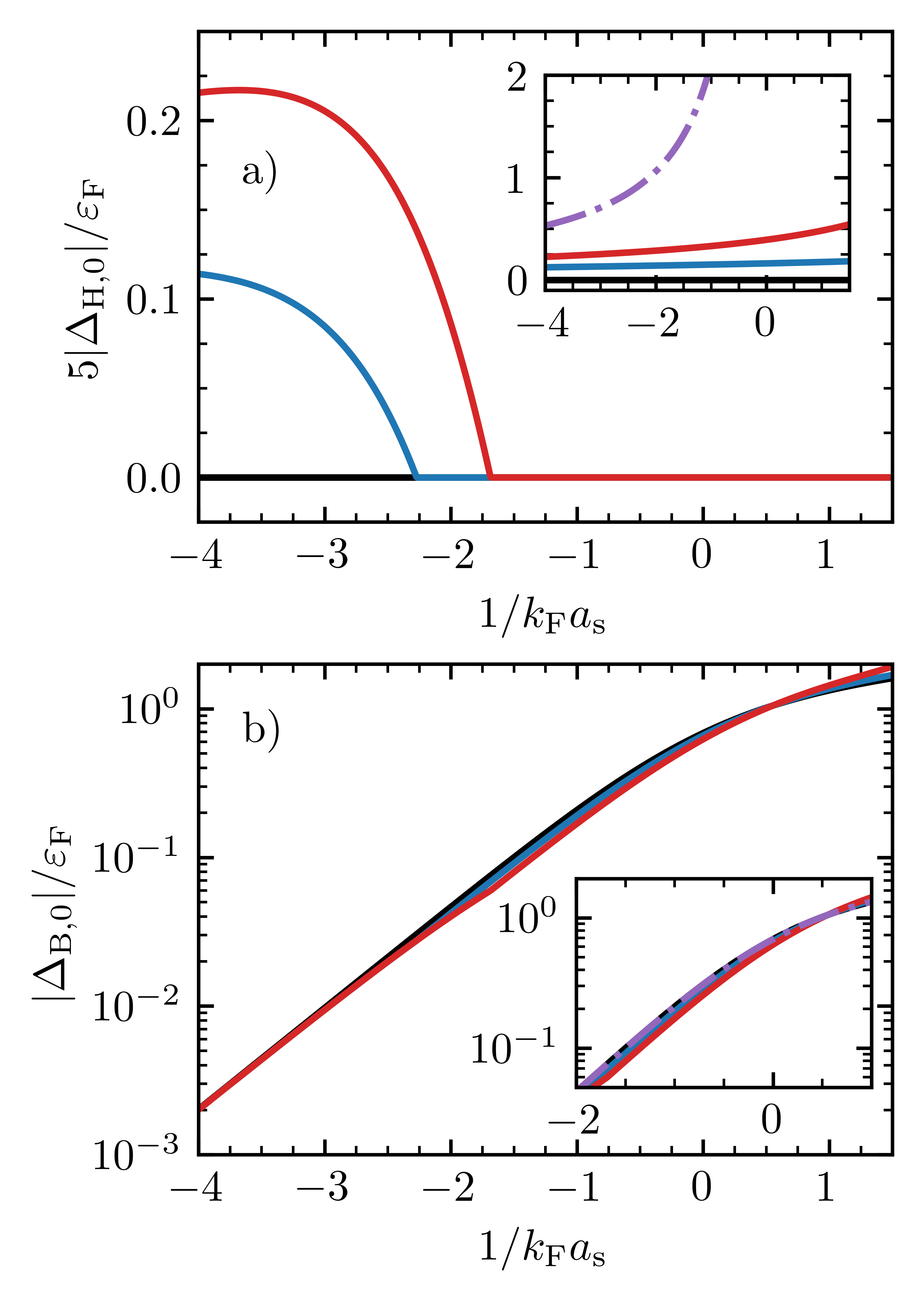}
\caption{Order parameters $\vert\Delta_{{\rm H},0}\vert/\varepsilon_{\rm F}$ and 
$\vert\Delta_{{\rm B},0}\vert/\varepsilon_{\rm F}$ versus $1/k_{\rm F} a_{\rm s}$, 
for various effective range parameters:
$k_{\rm F}r_{\rm e} = 0$ (solid black line), and $0.0625$ (solid blue line) and $0.1535$ (solid red line). 
Panel a) shows the Hartree order parameter 
$\vert\Delta_{{\rm H},0}\vert/\varepsilon_{\rm F}$
and panel b) depicts the superfluid order parameter 
$\vert\Delta_{{\rm B},0}\vert/\varepsilon_{\rm F}$. 
The main figures display the results from our WHFB method, while the insets reveal the behavior predicted by theories with equally-weighted Hartree and Bogoliubov channels. 
The dash-dotted purple line represents the results without proper many-body scattering renormalization, while the solid
black, blue and red lines include the many-body scattering renormalization within the equally-weighted approach.
}
\label{fig:SCESol1Bogo}
\end{figure}

In Fig.~\ref{fig:SCESol1Bogo}, we show the dimensionless modulus of the order parameters $\vert\Delta_{\rm H,0}\vert/\varepsilon_{\rm F}$ and $\vert\Delta_{\rm B,0}\vert/\varepsilon_{\rm F}$ versus $1/k_{\rm F} a_{\rm s}$ for different theories. The main figures show the results from WHFB method, while the insets reveal the behavior predicted by theories with equally-weighted Hartree and Bogoliubov channels~\cite{Kleinert-2011,Pitaevskii-2016,Pethick-2008}. 
The parameters used are 
$k_{\rm F}r_{\rm e} = 0$ (solid black line), and $0.0625$ (solid blue line) and $0.1535$ (solid red line). The dash-dotted purple line represents the results without proper many-body scattering renormalization, while the solid
black, blue and red lines include the many-body scattering renormalization within the equally-weighted approach. 

In Fig.~\ref{fig:SCESol1Bogo}a, the main plots describes $\vert\Delta_{\rm H,0}\vert/\varepsilon_{\rm F}$ versus $1/k_{\rm F} a_{\rm s}$ using our WHFB method. Note the factor of 5 (five) on the vertical scale both in the main and inset plots. 
As seen in the main plot, $\vert\Delta_{\rm H,0}\vert/\varepsilon_{\rm F}$ vanishes for $k_{\rm F} r_{\rm e} = 0$ (solid black line) for all values of $1/k_{\rm F}a_{\rm s}$, but it is non-zero in some region of
$1/k_{\rm F}a_{\rm s}$ as $k_{\rm F} r_{\rm e}$ increases.
Notably, for each value of  $1/k_{\rm F} a_{\rm s}$, it is straightforward to read out the value of  $(k_{\rm F}r_{\rm e})_{\rm c}$ (see phase 
diagram in Fig.~\ref{fig:ZeroTPhase}) 
beyond which $\vert\Delta_{\rm H,0}\vert/\varepsilon_{\rm F}$ vanishes. 
Physically this means that there is no Hartree shift of the chemical potential at the saddle-point level and
that the pairing (Bogoliubov) channel fully controls the saddle-point physics when $k_{\rm F}r_{\rm e} < (k_{\rm F}r_{\rm e})_{\rm c}$. 
In other words, below $(k_{\rm F}r_{\rm e})_{\rm c}$, any renormalization of the chemical potential must arise from fluctuations.

In Fig.~\ref{fig:SCESol1Bogo}a, the main plots also reveal that the Hartree order parameter 
$\vert\Delta_{\rm H,0}\vert/\varepsilon_{\rm F}$ vanishes continuously, indicating that the saddle-point description predicts a continuous HSF-SSF phase transition. Defining $\eta = 1/k_F a_s$, the Hartree order parameter
behaves near the transition point $\eta_{\rm c}$ as 
$\vert \Delta_{\rm H,0} \vert \sim (\eta_{\rm c} - \eta)^\beta$, 
for $\eta < \eta_{\rm c}$, where $\beta = 1$. This exponent is different from the standard Gaussian fixed point $(\beta = 1/2)$ of the conventional Ginzburg-Landau $\phi^4$ theory, because of the non-analytic term 
$2|\Delta_{{\rm B},0}||\Delta_{{\rm H},0}|/g$ that arises in the thermodynamic potential $\Omega_0$ of Eq.~(\ref{eqn:saddle-point-thermodynamic-potential}) using the relation in Eq.~(\ref{eqn:non-analytical-coupling}).
Furthermore, a detailed analysis reveals that, the isothermal compressibility 
$\kappa_{\rm T} = (\partial n/\partial \mu)/n^2\vert_{T,V}$ exhibits a discontinuity, 
thus producing a critical exponent $\gamma = 0$,  thermodynamically confirming the existence of a continuous phase transition. This discontinuity confirms that this transition belongs to a different universality class than the conventional Ginzburg-Landau $\phi^4$ theory, where the corresponding susceptibility diverges ($\gamma = 1$). A renormalization group analysis is necessary to obtain critical exponents beyond the saddle-point approximation, but we leave this effort for a future publication.

In Fig.~\ref{fig:SCESol1Bogo}a, the inset shows
results using standard methods where equal weights $h_0 = b_0 = 1$ $(g_{\rm H} = g_{\rm B} = g)$ are arbitrarily chosen~\cite{Pitaevskii-2016,Kleinert-2011,Kleinert-2016}.
The dash-dotted magenta line represents the solution without proper effective range renormalization~\cite{Pitaevskii-2016}, leading to a divergence in $\vert \Delta_{\rm H, 0} \vert$ at unitarity $(1/k_{\rm F} a_{\rm s} = 0)$. The results shown by the solid black, blue and red lines use a variational perturbation theory~\cite{Kleinert-2011,Kleinert-2016} and include the proper effective range renormalization, but still produce divergences 
in the Hartree contribution, now located at $1/k_{\rm F} a_{\rm s} = (8/\pi^2) (1/k_{\rm F} r_{\rm e})$.
By comparing Fig.~\ref{fig:SCESol1Bogo}a main and inset, it is evident that the 
self-consistent partitioning and regularization method within our WHFB approach solves this long-standing divergence in $\vert\Delta_{\rm H,0}\vert$ and provides physically acceptable results with clear interpretations.

In Fig.~\ref{fig:SCESol1Bogo}b, the main plot shows
$\vert\Delta_{\rm B,0}\vert/\varepsilon_{\rm F}$
versus $1/k_{\rm F}a_{\rm s}$. 
At fixed $1/k_{\rm F} a_{\rm s}$,
increasing the effective range parameter $k_{\rm F} r_{\rm e}$ reduces 
$\vert\Delta_{\rm B,0}\vert/\varepsilon_{\rm F}$
by a small amount for $1/k_{\rm F} a_{\rm s} < 0.55$. 
In contrast, increasing $k_{\rm F} r_{\rm e}$ enhances 
$\vert\Delta_{\rm B,0}\vert/\varepsilon_{\rm F}$ by a small amount for $1/k_{\rm F}a_{\rm s} > 0.55$. 
In Fig.~\ref{fig:SCESol1Bogo}b, the inset reveals results for $\vert\Delta_{\rm B,0}\vert/\varepsilon_{\rm F}$ versus $1/k_{\rm F} a_{\rm s}$ using the standard method, where $g_{\rm H} = g_{\rm B} = g$ $(h_0 = b_0 = 1)$. For $\vert\Delta_{\rm B,0}\vert/\varepsilon_{\rm F}$, the results of the two methods differ only by a few percent, because $\vert\Delta_{\rm B,0}\vert/\varepsilon_{\rm F}$ is much less sensitive to the renormalization of the chemical potential $\mu$ associated with the Hartree order parameter $\Delta_{\rm H,0}$, as is discussed in the next two sections.

From Fig.~\ref{fig:SCESol1Bogo}b, we can also extract important information in the weak coupling regime. For instance, the semilog plots show the exponential behavior of $\vert\Delta_{\rm B,0}\vert/\varepsilon_{\rm F}$ versus $1/k_{\rm F} a_{\rm s}$, 
that is, $\vert\Delta_{\rm B,0}\vert/\varepsilon_{\rm F} \sim 
e^{-1/\vert k_{\rm F}a_{\rm s}\vert}$ when 
$1/ k_{\rm F}a_{\rm s} \ll -1 $.
In this regime, the HSF-SSF phase boundary, given in Eq.~(\ref{eqn:HSF-SSF-condition}), satisfies the condition 
$(k_{\rm F}r_{\rm e})_{\rm c} \ll |k_{\rm F}a_{\rm s}|$.
When $k_{\rm F}r_{\rm e} < (k_{\rm F}r_{\rm e})_{\rm c}$ (see Fig.~\ref{fig:ZeroTPhase}), the SSF phase gives rise to the hierarchy of dimensionless length scales 
$k_{\rm F}r_{\rm e} <  (k_{\rm F}r_{\rm e})_{\rm c} \ll \vert k_{\rm F}a_{\rm s}\vert $.
Therefore, in weak coupling, the standard superfluid phase 
only exists when $k_{\rm F}r_{\rm e} \ll \vert k_{\rm F}a_{\rm s}\vert \ll 1$.
However, for $k_{\rm F}r_{\rm e} > (k_{\rm F}r_{\rm e})_{\rm c}$ (see Fig.~\ref{fig:ZeroTPhase}), the HSF phase  
allows for two hierarchies of dimensionless length scales:
either $k_{\rm F}r_{\rm e} \ll \vert k_{\rm F}a_{\rm s}\vert \ll 1$
or $\vert k_{\rm F}a_{\rm s}\vert \ll k_{\rm F}r_{\rm e} \ll 1$.
This distinction between hierarchies of dimensionless length scales is useful for our discussion of asymptotic limits in Sec.~\ref{SubSec:4B} and~\ref{SubSec:4C}.

Having identified a major issue in the literature and provided a physical solution to the problem, next we take a closer look at the two different phases that emerge from the regularization and the WHFB theory: the Hartree superfluid and the standard superfluid.
     
\subsection{Hartree superfluid (HSF)}\label{SubSec:4B}
	
The main characteristic of the HSF phase is the occurrence 
of a Hartree order parameter regularizing the chemical potential 
by applying the so called Hartree shift, ultimately 
lowering the chemical potential for weak interactions. 
Analyzing the many-body effective range, we can see that the Hartree order parameter captures the short distance physics of the system as it is strongly dependent on 
$k_{\rm F} r_{\rm e}$. 
In contrast, the superfluid order parameter $\Delta_{\rm B,0}$ is independent of $k_{\rm F} r_{\rm e}$ 
due to cancellations in the many-body effective range.
For effective ranges obeying $k_{\rm F}r_{\rm e} > (k_{\rm F}r_{\rm e})_{\rm c}$, 
the many-body UV cutoff in Eq.~(\ref{eqn:cutoff-comparison}) can be simplified using Eq.~(\ref{eqn:self-consistency-equation-hartree-full-v2}). This procedure leads to 
\begin{equation}\label{eqn:HSF-Cutoff}
\frac{k_{\rm c,B}}{k_{\rm F}} = \frac{2}{3}\frac{\varepsilon_{\rm F}}{\vert \Delta_{\rm B,0}\vert} - \frac{\pi}{2\vert k_{\rm F}a_{\rm s} \vert},
\end{equation}
which is independent of the two-body effective range $r_{\rm e}$. 
Due to Eq.~(\ref{eqn:HSF-Cutoff}), the corrections to the superfluid order parameter 
are only due to their coupling to the Hartree order parameter. 
In the HSF phase, the superfluid order parameter can be asymptotically solved for 
$\vert k_{\rm F} a_{\rm s} \vert \ll 1$ 
by cutting out the Fermi surface of the integral in Eq.~(\ref{eqn:self-consistency-equation-bogoliubov-full}) and using a constant density of states near $\varepsilon_{\rm F}$. This yields
	\begin{equation}
		\label{eqn:Bogoliubov-limit}
		\frac{\vert \Delta_{\rm B,0} \vert}{\varepsilon_{\rm F}} = 
		\frac{8}{e^{2}}\left(1-\frac{12}{e^2}e^{-\frac{\pi}{2\vert k_{\rm F}a_{\rm s} \vert}}\right)\exp\left(-\frac{\pi}{2 \vert k_{\rm F} a_{\rm s}\vert}\right)\, ,
	\end{equation}
	which gives an additional exponentially weak correction 
    to the standard superfluid order parameter. 
	Depending on the proximity of the effective range parameter 
	$k_{\rm F}r_{\rm e}$ to the critical effective range parameter 
	$\left(k_{\rm F}r_{\rm e}\right)_{\rm c}$,  
    we have 
	two different asymptotic limits. The one close to the phase 
	boundary, governed by $k_{\rm F}r_{\rm e} \ll \vert k_{\rm F}a_{\rm s} \vert \ll 1$, 
	shows that the Hartree order parameter depends 
	on $k_{\rm F} r_{\rm e}$ and 
    $\vert k_{\rm F} a_{\rm s} \vert$ 
    as follows   
\begin{eqnarray}
			\frac{\Delta_{\rm H,0}}{\varepsilon_{\rm F}} = &&  - \frac{\pi}{6}k_{\rm F}r_{\rm e} \left[ 1 - \frac{\pi^2}{8}\left(\frac{k_{\rm F}r_{\rm e}}{\vert k_{\rm F}a_{\rm s}\vert}\right) + \mathcal{O}\left(\frac{k_{\rm F}r_{\rm e}}{\vert k_{\rm F}a_{\rm s} \vert}\right)^2\right] \nonumber \\ &&  + \frac{8}{e^{2}}\left(1-\frac{12}{e^2}e^{-\frac{\pi}{2\vert k_{\rm F}a_{\rm s} \vert}}\right)\exp\left(-\frac{\pi}{2 \vert k_{\rm F} a_{\rm s}\vert}\right)\, . \label{eqn:Hartree-limit-inter}
\end{eqnarray}

As the interaction gets weaker, $\Delta_{\rm H, 0}$ looses its sensitivity to $k_{\rm F} r_{\rm e}$, as now the scattering parameter becomes the smallest length scale 
in the gas. 
In this regime, we enter the region  
$\vert k_{\rm F}a_{\rm s} \vert \ll k_{\rm F}r_{\rm e} \ll 1$, 
leading to the asymptotic expansion 
	\begin{eqnarray}
			\frac{\Delta_{\rm H,0}}{\varepsilon_{\rm F}} = && -\frac{4}{3\pi}\vert k_{\rm F} a_{\rm s} \vert
			\left[ 1 - \frac{8}{\pi^2}\frac{\vert k_{\rm F} a_{\rm s}\vert}{k_{\rm F} r_{\rm e}}
			+ \mathcal{O}\left(\frac{\vert k_{\rm F} a_{\rm s}\vert}{k_{\rm F} r_{\rm e}}\right)^2 \right] \nonumber \\ && + \frac{8}{e^{2}}\left(1-\frac{12}{e^2}e^{-\frac{\pi}{2\vert k_{\rm F}a_{\rm s} \vert}}\right)\exp\left(-\frac{\pi}{2 \vert k_{\rm F} a_{\rm s}\vert}\right). \label{eqn:Hartree-limit-BCS}
	\end{eqnarray}

The results in Eqs.~(\ref{eqn:Hartree-limit-inter}) and~(\ref{eqn:Hartree-limit-BCS}), arise from a series expansion of Eq.~(\ref{eqn:self-consistency-equation-hartree-full-v2}) in $k_{\rm F} r_{\rm e}/\vert k_{\rm F} a_{\rm s} \vert$ and  $\vert k_{\rm F} a_{\rm s} \vert/k_{\rm F} r_{\rm e}$, respectively. 
    
Since the Hartree order parameter contributes to the shift of the chemical potential $\mu$ from the Fermi energy 
$\varepsilon_{\rm F}$,
we analyze next 
$\mu/\varepsilon_{\rm F}$
in general, and in 
the two asymptotic limits
discussed above.

\begin{figure}[t]
\centering\includegraphics[width=.95\linewidth]{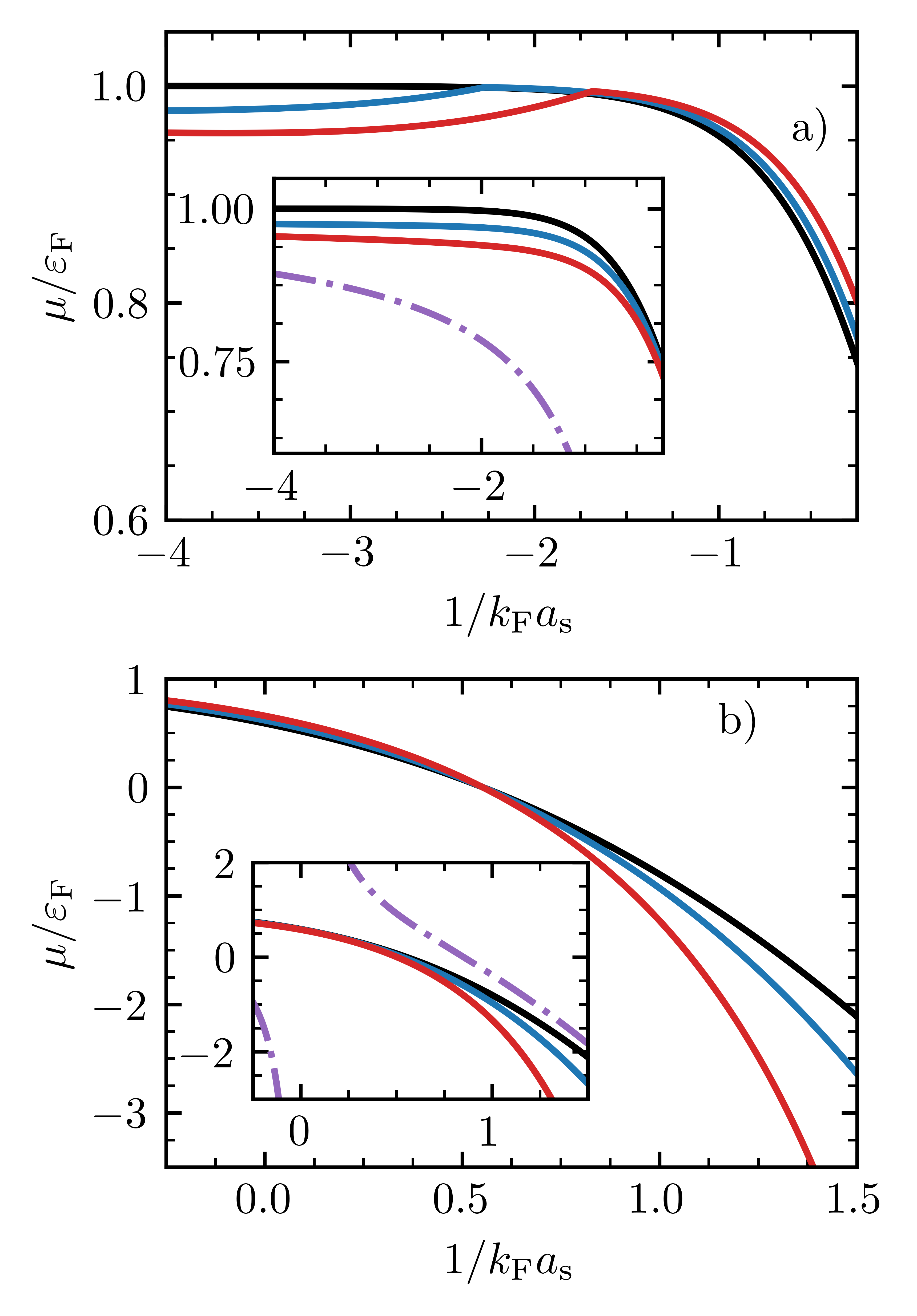}
\caption{Chemical potential 
$\mu/\varepsilon_{\rm F}$ versus
scattering parameter $1/k_{\rm F}a_{\rm s}$
for different effective range parameters 
$k_{\rm F}r_{\rm e} = 0$ (solid black line), $0.0625$ (solid blue line) 
and $0.1535$ (solid red line). In panel a), the region from the weakly interacting regime $1/k_{\rm F}a_{\rm s} < -1$ up to the unitary point $1/k_{\rm F}a_{\rm s} = 0$
is shown. In panel b), the region from the unitary point towards the BEC regime, around $1/k_{\rm F}a_{\rm s} = 0.55$ is emphasized. 
In both panels, the insets show the standard approach (equal weights), highlighting 
the unphysical divergence of $\mu$ (dash-dotted magenta line).}
\label{fig:SCESol1Hart}
\end{figure}

In Fig.~\ref{fig:SCESol1Hart}, we show 
the dimensionless chemical potential $\mu/\varepsilon_{\rm F}$ versus $1/k_{\rm F} a_{\rm s}$ for effective range parameters $k_{\rm F}r_{\rm e} = 0$ (solid black line), and $0.0625$ (solid blue line) and $0.1535$ (solid red line). 
The main figures show the results from WHFB method, while the insets reveal the behavior predicted by theories with equally-weighted Hartree and Bogoliubov channels~\cite{Kleinert-2011,Pitaevskii-2016,Pethick-2008}. 
The parameters used are 
$k_{\rm F}r_{\rm e} = 0$ (solid black line), and $0.0625$ (solid blue line) and $0.1535$ (solid red line). The dash-dotted purple line represents the results without proper many-body scattering renormalization, while the solid
black, blue and red lines include the many-body scattering renormalization within the equally-weighted approach. 
The dash-dotted magenta line represents the solution without proper effective range renormalization~\cite{Pitaevskii-2016}, leading to a divergence in $\mu$ at unitarity $(1/k_{\rm F} a_{\rm s} = 0)$. The results shown by the solid black, blue and red lines use a variational perturbation theory~\cite{Kleinert-2011,Kleinert-2016} and include the proper effective range renormalization, but still produce divergences 
in the chemical potential, now located at $1/k_{\rm F} a_{\rm s} = (8/\pi^2) (1/k_{\rm F} r_{\rm e})$.

In Fig.~\ref{fig:SCESol1Hart}a, we show 
$\mu/\varepsilon_{\rm F}$ versus $1/k_{\rm F} a_{\rm s}$ for $1/k_{\rm F}a_{\rm s} < 0$, that is, from the weakly interacting regime to the unitary point. 
The regions of more negative scattering parameters
are identified with the HSF phase shown in Fig.~\ref{fig:ZeroTPhase}. Within the HSF phase there
are two asymptoptic limits depending on the relative strengths of $k_{\rm F} r_{\rm e}$ and $k_{\rm F} a_{\rm s}$. 
In the asymptotic limit of $k_{\rm F}r_{\rm e} \ll \vert k_{\rm F}a_{\rm s} \vert \ll 1$, the chemical potential is
\begin{equation}\label{eqn:Chemical_Potential_BCS}
\begin{split}
			\frac{\mu}{\varepsilon_{\rm F}} = 1 & - \frac{\pi}{6}k_{\rm F}r_{\rm e} \left[ 1 - \frac{\pi^2}{8}\frac{k_{\rm F}r_{\rm e}}{\vert k_{\rm F}a_{\rm s}\vert } + \mathcal{O}\left(\frac{k_{\rm F}r_{\rm e}}{\vert k_{\rm F}a_{\rm s}\vert }\right)^2\right].
\end{split}
\end{equation}

However, when 
$\vert k_{\rm F}a_{\rm s} \vert \ll k_{\rm F}r_{\rm e} \ll 1$, 
the chemical potential becomes
	\begin{equation}\label{MuBCS}
		\begin{split}
			\hspace*{-.25cm}\frac{\mu}
			{\varepsilon_{\rm F}}
			= 
			1 & -
			\frac{4\vert k_{\rm F} a_{\rm s} \vert}{3\pi}
			\left[ 1 - \frac{8}{\pi^2}\frac{\vert k_{\rm F} a_{\rm s}\vert }{k_{\rm F} r_{\rm e}}
			+ \mathcal{O}\left(\frac{\vert k_{\rm F} a_{\rm s} \vert}{k_{\rm F} r_{\rm e}}\right)^2 \right].
		\end{split}
	\end{equation}
	Having discussed the behavior of the zero-temperature 
	order parameters and the chemical potential as a function 
	of the scattering length and effective range, 
	we analyze next the Helmholtz free energy
	$\mathcal{F} = \Omega + \mu Vn$, where $\Omega$ stands for the 
	thermodynamic potential, $V$ represents the sample volume, and $n$ is 
	the particle density. At $T = 0$, the Helmholtz free energy 
	$\mathcal{F}$ becomes the ground-state energy 
    $\mathcal{E}$ for fixed $n$, 
	meaning that $\mathcal{F} \to \mathcal{E}$.
	The grand-canonical potential $\Omega$ 
    contains the non-analytic coupling term, 
	shown in Eq.~(\ref{eqn:non-analytical-coupling}). However, the Legendre transformation, using the equation of state $\mu (n)$, eliminates this term
    leading to
\begin{align}
\label{eqn:Ground-state-energy-exact}
\frac{\mathcal{E}}{\varepsilon_{\rm F}N} & = - \frac{3\pi}{8}\left[ \frac{\vert\Delta_{\rm B,0}\vert^2}{\varepsilon_{\rm F}^2k_{\rm F} a_{\rm s}} - \frac{\Delta_{\rm H,0}^2}{\varepsilon_{\rm F}^2}\left( \frac{1}{k_{\rm F}a_{\rm s}} - \frac{8}{\pi^2}\frac{1}{k_{\rm F}r_{\rm e}} \right) \right] \nonumber \\
& + \frac{\mu (n) - \Delta_{\rm H,0}}{\varepsilon_{\rm F}} - \frac{3}{4\varepsilon_{\rm F}^{5/2}}\int_0^\infty \mathrm{d}\epsilon \sqrt{\epsilon}\left(E - \widetilde{\xi} \right),
\end{align}
after writing the explicit expressions for the
weight parameters $h_0$ and $b_0$ given in 
Eqs.~(\ref{eqn:h0}) and (\ref{eqn:b0}).
Here, we used $\widetilde{\xi} = \epsilon - \mu (n) + \Delta_{\rm H,0} $ and 
$E = \sqrt{ (\widetilde{\xi})^2 + \vert\Delta_{\rm B,0}\vert^2}$ for the
Bogoliubov dispersion.

\begin{figure}[t]
\centering
\includegraphics[width=.95\linewidth]{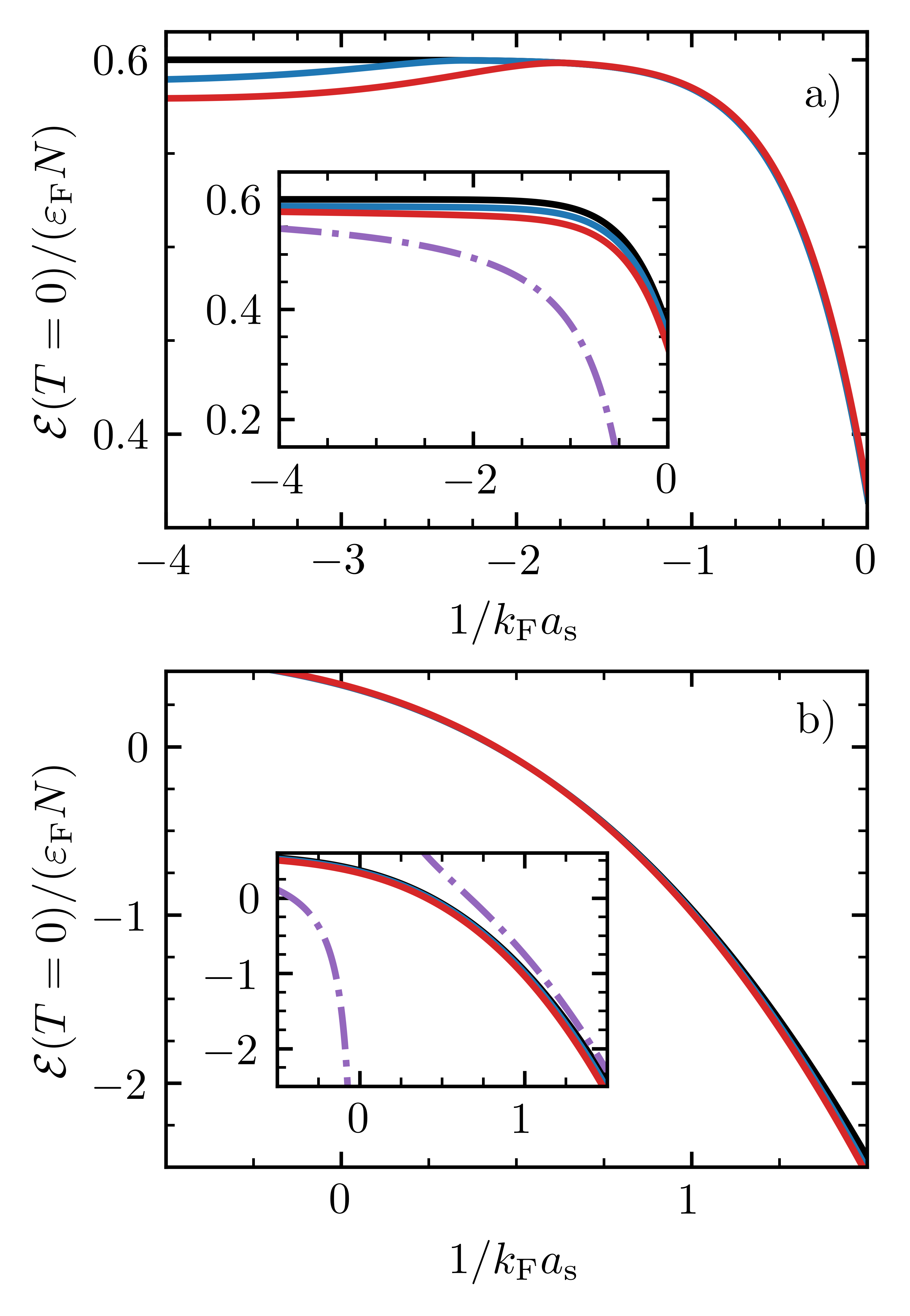}
\caption{Ground-state energy per particle 
$\mathcal{E}(T=0)/(\varepsilon_{\rm F}N)$ versus the scattering parameter $1/k_{\rm F}a_{\rm s}$
for different effective 
range parameters $k_{\rm F}r_{\rm e} = 0$ (solid black line), $0.0625$ (solid blue line) and $0.1535$ (solid red line). 
In panel a), the weakly interacting BCS region 
$1/k_{\rm F}a_{\rm s} < -1$ is shown as well as the approach to unitarity $(1/k_{\rm F}a_{\rm s} = 0)$ from 
the BCS side.
In panel b), the region around $1/k_{\rm F}a_{\rm s} = 0.55$, where $\mu =0$, is displayed and a glimpse of the strong coupling regime $1/k_{\rm F}a_{\rm s} > 1$ is also shown.
The insets in both panels illustrate the equally weighted standard theory, highlighting the
unphysical behavior (dash-dotted magenta line) it 
produces by neglecting the many-body effective 
range renormalization.}
\label{fig:Energy}
\end{figure}

In Fig.~\ref{fig:Energy}, we show the ground-state energy per particle $\mathcal{E}(T=0)/(\varepsilon_{\rm F}N)$ versus the scattering parameter $1/k_{\rm F}a_{\rm s}$ for effective range parameters,
$k_{\rm F}r_{\rm e} = 0$ (solid black line), and $0.0625$ (solid blue line) and $0.1535$ (solid red line). In the insets, we depict the equally weighted 
$(h_0 = b_0 = 1)$ theories 
just like discussed in Figs.~\ref{fig:SCESol1Bogo} and~\ref{fig:SCESol1Hart}.
 
In Fig.~\ref{fig:Energy}a, we show 
$\mathcal{E}(T=0)/(\varepsilon_{\rm F}N)$ versus 
$1/k_{\rm F}a_{\rm s}$ 
for $1/k_{\rm F}a_{\rm s} < 0$, covering the region
from weak interactions through the unitary point 
$1/k_{\rm F}a_{\rm s} = 0$. 
In the HSF phase, where $k_{\rm F}r_{\rm e} > (k_{\rm F}r_{\rm e})_{\rm c}$, the ground state energy per particle is reduced due to the presence of the Hartree order parameter $\Delta_{\rm H,0}$. The reduction of the ground state energy is stronger for larger values of $k_{\rm F}r_{\rm e}$ and becomes zero at the HSF to SSF transition boundary $(k_{\rm F}r_{\rm e})_{\rm c}$ , where $\Delta_{\rm H,0} = 0$. 

For $k_{\rm F}r_{\rm e} < (k_{\rm F}r_{\rm e})_{\rm c}$ , $\Delta_{\rm H,0}$ is strictly zero and the system is in the SSF phase, where the ground state energy is slightly enhanced with increasing effective range. This small increase is caused by a reduction in the negative contribution connected to $\vert \Delta_{\rm B,0}\vert/\varepsilon_{\rm F}$ in 
Eq.~(\ref{eqn:Ground-state-energy-exact}).
The inset shows the results for equally weighted theories $(h_0 = b_0 = 1)$ with the same $k_{\rm F} r_{\rm e}$ as discussed in Figs.~\ref{fig:SCESol1Bogo} and~\ref{fig:SCESol1Hart}, producing similar divergences like those for $\Delta_{\rm H, 0}$ and $\mu$.

In Fig.~\ref{fig:Energy}b, we display
$\mathcal{E}(T=0)/(\varepsilon_{\rm F}N)$ versus 
$1/k_{\rm F}a_{\rm s}$ in the neighborhood 
$1/k_{\rm F}a_{\rm s} = 0.55$ where the chemical potential vanishes. For scattering parameters $1/k_{\rm F}a_{\rm s} > 0.55$ the effective range causes a slight decrease of the ground-state energy per particle as the chemical potential becomes negative.
The inset shows the results for equally weighted theories $(h_0 = b_0 = 1)$ with the same $k_{\rm F} r_{\rm e}$ parameters as the main figure.

For weak interactions
($\vert k_{\rm F}a_{\rm s} \vert \ll 1$), we 
use the asymptotic expression  
$\mu/\varepsilon_{\rm F} = 1 + \Delta_{\rm H,0}/\varepsilon_{\rm F}$, evaluate the integral in Eq.~(\ref{eqn:Ground-state-energy-exact}), 
and use the many-body effective range 
$r_{\rm e,B}$, identified in Eq.~(\ref{eqn:Many-Body-Cutoff}), to obtain 
\begin{eqnarray}
\frac{\mathcal{E}}{\varepsilon_{\rm F}N} = \frac{3}{5} && - \frac{3\pi}{8}\left( \frac{1}{\vert k_{\rm F}a_{\rm s}\vert} + \frac{8}{\pi^2}\frac{1}{k_{\rm F}r_{\rm e}}\right)\frac{\vert\Delta_{\rm H,0}\vert^2}{\varepsilon_{\rm F}^2} \nonumber \\ && - \frac{3}{8}\left(1-\frac{\pi}{2}k_{\rm F}r_{\rm e,B}\right)\frac{\vert\Delta_{\rm B,0}\vert^2}{\varepsilon_{\rm F}^2} \, , \label{eqn:Ground-state-energy-BCS-general}
\end{eqnarray}
which is valid for both HSF and SSF phases. 
The factor $3/5$ is the energy of the free 
Fermi gas, while the second term represents the 
Hartree shift, which lowers the energy. 
The last term shows the 
energy gain due to pairing with an altered 
prefactor, associated with $k_{\rm F}r_{\rm e,B}$. 

Focusing on the HSF phase, we obtain next analytical expressions for $\mathcal{E}$ in two asymptotic regimes. Using $k_{\rm c,B}/k_{\rm F}$ from Eq.~(\ref{eqn:HSF-Cutoff}), 
and the weak coupling limit expression of 
the superfluid order parameter in Eq.~(\ref{eqn:Bogoliubov-limit}),
we obtain
\begin{align}
\label{eqn:Ground-state-Energy-Intermediate}
\frac{\mathcal{E}}{\varepsilon_{\rm F}N} =  &
\left[ \frac{3}{5} - \frac{\pi}{12}\left(1 - \frac{\pi^2}{8}\frac{k_{\rm F}r_{\rm e}}{\vert k_{\rm F}a_{\rm s}\vert}\right)k_{\rm F}r_{\rm e} \right] + \mathcal{O}\left(\frac{k_{\rm F} r_{\rm e}}{\vert k_{\rm F} a_{\rm s}\vert}\right)^2 \nonumber \\ 
& - \frac{24}{e^4}\left\{1-\frac{12}{e^2} \left[ f(k_{\rm F}a_{\rm s})\right]^{1/2}\right\}^3 f(k_{\rm F}a_{\rm s}), 
\end{align}
in the asymptotic regime 
$k_{\rm F}r_{\rm e} \ll \vert k_{\rm F}a_{\rm s} \vert \ll 1$, where the expression for $\Delta_{\rm H,0}$ given in Eq.~(\ref{eqn:Hartree-limit-inter}) is valid.
Here, the function 
$f(k_{\rm F}a_{\rm s}) = \exp\left(-\frac{\pi}{\vert k_{\rm F}a_{\rm s}\vert}\right)$ describes an exponential correction.
Using Eq.~(\ref{eqn:Hartree-limit-BCS}) for the Hartree order parameter, instead of Eq.~(\ref{eqn:Hartree-limit-inter}),
we obtain a different expression for the second hierarchy of scales, that is, $\vert k_{\rm F}a_{\rm s} \vert \ll k_{\rm F}r_{\rm e} \ll 1$, 
leading to 
\begin{align}\label{eqn:Ground-State-Energy-Deep-BCS}
\frac{\mathcal{E}}{\varepsilon_{\rm F}N}
= & \left[ \frac{3}{5} - \frac{2}{3\pi}\left(1 - \frac{6}{\pi}\frac{\vert k_{\rm F}a_{\rm s} \vert}{k_{\rm F}r_{\rm e}}\right)
\vert k_{\rm F}a_{\rm s}\vert \right] + \mathcal{O}\left(\frac{\vert k_{\rm F} a_{\rm s} \vert}{k_{\rm F} r_{\rm e}}\right)^2 \nonumber \\
& - \frac{24}{e^4}\left\{1-\frac{12}{e^2} \left[ f(k_{\rm F}a_{\rm s})\right]^{1/2}\right\}^3 f(k_{\rm F}a_{\rm s})\,.
\end{align}
The analytical results displayed in Eqs.~(\ref{eqn:Ground-state-Energy-Intermediate}) and~(\ref{eqn:Ground-State-Energy-Deep-BCS}) 
agree very well with the numerical evaluations.

In summary, the HSF is a superfluid phase 
characterized by two separate order parameters, the Hartree order parameter 
	$\Delta_{\rm H,0}$, responsible for the renormalization of 
	the chemical potential and the superfluid order parameter 
	$\Delta_{\rm B,0}$, representing
	Cooper pairs that are responsible for  
	fermionic superfluidity. The Hartree order parameter reduces the energy 
	by lowering the chemical potential and, thus, this phase 
	corresponds to a stable minimum of the underlying 
	grand-canonical potential. The non-analytic coupling of the two 
	order parameters gives rise to the mechanism that allows the Hartree shift to vanish, converting $\Delta_{\rm H,0}$ into a true order parameter that characterizes the HSF to SSF transition.

	\subsection{Standard superfluid (SSF)}\label{SubSec:4C}
	The standard superfluid phase is the region in the 
	phase diagram where the interaction strength becomes 
	sufficiently strong such that particle-hole contributions are  
suppressed and all interaction energy  
	is used to form Cooper pairs. Therefore, this phase is 
	characterized by a vanishing Hartree order parameter 
	$\Delta_{\rm H,0} = 0$ and non-zero 
	superfluid order parameter $\Delta_{\rm B,0}\neq 0$. 
	The vanishing of the Hartree 
	order parameter causes the many-body UV cutoff $k_{\rm c,B}$ in 
    Eq.~(\ref{express})
    to converge to the two-body UV cutoff $k_{\rm c}$ in Eq.~(\ref{eqn:two-body-cutoff}) resulting in
	\begin{equation}\label{eqn:SF-cutoff}
		\frac{k_{\rm c,B}}{k_{\rm F}} = \frac{4}{\pi k_{\rm F}r_{\rm e}}.
	\end{equation}
	Therefore, in the SSF phase, $\Delta_{\rm B,0}$ is directly 
	affected by $k_{\rm F} r_{\rm e}$. Since, the SSF phase exists
    only for $k_{\rm F} r_{\rm e} < \left(k_{\rm F} r_{\rm e}\right)_{\rm c}$, the only asymptotic regime reachable is
	$k_{\rm F}r_{\rm e} \ll \vert k_{\rm F}a_{\rm s} \vert \ll 1$, 
    leading to 
	\begin{equation}
		\label{eqn:A_for_SF}
		\frac{\vert \Delta_{\rm B,0} \vert}{\varepsilon_{\rm F}} = 
		\frac{8}{e^{2}}\exp\left(-\frac{\pi}{4}k_{\rm F}r_{\rm e}\right)\exp\left(-\frac{\pi}{2\vert k_{\rm F} a_{\rm s}\vert}\right).
	\end{equation}
	The relation above tells us that $k_{\rm F} r_{\rm e}$ 
	reduces the superfluid order parameter exponentially 
	when considering weak interactions, in contrast
    to the results of Eq.~(\ref{eqn:Bogoliubov-limit}) for 
    HSF phase, where $\vert \Delta_{\rm B,0} \vert/\varepsilon_{\rm F}$
is independent of $k_{\rm F} r_{\rm e}$ in the same asymptotic regime.    
Using Eq.~(\ref{eqn:A_for_SF}), we approach the HSF-SSF phase boundary from the right (see Fig.~\ref{fig:ZeroTPhase}) and use the continuity of 
$\Delta_{\rm B,0}$ to obtain
	\begin{equation}
		\label{eqn:critical_scattering_parameter}
		\left(k_{\rm F}r_{\rm e}\right)_{\rm c} = \frac{48}{\pi e^2} e^{-\frac{\pi}{2\vert k_{\rm F}a_{\rm s} \vert}},
	\end{equation}
describing an analytical approximation for the numerical 
phase boundary shown in Fig.~\ref{fig:ZeroTPhase}.
The analytical expression in Eq.~(\ref{eqn:critical_scattering_parameter})
is represented by the dash-dotted line in Fig.~\ref{fig:ZeroTPhase}.
This shows that the phase boundary $(k_{\rm F} r_{\rm e})_c$ 
has an exponential dependence on $1/\vert k_{\rm F} a_{\rm s} \vert$ in the weakly interacting limit 
$\vert k_{\rm F} a_{\rm s}\vert \ll 1$.

Since there is no Hartree shift in the SSF phase, 
the only correction to the weak-coupling 
chemical potential arises from the superfluid order parameter and hence is exponentially small, that is, 
$\mu = \varepsilon_{\rm F} - \mathcal{O}(\vert \Delta_{\rm B,0} \vert^2/\varepsilon_{\rm F})$. Using Eq.~(\ref{eqn:A_for_SF}), it is clear that deviations from 
$\varepsilon_{\rm F}$ are exponential in 
$k_{\rm F} r_{\rm e}$ and $1/\vert k_{\rm F} a_{\rm s} \vert$ separately, in sharp contrast with the polynomial corrections in 
the ratios 
$k_{\rm F} r_{\rm e}/\vert k_{\rm F} a_{\rm s} \vert$
or $\vert k_{\rm F} a_{\rm s} \vert/k_{\rm F} r_{\rm e}$
found for the HSF phase (see Eqs.~(\ref{eqn:Chemical_Potential_BCS}) and~(\ref{MuBCS})).

Lastly, we use the same general 
expression for the ground-state energy derived in 
Eq.~(\ref{eqn:Ground-state-energy-BCS-general}) and insert the 
	SSF asymptotic behavior of the superfluid order parameter 
	resulting in
	\begin{align}
		\frac{\mathcal{E}}{\varepsilon_{\rm F}N} = \frac{3}{5} - \frac{24}{e^4}\left(1- \frac{\pi}{2}k_{\rm F}r_{\rm e}\right) \exp\left(-\frac{\pi}{2} k_{\rm F}r_{\rm e}\right)f(k_{\rm F}a_{\rm s}),
	\end{align}
where the function 
$f(k_{\rm F}a_{\rm s}) = \exp\left(-\frac{\pi}{\vert k_{\rm F}a_{\rm s}\vert}\right)$ describes an exponential correction.
Here, the energy is only lowered due to pairing, 
which is slightly suppressed by the finite effective range, as can be seen in Fig.~\ref{fig:Energy}a.
	
In summary, in the weak coupling $\vert k_{\rm F} a_{\rm s} \vert \ll 1$ regime, the SSF phase possesses a superfluid order parameter that is more affected by the effective range
in comparison to the HSF phase. Furthermore, the SSF phase has a higher ground-state energy and a larger chemical 
potential when compared to the HSF phase. These differences
in sensitivity to the effective range and scattering parameters, 
allow for experiments that can distinguish the two phases. 
  
\subsection{Unitary gas}\label{SubSec:4D}
	
The unitary point ($1/k_{\rm F}a_{\rm s} = 0$) is a place of great theoretical and experimental interest due to the concept of universality, which is well known in the realm of ultracold dilute gases. 
A main difference between standard theories of superfluidity for ultracold fermions and theories of superfluidity for nuclear matter is that the interaction range plays just a minor role 
in ultracold fermions because they are dilute, while the densities in nuclear systems are sufficiently large for the interaction range to be at least of the same order of the interparticle spacing. 

The standard approach outlined in Refs.~\cite{Widera-2023,Timmermans-2011,Pitaevskii-2016} 
investigates the role of particle-hole interactions 
by using the weak coupling relation $g = 4\pi a_{\rm s}/m$,
leading to unphysical singularities at unitarity. 
Such divergences occur not only in the Hartree order parameter, but also in the chemical potential and the ground-state 
energy (see dash-dotted magenta line in the insets of Figs.~\ref{fig:SCESol1Bogo},\ref{fig:SCESol1Hart},\ref{fig:Energy}). 
Note that after unitary $(1/k_{\rm F}a_{\rm s} > 0)$, 
the dash-dotted magenta lines approach the zero-range curve 
(solid black line) from above. This results in a chemical 
potential $\mu$ and a ground-state energy $\mathcal{E}$ which 
have no lower bound when $1/k_{\rm F} a_{\rm s} \to 
0^{-}$ and no upper bound when $1/k_{\rm F} a_{\rm s} \to 
0^{+}$.

Our approach resolves the issues discussed above, because it 
considers the proper renormalization of the interaction by 
taking into account the effective range
and introduces a self-consistent weighting of the interaction channels. Both steps are necessary, because when only the 
effective range renormalization is considered for non-self-consistent weights of the Hartree and Bogoliubov channels 
then a singular point still occurs  
at $k_{\rm F} a_{\rm s} = (\pi^2/8) k_{\rm F}r_{\rm e}$. 
This value of the $k_{\rm F} a_{\rm s}$ corresponds to the infinite attraction limit in Eq.~(\ref{eqn:Scattering-length-explicit}), which cannot be exceeded. 
The introduction of
self-consistent weights solves this problem at unitary and beyond. 

For example, at unitarity, instead of a diverging Hartree order parameter, one gets a result which is fully 
determined by the effective range
\begin{equation}\label{eqn:Hartree-Unitarity}
\frac{\Delta_{\rm H,0}}{\varepsilon_{\rm F}} = \mathrm{min}\left( 0 ; -\frac{\pi}{6}k_{\rm F}r_{\rm e} + \frac{\vert \Delta_{\rm B,0} \vert}{\varepsilon_{\rm F}} \right),
\end{equation}
as described in Eq.~(\ref{eqn:self-consistency-equation-hartree-full-v2}). A non-trivial solution for $\Delta_{\rm H, 0}$ at unitarity only occurs for sufficiently large values of $k_{\rm F} r_{\rm e}$, which do not occur in $^6\mathrm{Li}$ or $^{40}\mathrm{K}$ for experimentally achievable densities. As a consequence of these experimental 
constraints, the only possible solution of 
Eq.~(\ref{eqn:Hartree-Unitarity}) is $\Delta_{\rm H, 0} = 0$ yielding a SSF phase at unitarity. Since the effective range is
essential in resolving singular issues, its effects are real 
in spite of the smallness of $k_{\rm F}r_{\rm e}$.  

For dilute unitary Fermi gases, as suggested by Zhang and Leggett in 2009~\cite{Legget-2009}, 
$k_{\rm F}r_{\rm e}$ is too small for its effects to be detected by current experimental setups, however $k_{\rm F}r_{\rm e}$ is very relevant for higher densities such as for nuclear matter or neutron stars near and away from unitarity. Therefore, for dilute Fermi gases at unitarity, where $k_{\rm F} r_{\rm e} \ll 1$, one finds the concept of quasi-universality, that is, 
all gases behave the same way, irrespective of the atomic species.

\subsection{Towards strong coupling}\label{SubSec:4E}
	
After crossing the unitary point, starting from weak interactions, 
the scattering length switches sign from negative to positive, that is, beyond unitarity the relation $k_{\rm F}a_{\rm s} > 0$ is satisfied.
As seen in the phase diagram of Fig.~\ref{fig:ZeroTPhase}, the SSF phase exists at unitarity $(1/k_{\rm F} a_{\rm s} = 0)$ and beyond for any $k_{\rm F} r_{\rm e}$. 
Furthermore, from Fig.~\ref{fig:Scattering-lengths},
we can see that there is only one asymptotic regime
$(k_{\rm F}r_{\rm e} \ll k_{\rm F}a_{\rm s} \ll 1)$, because 
the scattering length $a_s$ can only approach the background scattering length from above, that is, $k_{\rm F} a_{\rm s} \ge (\pi^2/8) k_{\rm F}r_{\rm e}$. 

The approach to strong coupling manifests itself also through the
chemical potential $\mu$, which changes sign from positive to 
negative approximately at $(1/k_{\rm F}a_{\rm s}) = 0.55$.
This result depends only weakly on the effective range parameter $k_{\rm F} r_{\rm e}$. For $(1/k_{\rm F}a_{\rm s}) < 0.55$, increasing the effective range enhances the degeneracy of the momentum distribution and augments 
the chemical potential. 
While for $(1/k_{\rm F}a_{\rm s}) > 0.55$,
increasing the effective range reduces the degeneracy of the momentum distribution and lowers the chemical potential.
In Fig.~\ref{fig:SCESol1Hart}b, we show the decrease of the chemical potential towards a more negative value, the change in degeneracy of the Fermi gas.

In the strong coupling regime, where $1/k_{\rm F}a_{\rm s} \gg 1$,  
a larger effective range facilitates the formation of two-body bound states 
leading to a slight increase in the magnitude of the superfluid order parameter
\begin{equation}
\label{eqn:Bogoliubov-BEC-limit}
\hspace*{-.225cm}\frac{\vert \Delta_{\rm B,0} \vert}{\varepsilon_{\rm F}}
= \frac{4}{\sqrt{3\pi}}
\frac{1}{\sqrt{k_{\rm F}a_{\rm s}}}\left[1 + \frac{1}{4}\frac{k_{\rm F}r_{\rm e}}{k_{\rm F}a_{\rm s}} + \mathcal{O}\left(\frac{k_{\rm F}r_{\rm e}}{k_{\rm F}a_{\rm s}}\right)^2\right],
\end{equation}
as evidenced by the correction proportional to 
$k_{\rm F}r_{\rm e}/k_{\rm F} a_{\rm s}$. This analytical expression arises from a expansion in the parameter $k_{\rm F}r_{\rm e}/k_{\rm F} a_{\rm s} \ll 1$
at Eq.~(\ref{eqn:self-consistency-equation-bogoliubov-full})
for the asymptotic regime $k_{\rm F}r_{\rm e} \ll k_{\rm F}a_{\rm s} \ll 1$.

In leading order, the result in Eq.~(\ref{eqn:Bogoliubov-BEC-limit})
shows proportionality to the inverse square-root of $k_{\rm F} a_{\rm s}$ obtained from a zero-range theory plus a positive correction due to the effective range $r_{\rm e}$, which enhances the binding energy of pairs. For zero effective range, the two-body binding energy is $E_{\rm b} = 1/ma_{\rm s}^2$, and the chemical potential $\mu = -E_{\rm b}/2$
in the strong coupling regime. However, for non-zero effective range, the two-body binding energy is altered to
\begin{equation}
		\label{eqn:Binding-energy}
		\frac{E_{\rm b}}{\varepsilon_{\rm F}} = \frac{1}{(k_{\rm F}a_{\rm s})^2}\left( 1 + \frac{k_{\rm F}r_{\rm e}}{k_{\rm F}a_{\rm s}} + \mathcal{O}\left(\frac{k_{\rm F}r_{\rm e}}{k_{\rm F}a_{\rm s}}\right)^2 \right),
\end{equation}
which is in excellent agreement with the numerical calculation for strong-coupling shown in Fig.~\ref{fig:SCESol1Hart}b.
Furthermore, the effective range also modifies the chemical potential to 
	\begin{equation}
		\label{eqn:ChemicalPotential-BEC-limit}
		\frac{\mu}{\varepsilon_{\rm F}}
		= 
		- \frac{E_b}{2 \varepsilon_{\rm F}}
		+ \frac{2}{3\pi}k_{\rm F}a_{\rm s}\left( 1 + \frac{k_{\rm F}r_{\rm e}}{k_{\rm F}a_{\rm s}} \right) + \mathcal{O}\left(\frac{k_{\rm F}r_{\rm e}}{k_{\rm F}a_{\rm s}}\right)^2.
	\end{equation}

To complete our discussion of ground-state properties,
we analyze next the pair size $\xi_{\rm pair}$, which is a measure of the strength of the attractive interactions.

\section{Pair Size}
\label{sec:PairSize}
	
The pair size $\xi_{\rm pair}$ characterizes the extent of the Cooper pair 
wavefunction $\chi ({\bf r})$ with zero center-of-mass momentum, 
where ${\bf r}$ is the relative coordinate. So, generally, the pair size is defined in real space according to
\begin{equation}
\xi_{\rm pair}^2 = \frac{\displaystyle\int d{\bf r} \overline{\chi} ({\bf r}) r^2 \chi({\bf r})}{\displaystyle
\int d{\bf r} \overline{\chi} ({\bf r}) \chi({\bf r})},
\end{equation}
which in momentum space becomes
\begin{equation}
\label{eqn:xi-pair-momentum-space}
\xi_{\rm pair}^2 = 
- \frac{\displaystyle\int d{\bf k} \overline{\Phi}_{\bf k} \nabla_{\bf k}^2\Phi_{\bf k}} {\displaystyle\int d{\bf k}\overline{\Phi}_{\bf k} \Phi_{\bf k}}.
\end{equation}
Here, $\Phi_{\bf k} =\Delta_{{\rm B},0}/(2E_\mathbf{k})$ is the Fourier transform of the Cooper pair wavefunction $\chi ({\bf r})$~\cite{Strinati-1994,SadeMelo-1997}, and 
$\Delta_{{\rm B},0}$ is the zero-temperature pairing amplitude that appears in the superfluid order parameter equation Eq.~(\ref{eqn:self-consistency-equation-bogoliubov-full}).
\begin{figure}[t]
\centering
\includegraphics[width=.95\linewidth]{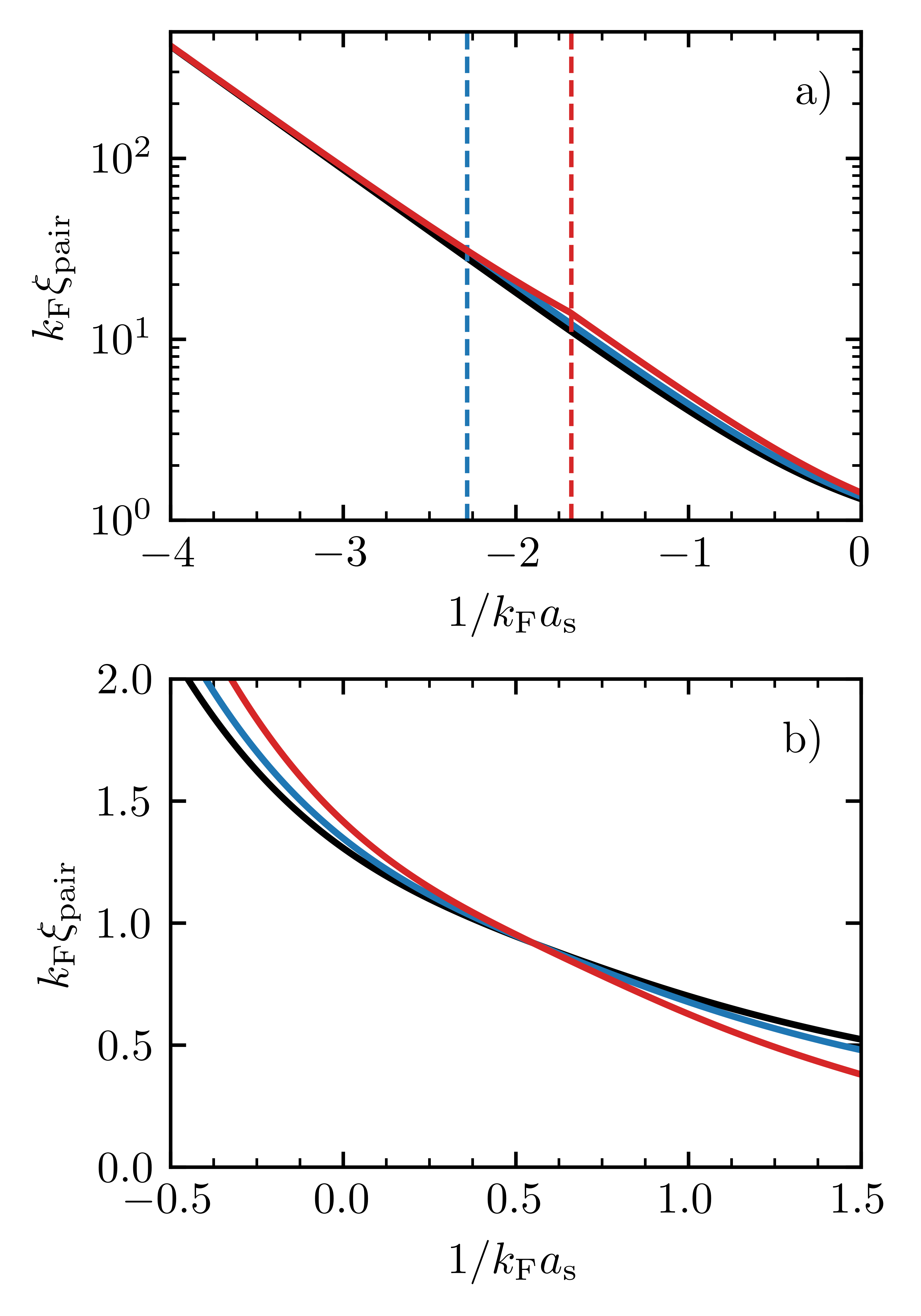}
\caption{Pair size $k_{\rm F} \xi_{\rm pair}$ versus scattering parameter $1/k_{\rm F} a_{\rm s}$
for effective ranges $k_{\rm F}r_{\rm e} = 0$ (solid black line), $0.0625$ (solid blue line) and $0.1535$ (solid red line).
Panel a) emphasizes the weakly interacting BCS region
$1/k_{\rm F} a_{\rm s} < -1$ and the region close
to unitarity when approached from the BCS side.
The vertical axis is shown in logarithmic scale.
The HSF-SSF phase boundary for given $k_{\rm F}r_{\rm e}$
is displayed as vertical
dashed blue ($k_{\rm F}r_{\rm e} = 0.0625$) and dashed red ($k_{\rm F}r_{\rm e} = 0.1535$) lines.
Panel b) shows the region around $1/k_{\rm F} a_{\rm s} = 0.55$ $(\mu = 0)$ and the beginning of the BEC region $1/k_{\rm F} a_{\rm s} > 1$. 
The vertical axis in shown in linear scale.
}
\label{fig:EandXi}
\end{figure}

In Fig.~\ref{fig:EandXi}, we show the dimensionless pair size 
$k_{\rm F}\xi_{\rm pair}$ as a function of $1/k_{\rm F} a_{\rm s}$ for effective range parameters $k_{\rm F}r_{\rm e} = 0$ (solid black line), $0.0625$ (solid blue line) and $0.1535$ (solid red line). As seen in Figs.~\ref{fig:EandXi}a and b, the pair size $\xi_{\rm pair}$ is a monotonically decreasing function of $1/k_{\rm F} a_{\rm s}$ for fixed $k_{\rm F}r_{\rm e}$. In Fig.~\ref{fig:EandXi}a,
we plot $k_{\rm F} \xi_{\rm pair}$ versus $1/k_{\rm F} a_{\rm s}$ for  
$1/k_{\rm F} a_{\rm s} < 0$.
In Fig.~\ref{fig:EandXi}b, 
we analyze $k_{\rm F} \xi_{\rm pair}$ versus $1/k_{\rm F} a_{\rm s}$ in the neighborhood of $1/k_{\rm F} a_{\rm s} = 0.55$, where the chemical potential falls below the bottom of the band, that is, $\mu = 0$. Note that for $1/k_{\rm F} a_{\rm s} < 0.55$ $(\mu > 0)$, an increase of $k_{\rm F}r_{\rm e}$ also increases 
$k_{\rm F} \xi_{\rm pair}$, while for $1/k_{\rm F} a_{\rm s} > 0.55$ $(\mu < 0)$, an increase of $k_{\rm F}r_{\rm e}$ decreases 
$k_{\rm F} \xi_{\rm pair}$.

As shown in Appendix~\ref{Chapter:Pair-size-appendix},
we analytically obtain an expression for 
$\xi_{\rm pair}$ in terms of $\mu$, 
$\Delta_{\rm H, 0}$, $\vert \Delta_{\rm B, 0}\vert$ and $\varepsilon_{\rm F}$.
The result is  
\begin{equation}
\label{eqn:BCS-pair-size}
(k_{\rm F} \xi_{\rm pair})^2 = \frac{\varepsilon_{\rm F}}{4\sqrt{2}}\frac{5|\Delta_{\rm B,0}|^2 + 2\mu_{\rm H,0}\left[ \mu_{\rm H,0} + E_{\bf 0} \right]}{|\Delta_{\rm B,0}|^2 E_{\bf 0}},
\end{equation}
where $E_{\bf 0} = \sqrt{\mu_{{\rm H},0}^2 + |\Delta_{\rm B,0}|^2}$ is the 
quasiparticle energy given in Eq.~(\ref{eqn:Bogoliubov-dispersion}) at zero momentum and $\mu_{\rm H,0} = \mu - \Delta_{\rm H,0}$ is the shifted chemical potential. 
The analytical result in Eq.~(\ref{eqn:BCS-pair-size}) agrees perfectly with the direct numerical calculation of $\xi_{\rm pair}$, from Eq.~(\ref{eqn:xi-pair-momentum-space}), plotted in Fig.~\ref{fig:EandXi}.

We now use the analytical expression in Eq.~(\ref{eqn:BCS-pair-size}) to discuss asymptotic limit of $k_{\rm F} \xi_{\rm pair}$.
For weak interactions, 
$(1/k_{\rm F}a_{\rm s} \ll -1$), 
we determine asymptotic expansions depending on whether the system is either in the HSF or in the SSF phase, seen in Fig.~\ref{fig:ZeroTPhase}.
In the HSF phase, where $k_{\rm F} r_{\rm e} \ge \left(k_{\rm F} r_{\rm e}\right)_{\rm c}$, we obtain in both hierarchies of scales, that is either for $k_{\rm F}r_{\rm e} < \vert k_{\rm F}a_{\rm s} \vert$ or $\vert k_{\rm F}a_{\rm s} \vert < k_{\rm F}r_{\rm e}$, the same asymptotic limit, since the superfluid order parameter does not depend on $k_{\rm F}r_{\rm e}$, as shown in Eq.~(\ref{eqn:Bogoliubov-limit}).
The resulting asymptotic expansion of the pair size is
\begin{eqnarray}
k_{\rm F} \xi_{\rm pair} = && 
\frac{e^2}{8\sqrt{2}}\left[1+\frac{12}{e^2}\exp\left(-\frac{\pi}{2\vert k_{\rm F}a_{\rm s} \vert}\right)\right] \nonumber \\ && \times\exp\left(+\frac{\pi}{2\vert k_{\rm F}a_{\rm s}\vert}\right),
\end{eqnarray}
which is independent of $k_{\rm F} r_{\rm e}$, but grows exponentially with $1/\vert k_{\rm F} a_{\rm s} \vert$ when approaching $k_{\rm F}a_{\rm s} \to 0^-$. Therefore $k_{\rm F} \xi_{\rm pair} \gg 1$.
In contrast, for the SSF phase, where
$k_{\rm F} r_{\rm e} \le \left(k_{\rm F} r_{\rm e}\right)_{\rm c}$, we obtain
\begin{equation}
k_{\rm F} \xi_{\rm pair} = 
\frac{e^2}{8\sqrt{2}}\exp\left(\frac{\pi}{4}k_{\rm F}r_{\rm e}\right)\exp\left(+\frac{\pi}{2\vert k_{\rm F}a_{\rm s}\vert}\right).
\end{equation}
Note that again $k_{\rm F} \xi_{\rm pair}$ grows exponentially with $1/\vert k_{\rm F} a_{\rm s}\vert$ when $k_{\rm F}a_{\rm s} \to 0^+$, 
but also contains an exponential
dependence on $k_{\rm F} r_{\rm e}$.
The separation between these two asymptotic solutions, and as such the HSF and SSF phases,
is shown in Fig.~\ref{fig:EandXi}a by the vertical
dashed blue ($k_{\rm F}r_{\rm e} = 0.0625$) and the dashed red ($k_{\rm F}r_{\rm e} = 0.1535$) lines.
In the neighborhood of $1/k_{\rm F} a_{\rm s} = 0.55$,
where $\mu = 0$, the pair size obeys the relation
$k_{\rm F}\xi_{\rm pair} = {\cal O} (1)$. For strong coupling (BEC regime), $1/ k_{\rm F} a_{\rm s} \gg 1$, we obtain 
\begin{equation}
k_{\rm F} \xi_{\rm pair} = \frac{k_{\rm F}a_{\rm s}}{\sqrt{2}}\left(1 - \frac{1}{2}\frac{k_{\rm F}r_{\rm e}}{k_{\rm F}a_{\rm s}}\right).
\end{equation}
In this limit,
$k_{\rm F} \xi_{\rm pair}$ tends to 
zero linearly with $k_{\rm F} a_{\rm s}$, 
has a small correction proportional to 
$k_{\rm F} r_{\rm e}/ k_{\rm F} a_{\rm s}$
and is always small, that is, $k_{\rm F} \xi_{\rm pair} \ll 1$. We emphasize that all the asymptotic analysis discussed above agrees well with the numerical results in the appropriate regimes.

We have discussed the consequences of the effective range and the interplay of the Hartree and Bogoliubov channels on several ground-state $(T = 0)$ properties including the chemical potential, order parameters, free energy, momentum distribution and pair size. Thus, next, we present results of the effects of the effective range and of the Hartree and Bogoliubov channels on the finite temperature $(T \ne 0)$ phase diagrams of interacting fermions.

\section{Finite-Temperature Phase Diagram}
\label{sec:FiniteTemperature}
    
	In the previous sections, we discussed the necessity of
	an effective scattering range to describe 
	the simultaneous effects of the Hartree and Bogoliubov 
	channels on the interacting Fermi gas. The method of implementing a partitioning of the interaction into Hartree and Bogoliubov channels fixes uncontrolled approaches that either ignore divergences~\cite{Pitaevskii-2016} or arbitrarily separate the interactions in equally weighted channels~\cite{Stoof-2008,Timmermans-2011,Widera-2023,Kleinert-2011}.
	As as consequence of our WHFB approach, 
	the Hartree channel has a 
	true order parameter $\Delta_{\rm H, 0}$ that can vanish as 
	$1/k_{\rm F} a_{\rm s}$ changes, and the theory has a self-consistently determined partitioning of the interactions without 
	unphysical divergences.
\begin{figure}[t]
\centering
\includegraphics[width=.95\linewidth]{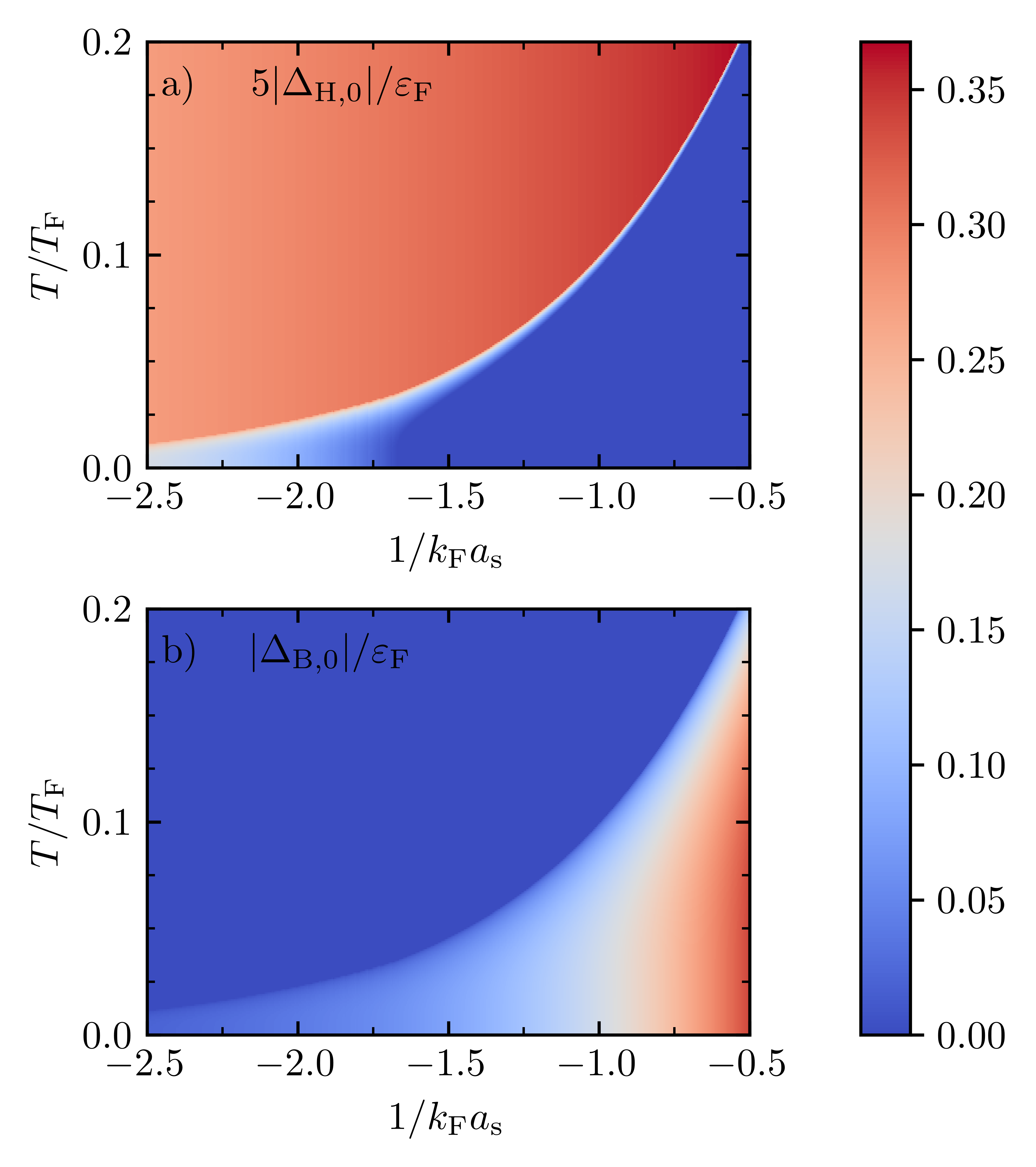}
\caption{Density plots of 
$\vert \Delta_{{\rm H},0} \vert/\varepsilon_{\rm F}$ and $\vert \Delta_{\rm B,0} \vert /\varepsilon_{\rm F}$
in the temperature $T/T_{\rm F}$ versus scattering parameter $1/k_{\rm F}a_{\rm s}$ plane for an effective range parameter of $k_{\rm F}r_{\rm e} = 0.1535$. Panel a) shows the Hartree order parameter $5|\Delta_{{\rm H},0}|/\varepsilon_{\rm F}$ while panel b) illustrates the superfluid order parameter $\vert \Delta_{\rm B,0} \vert/\varepsilon_{\rm F}$. The legend shows the range of values taken by $5 \vert \Delta_{{\rm H},0} \vert /\varepsilon_{\rm F}$ and $ \vert \Delta_{\rm B,0} \vert/\varepsilon_{\rm F}$.
}
\label{fig:FiniteTBogo}
\end{figure}
\begin{figure}[tb]
\centering
\includegraphics[width=.95\linewidth]{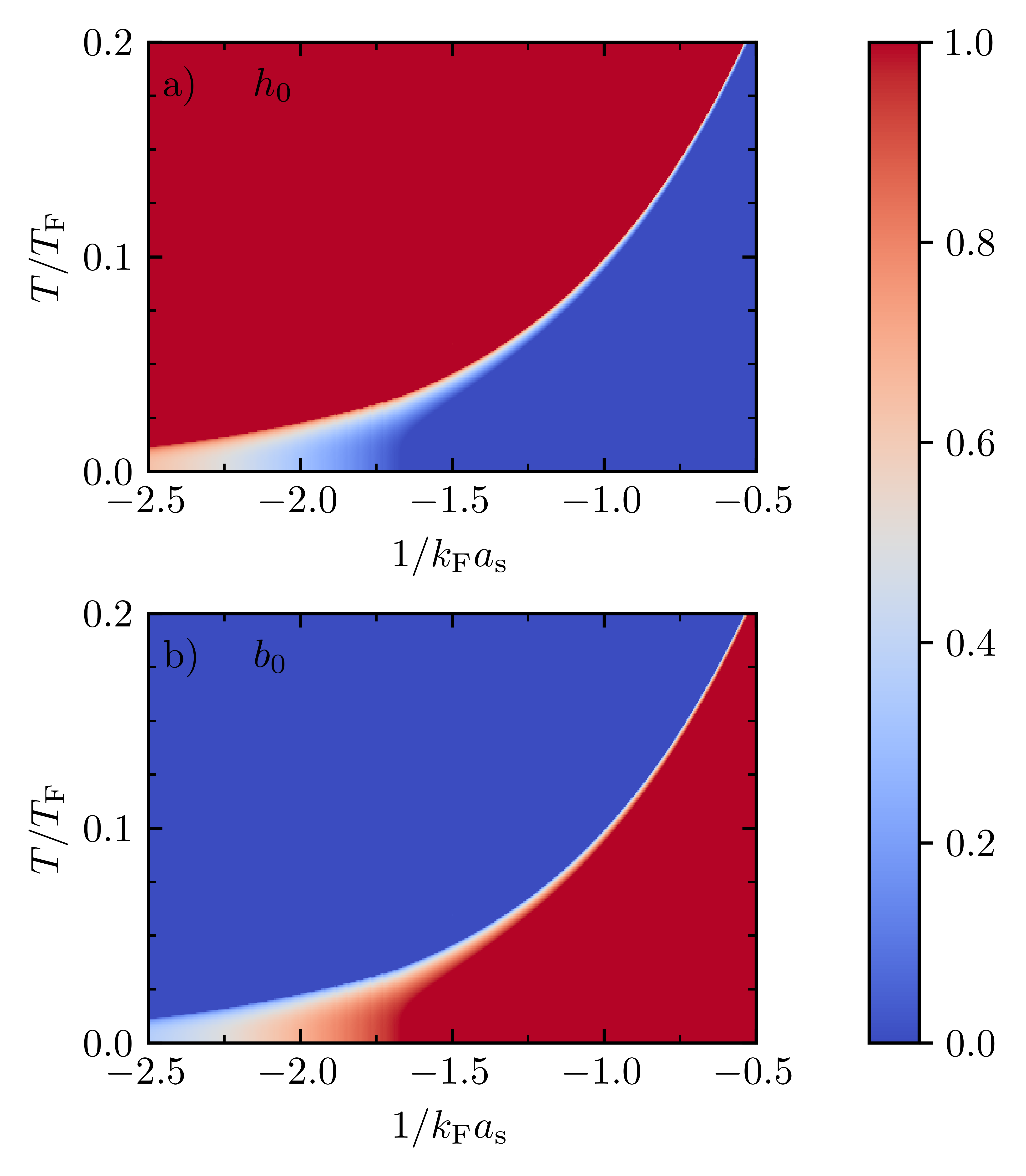}
\caption{Density plots of the Hartree weight $h_0$  and the Bogoliubov weight $b_0$ in the temperature $T/T_F$ versus scattering parameter $1/k_{\rm F}a_{\rm s}$ plane for an effective range parameter of $k_{\rm F}r_{\rm e} = 0.1535$. Panel a) shows $h_0$, while 
panel b) illustrates $b_0$. The legend shows the range of values taken by $h_0$ and $b_0$.
}
\label{fig:FiniteTHart}
\end{figure}

To illustrate non-zero temperature effects, we 
set the dimensionless effective range parameter to  $k_{\rm F}r_{\rm e} = 0.1538$. For an effective range
$r_{\rm e} = 87 a_0$ corresponding to $^6{\rm Li}$~\cite{Hutson-2014}, the choice $k_{\rm F}r_{\rm e} = 0.1538$ corresponds to a density of $n \approx 10^{15}/{\rm cm}^3$, which has not been achieved experimentally yet. For an effective range $r_{\rm e} = 214 a_0$, the parameter $k_{\rm F} r_{\rm e} = 0.1538$ gives a density of $n \approx 8 \times 10^{13}/{\rm cm}^3$. Our choice of parameters helps visualizing the generic effects on non-zero temperature. Similar effects occur for smaller values of $k_{\rm F} r_{\rm e}$, but they are pushed to more negative values of $1/k_{\rm F} a_{\rm s}$. Furthermore, our parameter choice highlights that our formalism builds a theoretical bridge between superfluid theories of ultracold atomic gases \cite{Leggett-1975,SchmittRink-1985,Engelbrecht-1993,SadeMelo-1997, Legget-2009}, where $k_{\rm F} r_{\rm e} \ll 1$, and of nuclei and neutron matter \cite{Soloviev-1958,Urban-2018,Clark-2019},
where $k_{\rm F} r_{\rm e} \sim 1$.
	
In Fig.~\ref{fig:FiniteTBogo}, we show density plots
of the order parameters $\vert \Delta_{{\rm H},0}\vert/\varepsilon_{\rm F}$ and $\vert \Delta_{{\rm B},0}\vert/\varepsilon_{\rm F}$
in the $T/T_{\rm F}$ versus $1/k_{\rm F} a_{\rm s}$ plane. Color maps of the Hartree order parameter
$\vert \Delta_{{\rm H},0}\vert/\varepsilon_{\rm F}$ 
are shown in panel a) and color maps of the Bogoliubov order parameter $\vert \Delta_{{\rm B},0}\vert/\varepsilon_{\rm F}$ are revealed in panel b). 
The color maps range from blue to red as indicated
in the legend.
In Fig.~\ref{fig:FiniteTBogo}a, $\vert \Delta_{{\rm H},0}\vert/\varepsilon_{\rm F}$ is non-zero at higher values of $T/T_{\rm F}$, above the critical line $T_{\rm H}/T_{\rm F}$, where $\vert \Delta_{{\rm H},0} \vert = 0$. While in Fig.~\ref{fig:FiniteTBogo}b , 
$\vert \Delta_{{\rm B},0}\vert/\varepsilon_{\rm F}$ is non-zero at lower values of $T/T_{\rm F}$, below the critical line $T_{\rm pair}/T_{\rm F}$, where $\vert \Delta_{{\rm B},0} \vert = 0$.
	
In Fig.~\ref{fig:FiniteTBogo}a, the Hartree order parameter $\vert \Delta_{{\rm H}, 0}\vert$ is largest in the normal fluid phase above $T_{\rm pair}$, since there the Fermi system is fully dominated by particle-hole processes. However, as the temperature is lowered below $T_{\rm pair}$, particle-particle processes start
to dominate, since the interaction energy and the available momentum states are being used to form Cooper pairs leading to a suppression of $\vert \Delta_{{\rm H}, 0} \vert$. These processes eventually force $\vert \Delta_{{\rm H}, 0} \vert$ to vanish at temperatures below $T_{\rm H}$. For fixed $T$ and increasing $1/k_{\rm F} a_{\rm s}$, we notice that $\vert \Delta_{{\rm H}, 0} \vert$ decreases, becoming zero when $T_{\rm H}$ is crossed.
	
In Fig.~\ref{fig:FiniteTBogo}b, the superfluid order parameter $|\Delta_{{\rm B},0}|$ is largest at lowest temperatures, emerging below the Cooper pair formation temperature $T_{\rm pair}$, seen as the border of the dark-blue region. The phase between 
$T_{\rm pair}$ and $T_{\rm H}$, which we name
the Hartree Superfluid (HSF) is characterized by 
$\vert \Delta_{{\rm B},0} \vert \ne 0$ and $\vert \Delta_{{\rm H}, 0} \vert \ne 0$. while the phase
below $T_{\rm H}$, which we call Standard Superfluid (SSF), is characterized by 
$\vert \Delta_{{\rm B},0} \vert \ne 0$ and $\vert \Delta_{{\rm H},0} \vert = 0$. For fixed temperature $T$ and increasing $1/k_{\rm F} a_{\rm s}$, we notice that $\Delta_{{\rm B},0}$ increases, as particle-particle (pairing) correlations dominate below $T_{\rm H}$. 

In Fig.~\ref{fig:FiniteTHart}, we show density plots
of the weight factors $h_0$ and $b_0$ from Eqs.~(\ref{eqn:h0}) and~(\ref{eqn:b0}) in the $T/T_{\rm F}$ versus $1/k_{\rm F} a_{\rm s}$ plane. 
The temperature $T_{\rm pair}$ corresponds to the sharp edge of red (blue) region in Fig.~\ref{fig:FiniteTHart}a (Fig.~\ref{fig:FiniteTHart}b) and the temperature $T_{\rm H}$ corresponds
to the sharp edge of the red (blue) region in  Fig.~\ref{fig:FiniteTHart}b
(Fig.~\ref{fig:FiniteTHart}a).
These two plots show that the normal fluid is fully dominated by particle-hole processes above $T_{\rm pair}$, that particle-hole and particle-particle (pairing) processes compete between $T_{\rm pair}$
and $T_{\rm H}$, and that particle-particle (pairing processes) dominate below $T_{\rm H}$. These results, at the saddle point level, improve on the standard BCS theory describing only the pairing (Bogoliubov) channel~\cite{Bardeen-1957}, and provides
particle-hole corrections to saddle point results 
that are not considered in the Gorkov-Melik-Bakhudarov (GMB) theory~\cite{Gorkov-1961}, as discussed below.
	\begin{figure}[tb]
		\centering
		\includegraphics[width=.95\linewidth]{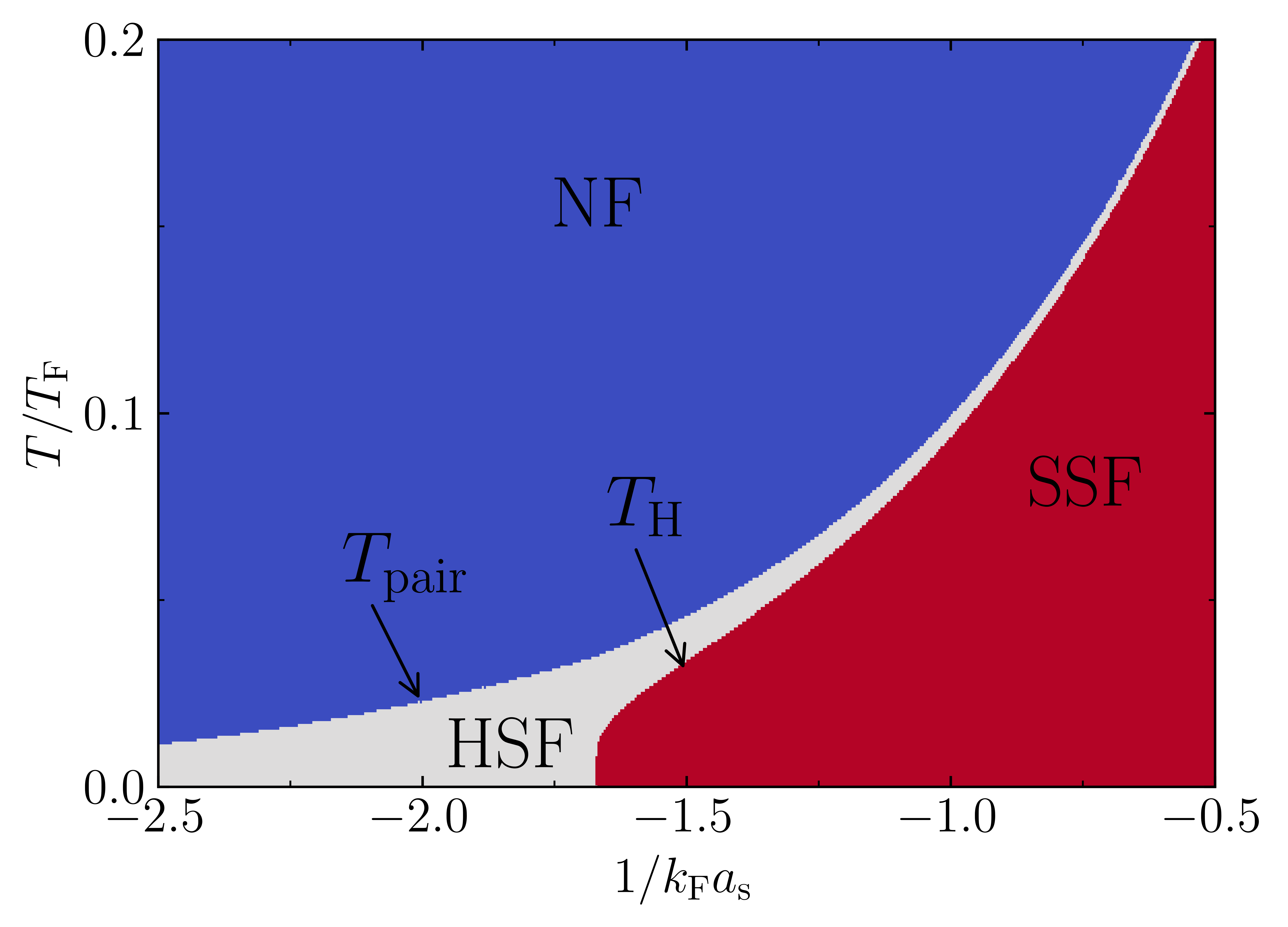}
        \caption{Phase diagram in the 
        temperature $T/T_{\rm F}$ versus scattering parameter $1/k_{\rm F}a_{\rm s}$ plane for the effective range parameter $k_{\rm F}r_{\rm e} = 0.1535$. Three separate phases emerge: Normal fluid (NF) (blue region), Hartree superfluid (HSF) (grey region) and standard superfluid (SSF) (red region). The pairing
        temperature $T_{\rm pair}$ and the Hartree temperature $T_{\rm H}$ are also indicated.}
		\label{fig:PhaseDiag}
	\end{figure}

In Fig.~\ref{fig:PhaseDiag}, we use the information contained in Figs.~\ref{fig:FiniteTBogo} and~\ref{fig:FiniteTHart} to determine the phase diagram shown in the $T/T_{\rm F}$ versus $1/k_{\rm F}a_{\rm s}$ plane. The temperatures $T_{\rm pair}$ and $T_{\rm H}$ versus $1/k_{\rm F}a_{\rm s}$ are indicated and the different saddle-point phases are color coded. 
The normal fluid (NF) phase at the upper left (blue) region is characterized by $\vert \Delta_{{\rm B} ,0} \vert = 0$ $(b_0 = 0)$ and non-vanishing 
$\vert \Delta_{{\rm H},0} \vert$ $(h_0 = 1)$. 
The standard superfluid (SSF) phase at the lower right (red) region is characterized by  
$\vert \Delta_{{\rm B},0} \vert \ne 0$  $(b_0 = 1)$ and 
$\vert \Delta_{{\rm H},0} \vert = 0$ $(h_0 = 0)$. 
The Hartree superfluid (HSF) phase shown in the gray
region has $\vert \Delta_{{\rm B},0} \vert \ne 0$  $(0 < b_0 < 1)$ and $\vert \Delta_{{\rm H},0} \vert \ne  0$ $(0 < h_0 < 1)$. The emergence of the HSF phase is a direct consequence of the partitioning of the interaction, which avoids the miscounting of states and fixes the unphysical divergences of the Hartree energy, chemical potential, 
and the ground state energy. 

\begin{figure}[tb]
\centering
\includegraphics[width=.95\linewidth]{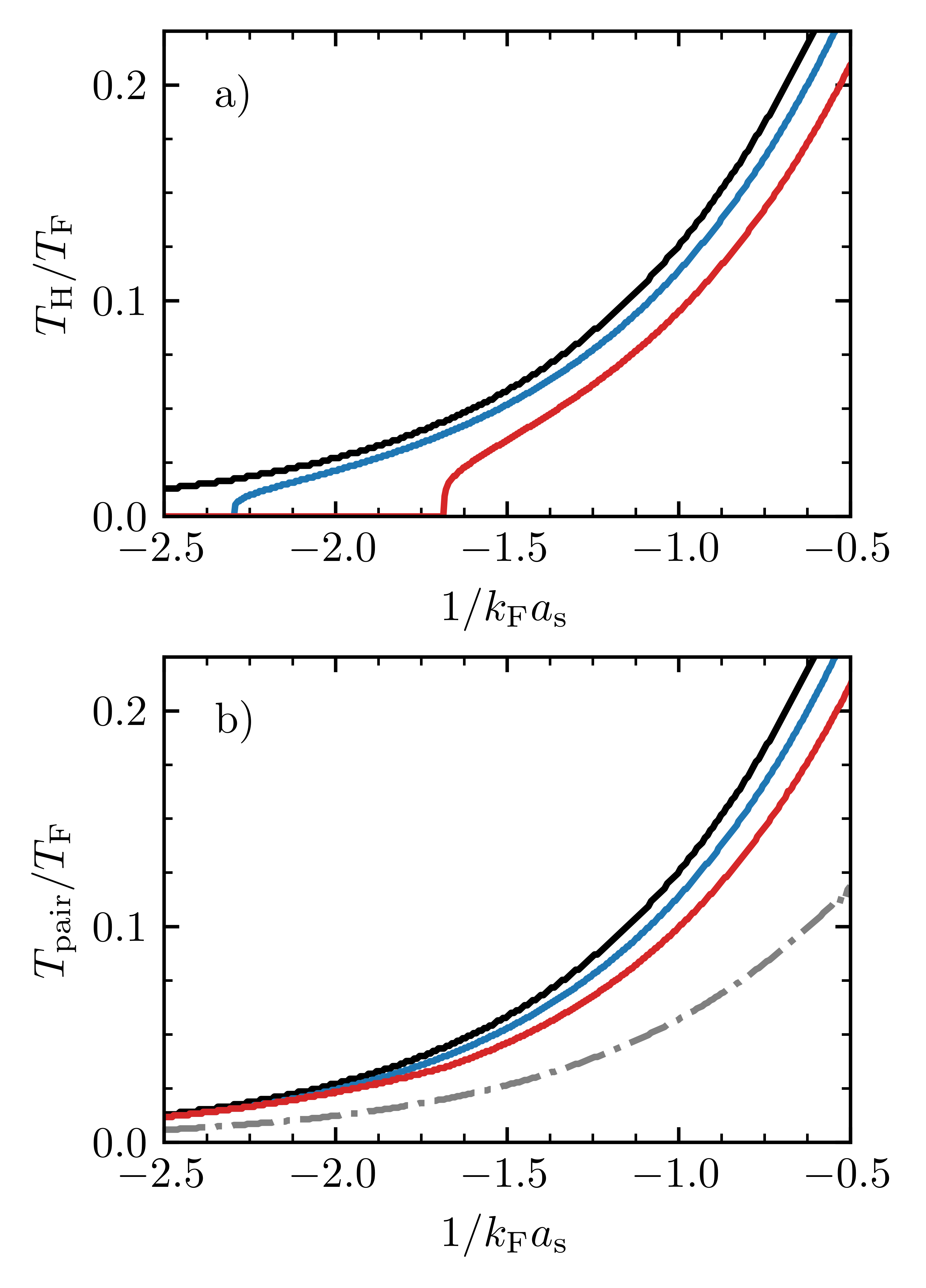}
\caption{Plots of the Hartree temperature $T_{\rm H}/ T_{\rm F}$ and pairing temperature $T_{\rm pair}/T_{\rm F}$ versus scattering parameter $1/k_{\rm F}a_{\rm s}$ for effective range parameters $k_{\rm F}r_{\rm e} = 0$ (solid black line), $0.0625$ (solid blue line) and $0.1535$ (solid red line).
Panel a) shows the HSF to SSF transition temperature $T_{\rm H}/T_{\rm F}$, while panel b) shows the NF to HSF pairing temperature $T_{\rm pair}/T_{\rm F}$. The dash-dotted gray line in panel b) shows the result obtained by Gorkov and Melik-Bhakudarov \cite{Gorkov-1961}.}
\label{fig:TC}
\end{figure}

In Fig.~\ref{fig:TC}, we show the temperatures 
$T_{\rm H}/T_{\rm F}$ and $T_{\rm pair}/T_{\rm F}$  versus $1/k_{\rm F} a_{\rm s}$ for various effective range parameters $k_{\rm F} r_{\rm e}$. The parameters used are: $k_{\rm F} r_{\rm e} = 0$
(solid black line), $k_{\rm F} r_{\rm e} = 0.0625$ (solid blue line) and $k_{\rm F} r_{\rm e} = 0.1535$ (solid red line). The general trend in these figures is that both $T_{\rm H}$ and $T_{\rm pair}$ decrease when $k_{\rm F} r_{\rm e}$ increases, that is, when density or the two-body effective range increases.
In Fig.~\ref{fig:TC}a, we reveal that 
$T_{\rm H}/T_{\rm F}$ vanishes below the critical
value of $1/k_{\rm F} a_{\rm s}$ obtained from
Eq.~(\ref{eqn:HSF-SSF-condition}), where the Hartree order parameter $\Delta_{\rm H,0}/\varepsilon_{\rm F}$
goes to zero.
In Fig.~\ref{fig:TC}b, we show that $T_{\rm pair}/
T_{\rm F}$ is reduced when $k_{\rm F} r_{\rm e}$ is increased. Furthermore, we show the correction to
$T_{\rm pair}/T_{\rm F}$ calculated by GMB~\cite{Gorkov-1961} including particle-hole fluctuations (dash-dotted gray line), but without considering the non-perturbative particle-hole corrections investigated in the partitioning method described here. The analysis performed by GMB includes particle-hole fluctuations about the BCS state, but ignores the non-perturbative Hartree-channel contribution at the saddle point. It is interesting to note that our WHFB method already captures non-perturbative corrections to $T_{\rm pair}$ due to particle-hole effects and thus serves as a better starting point for fluctuation calculations including particle-particle and particle-hole channels simultaneously. 

Having investigated finite temperature effects at the saddle-point level, we are ready to present our conclusions next.

\section{Conclusions}
\label{sec:conclusions}
We developed the Weighted Hartree-Fock-Bogoliubov (WHFB) method that can partition a given interaction into competing channels using a weight distribution 
determined by the minimization principle of the corresponding action. As an example of this concept, we investigated ultracold fermions with equal masses, balanced populations, and zero-ranged interactions partitioned in the particle-hole (Hartree) and particle-particle (Bogoliubov) channels. 

We solved a decades-long issue regarding divergences in the particle-hole channel. Using our method, we showed that these divergences 
can be eliminated by the weighted partitioning of the channels and the introduction of a many-body effective range. The partitioning and regularization procedures
have two important consequences. First, they lead to self-consistent relations between the 
Hartree and Bogoliubov order parameters. Second, they allow for the emergence of 
the Hartree superfluid as a new phase, where both the Hartree and Bogoliubov order parameters are non-zero, in contrast to the standard superfluid, where the Hartree order parameter is zero, but the Bogoliubov order parameter is not.

We demonstrated that non-perturbative corrections, due to the Hartree channel, emerge both in the normal state and in the 
Hartree superfluid, even at the saddle-point level, changing the critical temperature and the phase diagram for superfluidity. This finding was missed in the literature and can directly affect the particle-hole fluctuation corrections to the standard BCS pairing theory developed by Gorkov and Melik-Bakhudarov~\cite{Gorkov-1961}.

\section{Outlook}
\label{sec:outlook}
After describing non-perturbative effects in the particle-particle and particle-hole channels at the saddle-point level, it is natural to consider  fluctuations next.
It is well known that particle-particle fluctuations significantly improve the equation of state beyond the saddle-point level~\cite{Engelbrecht-1993}, when describing the full range of interaction strengths from the BCS to the BEC regimes.
Within the functional integral approach, pair fluctuations to the Gaussian order \cite{SadeMelo-1997} coincide with the diagrammatic theory of Nozières and Schmitt-Rink \cite{SchmittRink-1985}. However, the inclusion of non-perturbative particle-hole effects, through the Hartree channel, requires a modified fluctuation theory, around the saddle point, which must go beyond the Gorkov and Melik-Bakhudarov approach \cite{Gorkov-1961}.

Writing the superfluid order parameter as
$\Delta_{{\rm B}}(x) = \Delta_{{\rm B}, 0} + \eta_{\rm B} (x)$, and the Hartree order parameter as
$\Delta_{{\rm H}}(x) = \Delta_{{\rm H}, 0} + \eta_{\rm H} (x)$, where $\eta_{\rm B} (x)$ and $\eta_{\rm H} (x)$
are the fluctuations around the saddle point,
leads to the Gaussian fluctuation action
\begin{eqnarray}
\mathcal{S}_{\rm fluct}[\{\Delta_0\};&&\{\eta\}] = \int \mathrm{d}x\left[ \frac{|\eta_{\rm B}(\mathbf{x})|^2}{g_{\rm B}} + \frac{\eta_{\rm H}(\mathbf{x})^2}{g_{\rm H}} \right] \nonumber \\ && + \frac{1}{2}\int \frac{\text{d}x}{\beta V}\text{tr}\left\{\left[\mathbf{A}_0^{-1}\delta\mathbf{A}(\mathbf{x})\right]^2\right\} \text{.} \label{Eq:Gaussian-Fluctuations}
\end{eqnarray}
Here, the fluctuation matrix 
\begin{equation}
\delta\mathbf{A}(\mathbf{x}) = 
\begin{pmatrix}
\eta_{\rm H}(\mathbf{x}) & \hspace{.30cm}\eta_{\rm B}(\mathbf{x}) \\
\overline{\eta}_{\rm B}(\mathbf{x}) & -\eta_{\rm H}(\mathbf{x})
\end{pmatrix}
\end{equation}
includes both $\eta_{\rm B} (x)$ and $\eta_{\rm H} (x)$, with 
$\mathbf{A}_0^{-1}$ being the inverse propagator matrix of Eq.~(\ref{eqn:inverse-propagator-matrix}). 
The effects of simultaneous particle-particle and particle-hole fluctuations 
on the phase diagram, thermodynamic and collective mode properties will be described in a forthcoming publication.
Specifically, the question of whether the phase transition between Hartree and Standard superfluid phases survives the effects of fluctuations will be addressed. In addition, extensions of this theory for population and/or mass imbalanced systems will be considered in future work, as the main purpose of this paper was to introduce the partitioning and regularization method when two competing channels arise from the same interaction.

\section*{Acknowledgments}
We thank Corinna Kollath, Joshua Krauß, Marcos Alberto Gonçalves dos Santos Filho, Flavia Braga Ramos, Richard Schmidt and Sejung Yong for inspiring discussions.
We acknowledge financial support by the Deutsche Forschungsgemeinschaft (DFG, German
Research Foundation) via the Collaborative Research Center SFB/TR185 with Project No.\
277625399, which includes a Mercator Fellowship for one of us (C.A.R.S.d.M.).

\appendix

\section{Appendix: Calculation of Pair Size}
\label{Chapter:Pair-size-appendix}
	
In this appendix, we discuss the details on how to calculate the size of the Cooper pairs analytically at $T=0$. Our starting point is the definition of the pair size in momentum space, given in Refs.~\cite{SadeMelo-1997,Strinati-1994} as
	\begin{equation}\label{eqn:Cooper-pair-size}
		\xi_{\rm Pair}^2 = -\frac{\langle \Phi_{\mathbf{k}}| \boldsymbol{\nabla}_{\mathbf{k}}^2|\Phi_\mathbf{k}\rangle}{\langle\Phi_{\mathbf{k}}|\Phi_{\mathbf{k}}\rangle} = -\frac{\sum_{\mathbf{k}}\overline{\Phi}_{\mathbf{k}}\boldsymbol{\nabla}_{\mathbf{k}}^2\Phi_{\mathbf{k}}}{\sum_{\mathbf{k}}|\Phi_{\mathbf{k}}|^2},
	\end{equation}
	where we use the Cooper-pair wave function
	\begin{equation}
		\Phi_{\mathbf{k}} = \frac{\Delta_{\rm B,0}}{2E_{\mathbf{k}}}.
	\end{equation}
Here, $E_\mathbf{k}$ is the Bogoliubov dispersion as given in Eq.~(\ref{eqn:Bogoliubov-dispersion}). To simplify the notation, we define $\mu_{\rm H,0} = \mu - \Delta_{\rm H,0}$ as the Hartree-shifted chemical potential. To obtain $\xi_{\rm Pair}^2$, we need to calculate two summations 
	\begin{subequations}
		\begin{align}
			\langle \Phi_{\mathbf{k}}| \boldsymbol{\nabla}_{\mathbf{k}}^2|\Phi_\mathbf{k}\rangle & = \sum_{\mathbf{k}}\overline{\Phi}_{\mathbf{k}}\boldsymbol{\nabla}_{\mathbf{k}}^2\Phi_{\mathbf{k}}, \\
			\langle\Phi_{\mathbf{k}}|\Phi_{\mathbf{k}}\rangle & = \sum_{\mathbf{k}}|\Phi_{\mathbf{k}}|^2,
		\end{align}
	\end{subequations}
which are the expectation value of the relative position operator 
and the norm of the for the Cooper-pair wavefunction, respectively. These two expressions are calculated next using the thermodynamic limit 
defined in Section \ref{sec:Self-consistency-equations}.
	
	\subsection{Normalization factor}\label{App:A1}
	
The first and more straight-forward calculation is the evaluation of the norm of $\Phi_{\mathbf{k}}$, which in the thermodynamic limit is given by the integral
\begin{equation}
\langle \Phi_\mathbf{k} | \Phi_\mathbf{k} \rangle = \frac{|\Delta_{\rm B,0}|^2}{4} \int \frac{\mathrm{d}^3k}{\left(\frac{k^2}{2m} - \mu_{\rm H,0}\right)^2 + |\Delta_{\rm B,0}|^2}.
\end{equation}
Given that the dispersion $k^2/2m$ depends only on the modulus of the momentum, we perform the angular integration and reduce 
$\langle \Phi_\mathbf{k} | \Phi_\mathbf{k} \rangle$ to a one-dimensional integral. This is achieved with the
substitution $u^2 = k^2/(2m)$, leading to
\begin{equation}\label{eqn:BCS-state-norm}
\langle \Phi_\mathbf{k} | \Phi_\mathbf{k} \rangle = \int_{-\infty}^\infty \mathrm{d}u \frac{|\Delta_{\rm B,0}|^2\sqrt{2}\pi(m)^{3/2} u^2}{(u^2 - \mu_{\rm H,0})^2 + |\Delta_{\rm B,0}|^2},
\end{equation}
where we used the fact that the integrand is even in the variable $u$,  extended the original integration domain $[0,\infty)$ to 
$\mathbb{R}$, and divided the whole expression by $2$. 

To compute the integral above, we use complex analysis techniques.  First,
we factorize the denominator to get a simple expression in terms of its complex roots $\pm \gamma, \pm \overline{\gamma}$. These roots are obtained by using De Moivre's formula and are represented by 
\begin{equation}
\gamma = |\gamma|e^{i\frac{\theta}{2}}.
\end{equation}
As an example, we discuss below the case for $\mu_{\rm H,0} > 0$. 
Here, the modulus is
\begin{equation}\label{eqn:Gamma_abs}
|\gamma| = (\mu_{\rm H,0}^2 + |\Delta_{\rm B,0}|^2)^{1/4},
\end{equation}
while the phase is 
\begin{equation}\label{eqn:Residue-angle}
\theta = \arctan\left(\frac{|\Delta_{\rm B,0}|}{\mu_{\rm H,0}}\right).
\end{equation}

Using the representation above, the roots of the denominator are given by the set $\mathcal{P} = \{\gamma,-\gamma,\overline{\gamma},-\overline{\gamma}\}$ and the integral becomes
\begin{equation}
		\langle \Phi_\mathbf{k} | \Phi_\mathbf{k} \rangle = \int_{-\infty}^\infty \mathrm{d}z \frac{|\Delta_{\rm B,0}|^2\sqrt{2}\pi(m)^{3/2} z^2}{(z^2-\gamma^2)(z^2-\overline{\gamma}^2)},
\end{equation}
where $z \in \mathbb{C}$ is the complex variable. This procedure 
describes an analytical continuation of $u \in \mathbb{R}$ 
to the complex plane. 

\begin{figure}[t]
		\centering
		\includegraphics[width=.95\linewidth]{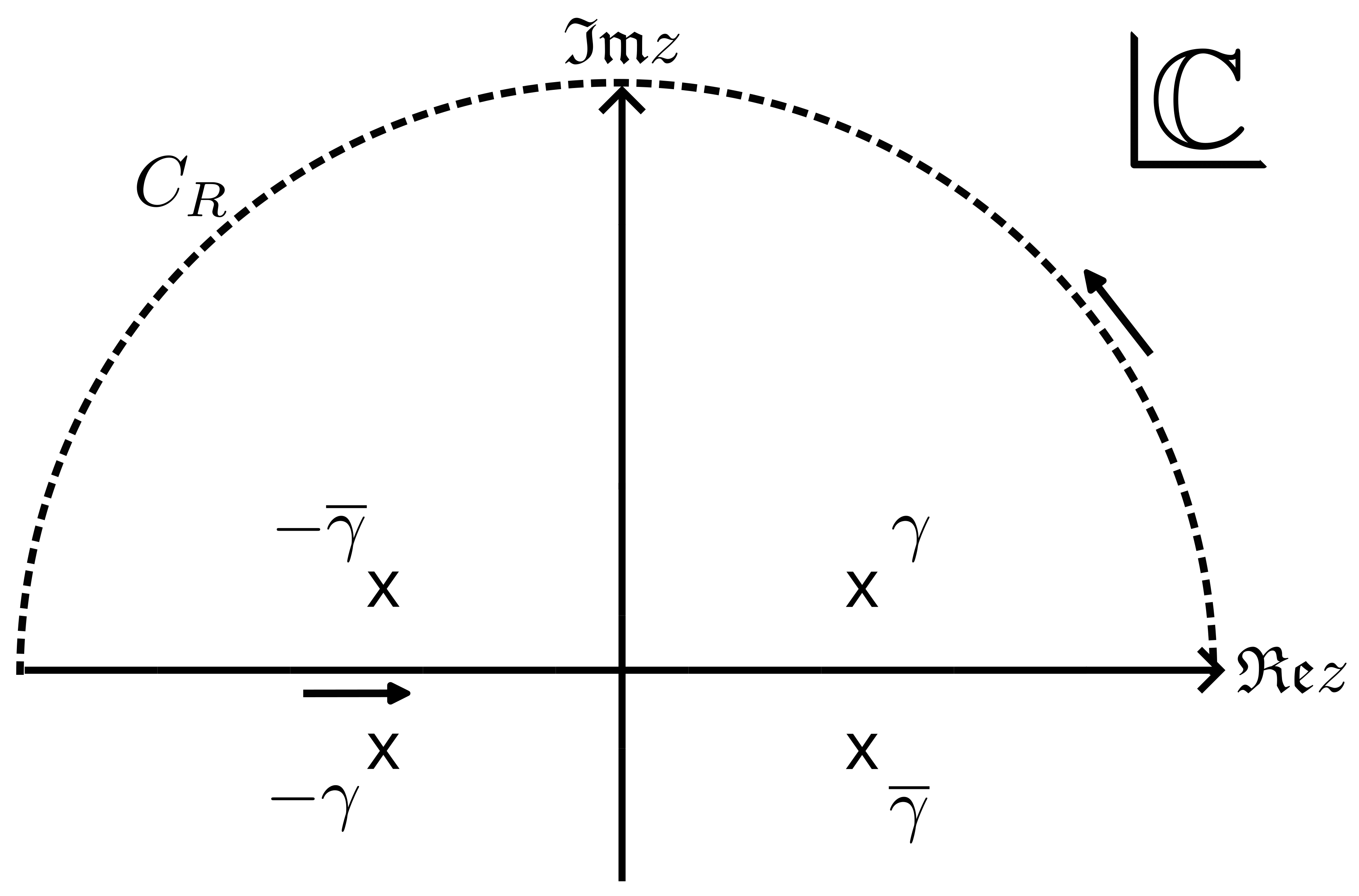}
		\caption{Illustration of the integration contour construction for a given $R > 0$ encircling the poles $\gamma$ and $-\overline{\gamma}$ in the upper half plane.}
		\label{fig:Integration-contour}
\end{figure}

For any function $f:\mathbb{C} \to \mathbb{C}$, 
we define a closed contour $\Gamma_{R} = [-R,R] \cup C_R$, illustrated in Fig.~\ref{fig:Integration-contour}, where $C_R$ is the upper half circle in the complex plane with radius $R > 0$.
The contour integral becomes
\begin{equation}
		\oint_{\Gamma_{R}} \mathrm{d}z f(z) = \int_{-R}^R \mathrm{d}z f(z) + \int_{C_R}\mathrm{d}z f(z).
\end{equation}
In the limit of $R \to \infty$, the integration along the infinite radius circle vanishes by Jordan's Lemma
since $f(z) \propto z^{-2}$, and the contribution from the poles gives
\begin{subequations}
\begin{align}
\int_\mathbb{R} \mathrm{d}z f(z) & = \oint_{\Gamma_{\infty}} \mathrm{d}z f(z)  \\ & = 2i\pi \sum_{z_0 \in \mathcal{P}_+} \mathop{\mathrm{Res}}_{z = z_0}\hspace{.05cm}f(z).
\end{align}
\end{subequations}
The last equality is due to Cauchy's residue theorem \cite{Stalker-1998}, where one can express any closed contour integral of a meromorphic function by the sum of the enclosed residues. The set $\mathcal{P}_+ = \{\gamma,-\overline{\gamma}\}$ includes all poles of the function $f$ with an imaginary part larger than 0, that is, the poles in the upper half plane. Note the use of the index function $1$, because we enclose our contour counter-clockwise. Using 
	\begin{equation}
		f(z) = \frac{z^2}{(z^2 - \gamma^2)(z^2-\overline{\gamma}^2)}
	\end{equation}
leads to the norm of the Cooper-pair wavefunction
\begin{equation}
		\hspace{-.3cm}\langle \Phi_\mathbf{k} | \Phi_\mathbf{k} \rangle = |\Delta_{\rm B,0}|^2\pi^2(2m)^{3/2} i \sum_{z_0 \in \mathcal{P}_+} \mathop{\mathrm{Res}}_{z = z_0}\hspace{.05cm} f(z).  
\end{equation}
Since the function $f (z)$ has only simple poles, calculating the residues is a straight-forward analytical task, leading to
	\begin{subequations}
		\begin{align}
			\mathop{\mathrm{Res}}_{z = \gamma}\hspace{.05cm} f(z) & = \frac{1}{2}\frac{\gamma}{\gamma^2 - \overline{\gamma}^2}, \\
			\mathop{\mathrm{Res}}_{z = -\overline{\gamma}}\hspace{.05cm}f(z) & = -\frac{1}{2}\frac{\overline{\gamma}}{\gamma^2 - \overline{\gamma}^2}.
		\end{align}
	\end{subequations}
By using the modulus and phase representation of the poles, we obtain
	\begin{equation}
		\langle \Phi_\mathbf{k} | \Phi_\mathbf{k} \rangle = |\Delta_{\rm B,0}|^2\sqrt{2}\pi(m)^{3/2} 2i\pi \frac{1}{4i\vert \gamma \vert\sin(\theta/2)}
	\end{equation}
for $\mu_{\rm H,0} > 0$.
As a final step, we use the expression of the phase from Eq.~(\ref{eqn:Residue-angle}) and apply trigonometric identities to evaluate $\sin (\theta/2)$ giving the final result
\begin{equation}\label{eqn:BCS-state-width}
\langle \Phi_\mathbf{k} | \Phi_\mathbf{k} \rangle = 2\pi^2\left(\frac{m}{2}\right)^{3/2}|\Delta_{\rm B,0}|\sqrt{\mu_{\rm H,0} + \vert\gamma\vert^2}.
\end{equation}
For $\mu_{\rm H,0} < 0$, the poles are rotated by a factor of 
$e^{-i\pi/2}$. As one pole is rotated out of the contour its conjugate is rotated into the contour, giving rise to an additional minus sign that is canceled by a minus sign in the residues, yielding the same result.
Note that $\langle \Phi_\mathbf{k} | \Phi_\mathbf{k} \rangle$ is 
always positive and that the argument inside the square root is always
positive, since $\vert \gamma \vert^2 > \mu_{\rm H,0}$.

\subsection{Relative position expectation value}
	
In contrast to the norm of the Cooper-pair wavefunction, the expectation value of the relative position operator involves a second spatial derivative and is more demanding to calculate. By partial integration, the surface term converges to zero upon integration over infinite three-dimensional momentum space, leading to
\begin{subequations}
\begin{align}
\langle \Phi_{\mathbf{k}}| \boldsymbol{\nabla}_{\mathbf{k}}^2|\Phi_\mathbf{k}\rangle & = \int \mathrm{d}^3k \overline{\Phi}_\mathbf{k}\boldsymbol{\nabla}_\mathbf{k}^2\Phi_\mathbf{k} \\
& = - \int \mathrm{d}^3k |\boldsymbol{\nabla}_\mathbf{k}\Phi_\mathbf{k}|^2. \label{eqn:Position_space_integral}
\end{align}
\end{subequations}
Using again spherical symmetry, the angular derivatives in the gradient vanish and we simplify the integral in Eq.~(\ref{eqn:Position_space_integral}) 
by considering only the radial derivative. 
Substituting again $u^2 = k^2/(2m)$ gives 
\begin{equation}
\langle \Phi_{\mathbf{k}}| \boldsymbol{\nabla}_{\mathbf{k}}^2|\Phi_\mathbf{k}\rangle = -2\sqrt{2m}\pi|\Delta_{\rm B,0}|^2 \int_{-\infty}^\infty du g(u),
\end{equation}
where the integrand is
\begin{equation}
g(u) = \frac{u^4(u^2 - \mu_{\rm H,0})^2}{[(u^2 - \mu_{\rm H,0})^2 + |\Delta_{\rm B,0}|^2]^3}.
\end{equation}
Notice that the denominator of $g (u)$ is the third power of the denominator in Eq.~(\ref{eqn:BCS-state-norm}) and, thus, has the same roots. However, in this case, this leads to third-order poles rather than simple poles. 

Through a similar procedure, as that outlined earlier, we replace the integral along the real line by an integral along the closed contour 
$\Gamma_R$ shown in Fig.~\ref{fig:Integration-contour}. Since 
the function $g(z) \propto z^{-6}$ when complex $z$ goes to infinity, 
we apply again Jordan's Lemma and Cauchy's residue theorem to calculate the
integral. The result is
\begin{equation}
\langle \Phi_{\mathbf{k}}| \boldsymbol{\nabla}_{\mathbf{k}}^2|\Phi_\mathbf{k}\rangle = -4i\pi^2\sqrt{2m}|\Delta_{\rm B,0}|^2 \sum_{z_0 \in \mathcal{P}_+} \mathop{\mathrm{Res}}_{z = z_0}\hspace{.05cm} g(z).
\end{equation}
The poles follow the same pattern as before with $\mathcal{P}_+ = \{\gamma,-\overline{\gamma}\}$ being the ones in the upper half plane. Because these are third-order poles, extra are is necessary. 
For an $\mathrm{n}^{\rm th}$-order pole of the function $g$ at $z_0 \in \mathbb{C}$, the residue is
\begin{equation}
\mathop{\mathrm{Res}}_{z = z_0}\hspace{.05cm} g(z) = \frac{1}{(n-1)!} \lim\limits_{z \to z_0}\frac{\mathrm{d}^{n-1}}{dz^{n-1}}\left[(z-z_0)^n g(z)\right].
\end{equation}
Using the expression above, the calculation of the residues reduces to taking derivatives and then evaluating the limit. As the expression for individual residues are quite long and give no physical insight,  
we do not write them down here. However, the sum of the residues of 
the two relevant poles has a simpler and shorter structure due to the symmetry $\gamma \longleftrightarrow -\overline{\gamma}$ in the residues.
For $\mu_{\rm H,0} > 0$, this analysis leads to
\begin{widetext}
\begin{align}
\sum_{z_0 =\gamma,-\overline{\gamma}} \mathop{\mathrm{Res}}_{z = z_0}\hspace{.05cm} g(z) = - \frac{|\gamma|^2(\gamma^2 + \overline{\gamma}^2 - 5|\gamma|^2) + 6|\gamma|^2\mu_{\rm H,0} - 3\mu_{\rm H,0}^2}{16|\gamma|^2(\gamma - \overline{\gamma})^5}.
\end{align}
Using the modulus and phase representation of $\gamma$ and $\overline{\gamma}$, we write $\gamma^2 + \overline{\gamma}^2 = 2\vert\gamma\vert^2 \cos(\theta)$ and $\gamma - \overline{\gamma} = 2i\vert \gamma\vert\sin(\theta/2)$, and use Eq.~(\ref{eqn:Residue-angle}) 
to eliminate the phase $\theta$ giving 
\begin{equation}
\sum_{z_0 =\gamma,-\overline{\gamma}} \mathop{\mathrm{Res}}_{z = z_0}\hspace{.05cm} g(z) = \frac{\sqrt{2}(\mu_{\rm H,0} + \vert\gamma\vert^2)^{5/2}}{128i\vert\gamma\vert^2|\Delta_{\rm B,0}|^5}(\vert\gamma\vert^2 - \mu_{\rm H,0})(5\vert\gamma\vert^2 - 3\mu_{\rm H,0}).
\end{equation}
Notice that the expression above is always a positive number divided by the imaginary unit $i$, since $\vert\gamma\vert^2 \ge \mu_{\rm H,0}$
as seen in Eq.~(\ref{eqn:Gamma_abs}).
The final result is then
\begin{equation}\label{eqn:Last}
\langle \Phi_{\mathbf{k}}| \boldsymbol{\nabla}_{\mathbf{k}}^2|\Phi_\mathbf{k}\rangle = -\frac{\pi^2}{16}\sqrt{m} \frac{(\mu_{\rm H,0} + \vert\gamma\vert^2)^{5/2}(\vert\gamma\vert^2 - \mu_{\rm H,0})(5\vert\gamma\vert^2 - 3\mu_{\rm H,0})}{\vert\gamma\vert^2 |\Delta_{\rm B,0}|^3},
\end{equation}
which is always negative. We mention in passing that the same result
is obtained for $\mu_{\rm H,0} < 0$.

Lastly, we use Eqs.~(\ref{eqn:BCS-state-width}) and~(\ref{eqn:Last}) to write the square of the Copper pair size 
$\xi_{\rm pair}^2$ given in Eq.~(\ref{eqn:Cooper-pair-size}). The expression obtained for $\xi_{\rm pair}^2$ is always positive, as expected, and the result for 
the dimensionless Cooper pair size $k_{\rm F}\xi_{\rm pair}$
is given in Eq.~(\ref{eqn:BCS-pair-size}) of the main text. 
\end{widetext}

\bibliographystyle{apsrev4-1custom}

\end{document}